\documentclass[smallextended]{svjour3}

\smartqed

\newcommand{\red}[1]{\textcolor{red}{#1}} 
 
\usepackage{mathdots}
\usepackage[switch]{lineno}
\usepackage{graphicx} 
\usepackage{booktabs}
\usepackage{mathdots}
\usepackage{amsmath}
\usepackage{amssymb}

\usepackage[misc]{ifsym}
\usepackage{url} 
\usepackage[T1]{fontenc}

\usepackage{ulem}

\RequirePackage{fix-cm}
\usepackage[section]{placeins}

\smartqed

\usepackage{color}

\usepackage{graphicx}
\usepackage[misc]{ifsym}

\usepackage{caption}
\usepackage{ulem}

\begin{document}

\title{CATCH: Chasing All Transients Constellation Hunters Space Mission}

\author{Panping Li\textsuperscript{1, 2} \and  Qian-Qing Yin\textsuperscript{1} \and  Zhengwei Li\textsuperscript{1} \and  Lian Tao\textsuperscript{1} \and  Xiangyang Wen\textsuperscript{1} \and  Shuang-Nan Zhang\textsuperscript{1,2} \and  Liqiang Qi\textsuperscript{1} \and  Juan Zhang\textsuperscript{1} \and Donghua Zhao\textsuperscript{3} \and  Dalin Li\textsuperscript{4} \and  Xizheng Yu\textsuperscript{4} \and  Qingcui Bu\textsuperscript{5} \and  Wen Chen\textsuperscript{6} \and  Yupeng Chen\textsuperscript{1} \and  Yiming Huang\textsuperscript{1,2} \and  Yue Huang\textsuperscript{1} \and  Ge Jin\textsuperscript{7} \and  Gang Li\textsuperscript{1} \and  Hongbang Liu\textsuperscript{8} \and Xiaojing Liu\textsuperscript{1} \and  Ruican Ma\textsuperscript{1,2} \and  Wenxi Peng\textsuperscript{1} \and Ruijing Tang\textsuperscript{9} \and  Yusa Wang\textsuperscript{1} \and  Jingyu Xiao\textsuperscript{1,2} \and  Shaolin Xiong\textsuperscript{1} \and  Sheng Yang\textsuperscript{1} \and  Yanji Yang\textsuperscript{1} \and  Chen Zhang\textsuperscript{3} \and  Tianchong Zhang\textsuperscript{1} \and Liang Zhang\textsuperscript{1} \and  Xuan Zhang\textsuperscript{1,10} \and  Haisheng Zhao\textsuperscript{1} \and Kang Zhao\textsuperscript{1,2} \and  Qingchang Zhao\textsuperscript{1,2} \and  Shujie Zhao\textsuperscript{1,2} \and  Xing Zhou\textsuperscript{1,2}}

\institute{\Letter{Lian Tao} \at
          \email{taolian@ihep.ac.cn}
          \\
          \Letter Xiangyang Wen \at 
          \email{wenxy@ihep.ac.cn}
          \\
          \and
           Panping Li \and Shuang-Nan Zhang \and Yiming Huang \and  Ruican Ma \and Jingyu Xiao \and  Kang Zhao \and  Qingchang Zhao \and Shujie Zhao \and Xing Zhou \at
           Key Laboratory for Particle Astrophysics, Institute of High Energy Physics, Chinese Academy of Sciences, Beijing 100049, China \\
           University of Chinese Academy of Sciences, Chinese Academy of Sciences, Beijing 100049, China 
          \and 
          Qian-Qing Yin \and Zhengwei Li \and Lian Tao \and Xiangyang Wen \and Liqiang Qi \and Juan Zhang \and Yupeng Chen \and Yue Huang \and  Gang Li \and Xiaojing Liu \and Wenxi Peng \and Yusa Wang \and Shaolin Xiong \and Sheng Yang \and Yanji Yang \and Tianchong Zhang \and Liang zhang\and Haisheng Zhao \at
           Key Laboratory for Particle Astrophysics, Institute of High Energy Physics, Chinese Academy of Sciences, Beijing 100049, China 
          \and
           Donghua Zhao \and Chen Zhang    \at
           National Astronomical Observatories, Chinese Academy of Sciences, Beijing 100101, China 
          \and 
           Dalin Li \and Xizheng Yu  \at
           Key Laboratory of Electronics and Information Technology for Space Systems, National Space Science Center, Chinese Academy of Sciences, Beijing 100190, China
          \and
           Qingcui Bu \at 
           Institut f\"ur Astronomie und Astrophysik, Kepler Center for Astro and Particle Physics, Eberhard Karls Universit\"at, Sand 1, 72076 T\"ubingen, Germany
          \and 
           Wen Chen   \at
           Innovation Academy for Microsatellites of Chinese Academy of Sciences, Shanghai 200135, China 
          \and 
           Ge Jin  \at
           North Night Vision Science \& Technology Research Institute Group Co., Ltd, Nanjing 211100, China 
          \and 
           Hongbang Liu \at 
           Guangxi University, Nanning 530004, China
          \and 
           Ruijing Tang \at 
          Beijing Jiaotong University, Beijing 100044, China
          \and 
           Xuan Zhang \at 
           Key Laboratory for Particle Astrophysics, Institute of High Energy Physics, Chinese Academy of Sciences, Beijing 100049, China\\
           Nanchang University, Nanchang 330031, china
        }


\authorrunning{Panping Li et al.}
\maketitle

\begin{abstract}
In time-domain astronomy, a substantial number of transients will be discovered by multi-wavelength and multi-messenger observatories, posing a great challenge for follow-up capabilities. We have thus proposed an intelligent X-ray constellation, the Chasing All Transients Constellation Hunters (\textit{CATCH}) space mission. Consisting of 126 micro-satellites in three types, \textit{CATCH} will have the capability to perform follow-up observations for a large number of different types of transients simultaneously. Each satellite in the constellation will carry lightweight X-ray optics and use a deployable mast to increase the focal length. The combination of different optics and detector systems enables different types of satellites to have multiform observation capabilities, including timing, spectroscopy, imaging, and polarization. Controlled by the intelligent system, different satellites can cooperate to perform uninterrupted monitoring, all-sky follow-up observations, and scanning observations with a flexible field of view (FOV) and multi-dimensional observations. Therefore, \textit{CATCH} will be a powerful mission to study the dynamic universe. Here, we present the current design of the spacecraft, optics, detector system, constellation configuration and observing modes, as well as the development plan.
\end{abstract}

\section{Introduction}
\label{sec:intro}

Time-domain astronomy, focusing on the evolution and changes of cosmic sources, is expected to bring great breakthroughs in modern astronomy. Multi-wavelength observing campaigns have thus aimed to dynamically picture the universe and study the physical processes underlying the evolution and changes of cosmic sources. With currently operating and planned survey telescopes, our catalog of transients is growing rapidly. The annual number of transient reports in the Transient Name Server (TNS)\footnote{\url{https://www.wis-tns.org/}} has increased from 7,546 in 2016 to 38,054 in 2021. In the future, it is expected that tens of thousands of transients will be discovered every day. For example, the Vera C. Rubin Observatory (also known as \textit{LSST}) is predicted to produce tens of millions of optical detection alerts every night \cite{kantor2014}, and only for Type Ia supernovae, the annual detections will be as high as 200,000 cases \cite{lsst2009}; the Square Kilometre Array (\textit{SKA}) will have an extremely high sensitivity, good angular resolution, and a large field-of-view (FOV) to survey the radio sky, and in the first phase of SKA-Mid (i.e. SKA1-mid), its weekly detection rate, for fast radio bursts (FRBs) and tidal disruption events (TDEs) only, will be 1--1000 cases \cite{fender2015}; the Einstein Probe (\textit{EP}) will detect 10--20 fast X-ray transients per day \cite{yuan2018} and will greatly enrich the X-ray transient family. In addition, multi-messenger (e.g. gravitational-wave and neutrino) observatories have also set their sights on detecting more gravitational wave and neutrino events. Around 2030, it is estimated that more than 10,000 gravitational wave events from binary neutron star mergers will be detected each year \cite{amati2018}.

In view of the large amounts of transients detected, follow-up observations are urgently needed in order to fully understand these transient phenomena. In particular, because transients are usually extreme objects or undergo extreme physical processes, X-ray radiation will typically be generated and will carry a lot of key information, hence the need for X-ray follow-up observations. Unfortunately, compared with other energy bands, X-ray observing resources are very scarce. In the foreseeable future, hopefully a dozen X-ray astronomical satellites will be in orbit (see Table~\ref{tab:1}), and they only focus on: 

\begin{itemize}
\item high sensitivities and large effective areas in order to observe the deep universe and weak sources, e.g. \textit{Chandra} \cite{Weisskopf2002}, \textit{XMM-Newton} \cite{Jansen2001}, Neutron Star Interior Composition ExploreR (\textit{NICER}) \cite{Gendreau2016}, \textit{Insight}-HXMT \cite{Zhang2020}, Advanced Telescope for High ENergy Astrophysics (\textit{ATHENA}) \cite{nandra2013}, enhanced X-ray Timing and Polarimetry mission (\textit{eXTP}) \cite{zhang2019}, Spectroscopic Time-Resolving Observatory for Broadband Energy X-rays (\textit{STROBE-X}) \cite{Ray2019} and \textit{Lynx} \cite{Gaskin2019}.

\item high spatial and spectral resolutions in order to have the most detailed understanding of the cosmos, e.g. \textit{Chandra}, X-Ray Imaging and Spectroscopy Mission (\textit{XRISM}) \cite{XRISM2020,XRISM2022}, Hot Universe Baryon Surveyor (\textit{HUBS}) \cite{Cui2020}, \textit{ATHENA} and \textit{Lynx}.

\item powerful survey capabilities in order to find more transients, e.g.  Neil Gehrels Swift Observatory (\textit{Swift}) \cite{Burrows2005}, Monitor of All-sky X-ray Image (\textit{MAXI}) \cite{Matsuoka2009}, \textit{eROSITA} \cite{Predehl2021}, \textit{EP}, Transient High-Energy Sky and Early Universe Surveyor (\textit{THESEUS}) \cite{amati2018} and Transient Astrophysics Probe (\textit{TAP}) \cite{Camp2019}.

\end{itemize}

None of these X-ray satellites can perform follow-up observations for a great deal of sources simultaneously; even though these observatories cooperate together, they can only track a very few number of transients.
\begin{table}[h]
\begin{center}
\caption{Comparisons between \textit{CATCH} and other X-ray telescopes}\label{tab:1}
\resizebox{.99\columnwidth}{!}{
\begin{tabular}{@{}cccccc@{}}
\toprule
Telescope & Energy Range  & Number of Follow-up Targets &  Launch date& mission status & Ref. \\
      &(keV) &    & (year)  \\
\midrule
\red{\textit{CATCH}} & \red{0.3--10} & \red{100$+$}   & \red{around 2030} & \red{under study} &\red{ ... } \\
\noalign{\smallskip}\hline\noalign{\smallskip}
\textit{Chandra}  & \begin{tabular}[c]{@{}l@{}}ACIS: 0.2--10 \\ HRC: 0.1--10\end{tabular}    & 1  & 1999 & in operation  & \cite{Weisskopf2002}  \\ 
\noalign{\smallskip}\hline\noalign{\smallskip}
\textit{XMM-Newton}     & EPIC: 0.1--12    & 1    & 1999 & in operation & \cite{Jansen2001} \\
\noalign{\smallskip}\hline\noalign{\smallskip}
\textit{Swift}     & \begin{tabular}[c]{@{}l@{}}XRT: 0.2--10 \\ BAT: 15--150 \end{tabular}    & 1     & 2004  & in operation   &  \cite{Burrows2005} \\ 
\noalign{\smallskip}\hline\noalign{\smallskip}
\textit{MAXI}      & \begin{tabular}[c]{@{}l@{}}GSC: 2--30 \\ SSC: 0.5-12\end{tabular}   & 0 (Survey Telescope)  & 2009 & in operation   & \cite{Matsuoka2009} \\
\noalign{\smallskip}\hline\noalign{\smallskip}
\textit{NuSTAR}   & 3--79     & 1    & 2012 &in operation & \cite{Harrison2013} \\ 
\noalign{\smallskip}\hline\noalign{\smallskip}
\textit{Astrosat}     & \begin{tabular}[c]{@{}l@{}l@{}l@{}}SXT: 0.3--8 \\  LAXPC: 3--80 \\ CZTI: 10--100 \\ SSM: 2.5--10 \end{tabular}   & 1     & 2015 &in operation   & \cite{Singh2014} \\  
\noalign{\smallskip}\hline\noalign{\smallskip}
\textit{NICER}     & 0.2--12     & 1        & 2017& in operation  &  \cite{Gendreau2016} \\
\noalign{\smallskip}\hline\noalign{\smallskip}
\textit{Insight}-HXMT   & \begin{tabular}[c]{@{}l@{}l@{}}LE: 1--15 \\ ME: 5--30 \\ HE: 20--250 \end{tabular}     & 1        & 2017 & in operation  & \cite{Zhang2020} \\ 
\noalign{\smallskip}\hline\noalign{\smallskip}
\textit{eROSITA}    & 0.3--10    & 1     & 2019 &in operation  &  \cite{Predehl2021} \\
\noalign{\smallskip}\hline\noalign{\smallskip}
\textit{IXPE}    & 2--8    & 1     & 2021  &in operation &  \cite{Weisskopf2016} \\
\noalign{\smallskip}\hline\noalign{\smallskip}
\textit{SVOM}    & \begin{tabular}[c]{@{}l@{}}ECLAIRs: 4--120 \\ MXT: 0.2--10 \end{tabular}    & 1     & 2023 &approved  &  \cite{Bernardini2021} \\
\noalign{\smallskip}\hline\noalign{\smallskip}
\textit{EP}    & \begin{tabular}[c]{@{}l@{}}WXT: 0.5--4 \\ FXT: 0.3--8 \end{tabular}    & 1   & 2023 &approved  & \cite{yuan2018}  \\
\noalign{\smallskip}\hline\noalign{\smallskip}
\textit{XRISM}    & \begin{tabular}[c]{@{}l@{}}Resolve: 0.3--12 \\ Xtend: 0.4--12 \end{tabular}    & 1   & 2023 &approved  & \cite{XRISM2020,XRISM2022}  \\
\noalign{\smallskip}\hline\noalign{\smallskip}
\textit{eXTP}  & \begin{tabular}[c]{@{}l@{}l@{}l@{}}SFA: 0.5--10 \\ LAD: 2--30 \\ PFA: 2--8 \\ WFM: 2--50 \end{tabular}    & 1         & 2027 &under study &  \cite{zhang2019} \\
\noalign{\smallskip}\hline\noalign{\smallskip}
\textit{TAP}     &  \begin{tabular}[c]{@{}l@{}l@{}l@{}}XRT: 0.2--10 \\ WFI: 0.4--4 \end{tabular}& 1     & 2029& proposed & \cite{Camp2019} \\
\noalign{\smallskip}\hline\noalign{\smallskip}
\textit{HUBS}    & 0.1-2    & 1        & around 2030 & under study & \cite{Cui2020} \\ 
\noalign{\smallskip}\hline\noalign{\smallskip}
\textit{STROBE-X}    & \begin{tabular}[c]{@{}l@{}l@{}}XRCA: 0.2--12 \\  LAD: 2--30 \\ WFM: 2--50 \end{tabular}    & 1        & 2031 & proposed  & \cite{Ray2019} \\ 
\noalign{\smallskip}\hline\noalign{\smallskip}
\textit{ATHENA}      & \begin{tabular}[c]{@{}l@{}}X-IFU: 0.2--12 \\  WFI: 0.2--15\end{tabular}   & 1     & 2035  & approved  & \cite{nandra2013}; \cite{Barret2018}  \\ 
\noalign{\smallskip}\hline\noalign{\smallskip}
\textit{Lynx}     &  0.2--10    & 1     &  ? & proposed & \cite{Gaskin2019} \\
\noalign{\smallskip}\hline\noalign{\smallskip}
\textit{THESEUS}  &   \begin{tabular}[c]{@{}l@{}l@{}}SXI: 0.3--6 \\  XGIS: 2--20000 \end{tabular}     & 1     & ? & proposed & \cite{amati2018}\\
\bottomrule
\end{tabular}
}
\end{center}
\footnotemark{\textit{CATCH} is marked in red.}

\end{table}


There are additional challenges in the field of time-domain astronomy. For instance, most X-ray telescopes, such as \textit{Swift}, \textit{MAXI}, \textit{EP}, and \textit{eXTP} operate in low-earth orbits, and so their observations would necessarily be interrupted when the satellites enter the South Atlantic Anomaly (SAA) or the Earth shadow. Therefore, these space telescopes can only perform observations in part of the orbit, and the observation in every orbit will be split into several short segments; unfortunately, some transient phenomena (e.g. X-ray bursts associated with FRBs), which reveal rich physical mechanisms and bring important scientific discoveries, may be missed.

Moreover, some transient events appear over very short time scales. Although their occurrences are always expected, the exact moment is still unpredictable, e.g. glitches in pulsars \cite{Espinoza2011} and magnetars \cite{kaspi2017}. Thus, long-term continuous monitoring is required in order to catch these fast transient phenomena. However, that will consume a lot of observing resources and is difficult to be done at present.  

In addition, some transient events, such as gravitational-wave and high-energy neutrino events, are localized in a large irregular region. For example, the gravitational-wave event GW170817 is located in an elongated region of 28\,deg$^2$ \cite{Abbott2017}. Previous X-ray narrow FOV telescopes can only cover the sky region with a fixed FOV, and thus need to scan backwards and forwards in order to piece the source region and search for the counterpart, which is inefficient when dealing with fast transients. Although all-sky monitors are capable of covering a large sky region, the relatively low sensitivity will limit the searching depth.   

Further, in order to thoroughly understand the physical mechanisms accounting for transient phenomena, multi-dimensional (imaging, spectroscopy, timing and polarization) observations are demanded. But the current telescopes can not perform simultaneous multi-dimensional detections. For instance, although the Imaging X-ray Polarimetry Explorer (\textit{IXPE}) \cite{Weisskopf2016} has opened a new window in measuring X-ray polarization from cosmic sources, it is not suitable to carry out good spectral and timing measurements. Among the planned missions, only the \textit{eXTP} can do spectral, timing, and polarization measurements, but it will not be able to track a lot of transients simultaneously.

As discussed above, the capabilities of unbiased follow-up samples of all types of transients (including unknown types), uninterrupted and long-term monitoring, flexible FOV and multi-dimensional (i.e. imaging, spectral, timing and polarization) detection will play an important role in the era of time-domain astronomy, but both the currently operating and planned X-tay telescopes do not have or aim at these capabilities. Thus, we propose a new X-ray mission, i.e. the Chasing All Transients Constellation Hunters (\textit{CATCH}) space mission. With the cooperation of over a hundred satellites in the constellation, \textit{CATCH} can chase a lot of transients simultaneously, and perform uninterrupted, long-term monitoring observations. Moreover, it has a flexible FOV to search for multi-messenger counterparts, and is capable to simultaneously perform imaging, spectral, timing, and polarization observations. Thus, \textit{CATCH} will be a powerful mission for our understanding of the dynamic universe. 

In Section~\ref{sec:des} of this paper, we will introduce the current design, including that of the spacecraft, X-ray optics, focal plane detector, constellation configuration and observing modes, as well as details of the observing modes. We will then present a summary in Section~\ref{sec:sum}.

\section{X-ray constellation}
\label{sec:des}

Considering the number of transients reported in the TNS and the demand for multi-dimensional follow-up observations (more details will be presented in a following paper), \textit{CATCH}, as it is currently designed, will consist of 126 satellites in three types (Table~\ref{tab:types}), integrated into two types of spacecrafts (Table~\ref{tab:spacecraft}). Type-A satellites are designed to perform timing monitoring for a substantial fraction of all types of transients. Based on their measurements, type-B satellites will do further imaging, spectral and timing observations, and type-C satellites will obtain the measurements of X-ray polarization.

Type-A satellites will use type-I spacecraft, and type-B and C satellites will adopt type-II spacecraft (see Section~\ref{sec:satellite}). For type-A satellites, the Micro Pore Optics (MPO) will be used as the optics systems, while for type-B and C satellites, the X-ray optics is developed based on the conventional Wolter-I geometry, but is more lightweight (see Section~\ref{sec:opt}). As described in Section~\ref{sec:det}, the focal plane module of type-A satellites consists of a 4-pixel Silicon Drift Detector (SDD) array, in order to better estimate the background. A large SDD array or pn-charge coupled device (pn-CCD) will be used as the focal plane detector of type-B satellites, and the Gas Micro Plate Detector (GMPD) or Gas Pixel Detector (GPD) system will be used for X-ray polarization measurements on-board type-C satellites. The satellites in the constellation will be deployed into three near-Earth orbits with low inclination (Section~\ref{sec:orb}). Under control of the intelligent control system, different satellites are able to cooperate to carry out follow-up observations in four modes (Section~\ref{sec:obs_mode}). 
\begin{figure}[h]
\centering
\includegraphics[width=\textwidth]{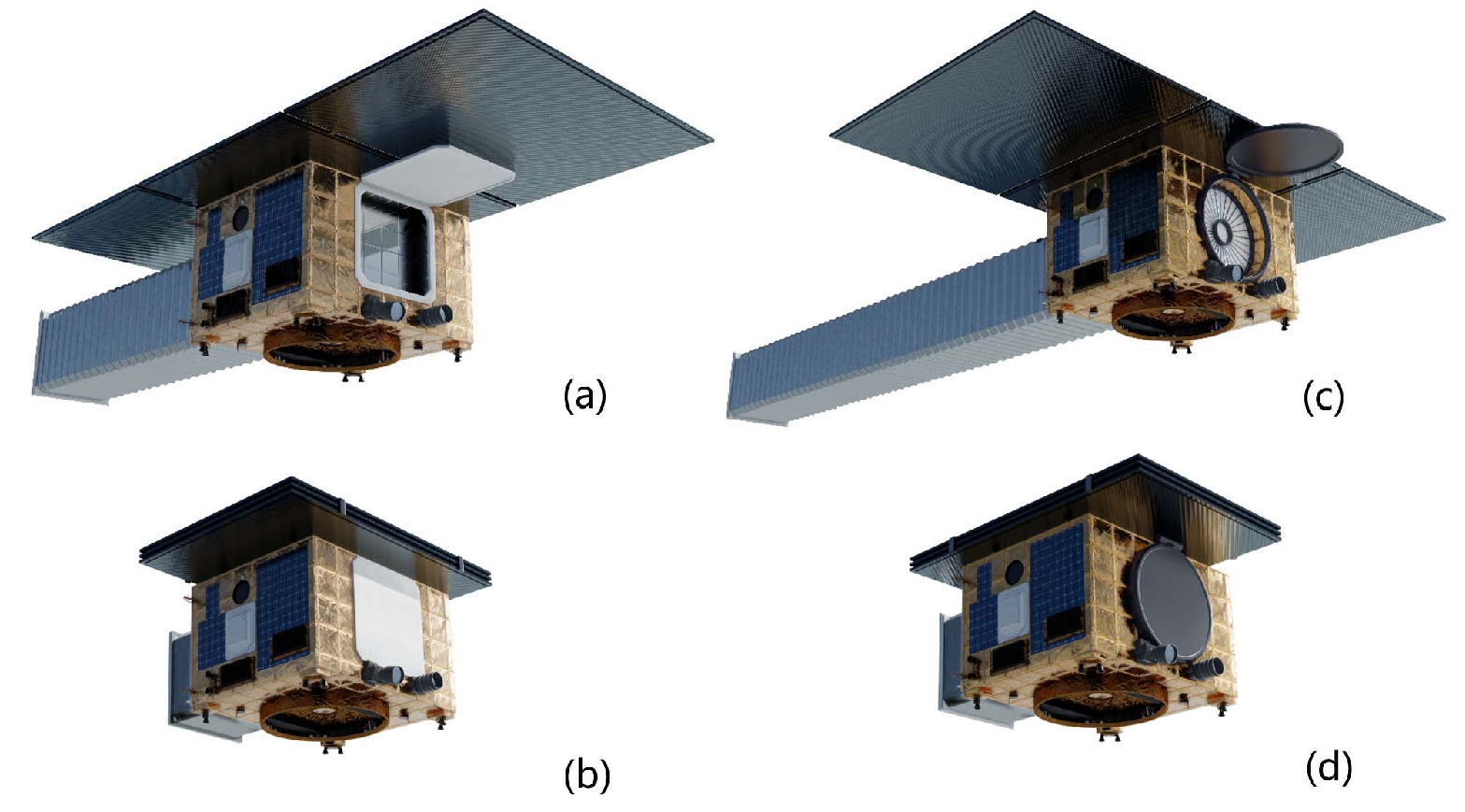} 
\caption{Diagrams of satellites of different types. Left panels: the deployed (top) and stowed (bottom) configurations of type-A satellites. Right panels: the deployed (top) and stowed (bottom) configurations of type-B and C satellites.
\label{FIG:1}}
\end{figure}
\begin{table}
\begin{center}
\caption{Parameters of three types of satellites in \textit{CATCH}}
\label{tab:types}
\resizebox{.99\columnwidth}{!}{
\begin{tabular}{@{}lcccc@{}}
\toprule
Parameter & Type-A  & Type-B & Type-C \\
\midrule
Scientific capability  &  Timing &  Imaging, spectral, timing & Polarization \\
X-ray optics  & MPO  & Wolter-I & Wolter-I \\
Sensitive Energy (keV)   & 0.5--4  & 0.3--10 & 2--8 \\ 
Detector   & 4-pixel SDD array  & multi-pixel SDD array/pn-CCD & GMPD/GPD \\ 
Spacecraft & type I & type II & type II \\
Total weight (kg) & 25 & 35 & 40 \\
No. of satellites  &  108 & 9 & 9 \\
\noalign{\smallskip}\hline\noalign{\smallskip}
\multicolumn{4}{c}{X-ray optics} \\
\noalign{\smallskip}\hline\noalign{\smallskip}
Focal Length (cm)   & 100  & 200 & 300 \\ 
FOV   & $1^{\circ} \times 1^{\circ}$  & $0.5^{\circ} \times 0.5^{\circ}$ & $0.25^{\circ} \times 0.25^{\circ}$ \\ 
Diameter (cm)   & 20 & 20 & 25 \\ 
Effective area (cm$^2$)   & $>40$@1\,keV  & $>120$@1\,keV & $>120$@4.5\,keV \\ 
Angular resolution ($^\prime$)   & 5  & 2 & 1 \\
Optics Mass (kg) & 1 & 5 & 7 \\
\noalign{\smallskip}\hline\noalign{\smallskip}
\multicolumn{4}{c}{Focal plane detector} \\
\noalign{\smallskip}\hline\noalign{\smallskip}
Energy resolution   & 160\,eV@4\,keV & 160\,eV@4\,keV & ...  \\ 
Timing resolution   & $\leq 10\,\mu$s  & $\leq 10\,\mu$s & ...  \\ 
Timing accuracy & $\leq 2\,\mu$s & $\leq 2\,\mu$s  & ... \\
Sensitive area (mm$^2$) & central/total: 50/110 & 303  &  414 \\
pixel size & ...  & 0.75\,mm hexagon & ...  \\
pixel number & 4  & 256 & ...  \\ 
Position resolution &...& ...& $\leq 100\,\mu m$ \\ 
Modulation factor &...& ...& $\geq 0.4$@4.5\,keV \\ 
Detection efficiency &... & ...&$\geq 10\%$@4.0\,keV \\
Date Rate (GB/day) & 1.0 & 0.8 & 2.0 \\
Power (W) & 6.0 &7.5 & 12.0 \\
Detector Mass (kg)& 1.8 & 3.0 & 4.5 \\
\bottomrule
\end{tabular}
}
\end{center}
\footnotemark{The angular resolution is the full width at half maximum (FWHM) of the point spread function (PSF) of the optics. The sensitive area is the detector area that is sensitive to the incidence X-rays. The length of Wolter-I optics is 400\,mm, and some optimizations will be made in the future. For type-B and C} satellites, the parameters of the preferred detector system (i.e. multi-pixel SDD array and GMPD, respectively) are given. 
\end{table}


\subsection{Spacecraft} 
\label{sec:satellite}

\textit{CATCH} will adopt three-axis stabilized small spacecrafts in two types (Table~\ref{tab:spacecraft}), and both types have low mass and cost. Type-I spacecrafts are more compact with a mass of $\sim$15\,kg, which will be used for type-A satellites and are capable of responding quickly to pointing at celestial targets via momentum wheels, after receiving commands from the BeiDou Navigation Satellite System. Type-II spacecrafts are also able to quickly target sources, and support heavier payload, holding type-B and C satellites. As shown in Figure~\ref{FIG:1}, the optics module is attached to the spacecraft, and the detector systems is mounted on an extendable mast covered by a thermal shield. The mast will be deployed in-orbit, in order to achieve the 1\,m, 2\,m and 3\,m focal lengths of type-A, B, and C satellites, respectively. The thermal shield can reduce the deformation of the mast due to temperature variations. 

\begin{table}[h]
\begin{center}
\caption{Parameters of two types of spacecrafts}\label{tab:spacecraft}
\begin{tabular}{@{}lcc@{}}
\toprule
Parameter & Type-I  & Type-II \\
\midrule
Spacecraft weight & $\sim 15$\,kg & $\sim 25$\,kg \\
Allowable payload weight & $\leqslant 7 $\,kg & $\leqslant 15 $\,kg \\
Allowable power & $ \leqslant 10 $\,W & $\leqslant 20 $\,W \\
Pointing accuracy ($3 \sigma$) & $0.1^{\circ}$ & $0.1^{\circ}$ \\
Attitude stability ($3 \sigma$) & $0.01^{\circ}$/s & $0.01^{\circ}$/s \\
Attitude maneuverability & $45^{\circ}$/min & $45^{\circ}$/min \\
Telemetry rate & 10\,Mbps & 50\,Mbps \\
Corresponding satellite type & type-A & type-B and C\\
\bottomrule
\end{tabular}
\end{center}
\end{table}


\subsection{X-ray Optics}
\label{sec:opt}
MPO, LIGA-Micro-Slot-Optics (LMSO), and Ni-Co alloy coated Wolter-I type optics are three options of \textit{CATCH}'s optical system. These optics are light in weight and small in size. The MPO and LMSO are the candidate optics for type-A satellites, while the Ni-Co alloy coated Wolter-I focus optics is the candidate optics of type-B and C satellites with better angular resolution. The focal length, FOV, effective area, and angular resolution of the four types of satellites can be found in Table~\ref{tab:types}.

\begin{figure}[h]
\centering
\includegraphics[width=.8\columnwidth]{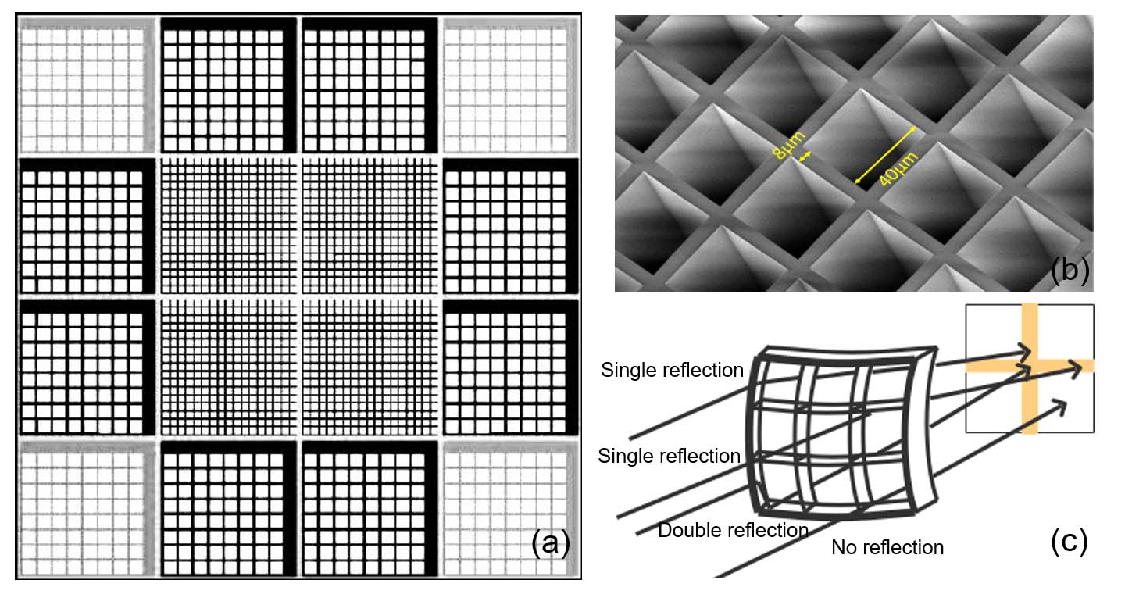} \\
\caption{Panel (a): design of the MPO module of type-A satellites. Three types of MPO pieces with different aspect ratios are used in order to improve the imaging focus performance. Panel (b): close-up of the MPO piece. The inner wall of the pore is iridium. Panel (c): schematic views of MPO focusing X-rays \cite{yuan22018}. The central bright spot is generated by two reflections of the sidewalls of the pores, while the cross arms is formed with single reflections. 
\label{FIG:2}}
\end{figure}

\subsubsection{MPO}

The MPO was first designed under the inspiration from the lobster eye imaging process \cite{Angel1979}, and is developed as a new model for X-ray all-sky monitors \cite{hudec2015}. Compared with the design of MPO optics used in all-sky monitors, e.g. \textit{EP}/WXT \cite{yuan2016,yuan22018}, the MPO optics of \textit{CATCH} has a longer focal length and a smaller FOV so as to increase the effective area and improve the sensitivity. These MPO optics are also known as the Lobster-Eye narrow field X-ray optics. The focal lengths of type-A satellites are designed to be 1\,m at present, and the FOVs are $1^{\circ}\times1^{\circ}$. The MPO is made of spherical glass lenses coated with iridium film, designed by the National Astronomical Observatories (NAOC) and manufactured by the North Night Vision Science \& Technology Research Institute Group (the same research and development teams as \textit{EP}/WXT). The thickness of the MPO piece is on the scale of a millimeter, and there are about one million square micro-pores arranged in a matrix on the glass sheet. The sidewall of the pores will focus soft X-ray photons via grazing incidence. A central bright spot is formed on the focal plane if the photons are reflected twice by the sidewall of the pores, and the cross-arms is produced by single reflections of the sidewall. As shown in Figure~\ref{FIG:2}, the type-A telescope consists of $4\times4$ MPO pieces. In addition, the MPO pieces in the center, edge, and corner of the MPO module have different aspect ratios to optimize on-axis performance. 

\begin{figure}[h]
\centering
\includegraphics[width=.6\columnwidth]{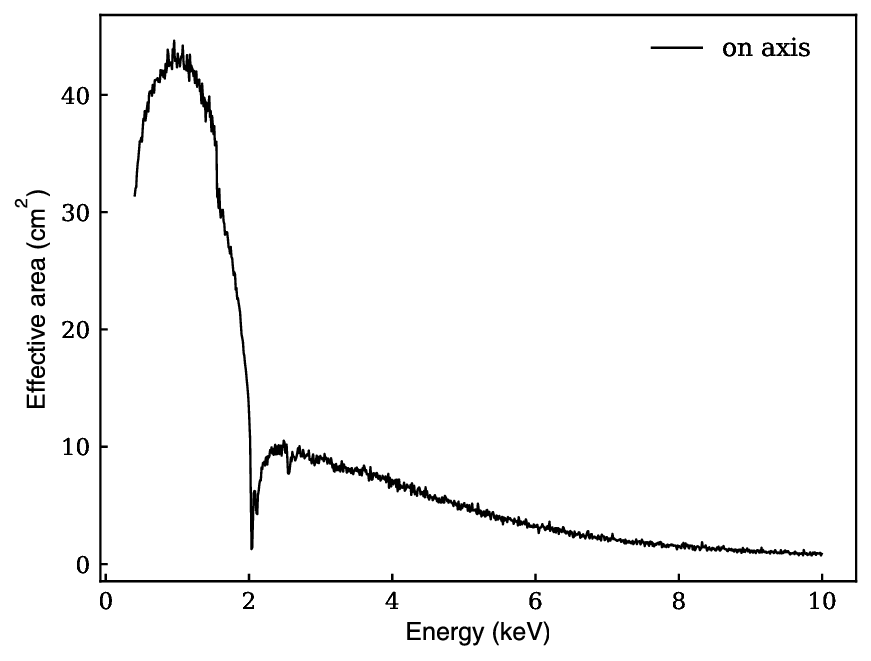}
\includegraphics[width=.6\columnwidth]{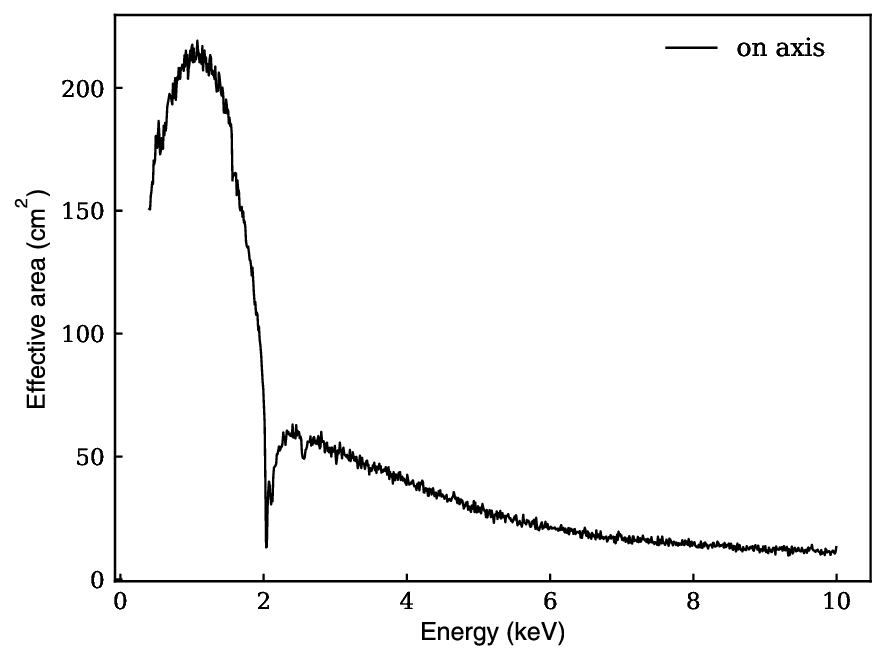}
\caption{Top and bottom panels: simulated on-axis effective area curves of type-A (1\,m focal length) satellites and type-A$^{+}$} (2\,m focal length) satellites, respectively.
\label{FIG:3}
\end{figure}

Based on the simulations in Geant4 \cite{zhao2014,zhao2017}, we estimate the on-axis effective area of the MPO optics onboard type-A satellites, as shown in Figure~\ref{FIG:3}. The effective area is $\sim40$\,cm$^2$ at 1\,keV. The designed angular resolution is 5$^\prime$. With this level of angular resolution, more than 97\% X-ray point sources from the \textit{ROSAT} sky map can be well resolved. The uniformity of the field of Lobster-Eye vision is also excellent. Moreover, if the cost of the MPO optics drops in the future, for better performance, we will consider to replace the MPO optics of type-A satellites with a $8\times8$ MPO array with a longer focal length of 2\,m. The simulated effective area of this updated version of type-A satellites (called as type-A$^{+}$ satellites) can be found in Figure~\ref{FIG:3}. The on-axis effective area is $\sim200$\,cm$^2$ at 1\,keV, much larger than that of type-A satellites.


\begin{figure}[h]
\centering
\includegraphics[width=.8\columnwidth]{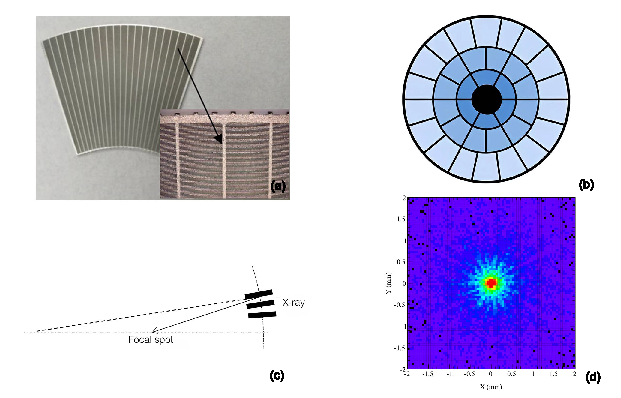} \\
\caption{Panel (a): preliminary design of the LMSO system. Panel (b): a LMSO piece and its micrograph. Panel (c): schematic views of LMSO focusing X-rays \cite{zhang2021}. Panel (d): simulated on-axis PSF of LMSO.
\label{FIG:4}}
\end{figure}

\subsubsection{LMSO}
An alternative to the optics of type-A satellites is the LMSO system, developed by the Institute of High Energy Physics (IHEP) and the NAOC of the Chinese Academy of Sciences, using X-ray lithography and electroplating technologies \cite{zhang2021}. LMSO is made of a planar metal lens with millimeter thickness. It is fan-shaped and covered with micro-slits of micron width that are long strips and arranged in an annular pattern. Figure~\ref{FIG:4} shows the three-circle layer design under the basic constraint that the aperture is not larger than 150\,mm. Multiple lenses are assembled into an approximate sphere, and X-ray photons are focused by a grazing incidence reflection on the side wall of the loop of the micro-slits in the lens, forming an approximately axis-symmetric PSF (panel (d) of Figure~\ref{FIG:4}). This is significantly different from conventional Wolter-I optics and MPO optics which require the photons to be focused by reflecting X-rays twice. Thus, LMSO has the advantages of a large effective area, compact size, and easy integration.

\begin{figure}[h]
\centering
\includegraphics[width=.6\columnwidth]{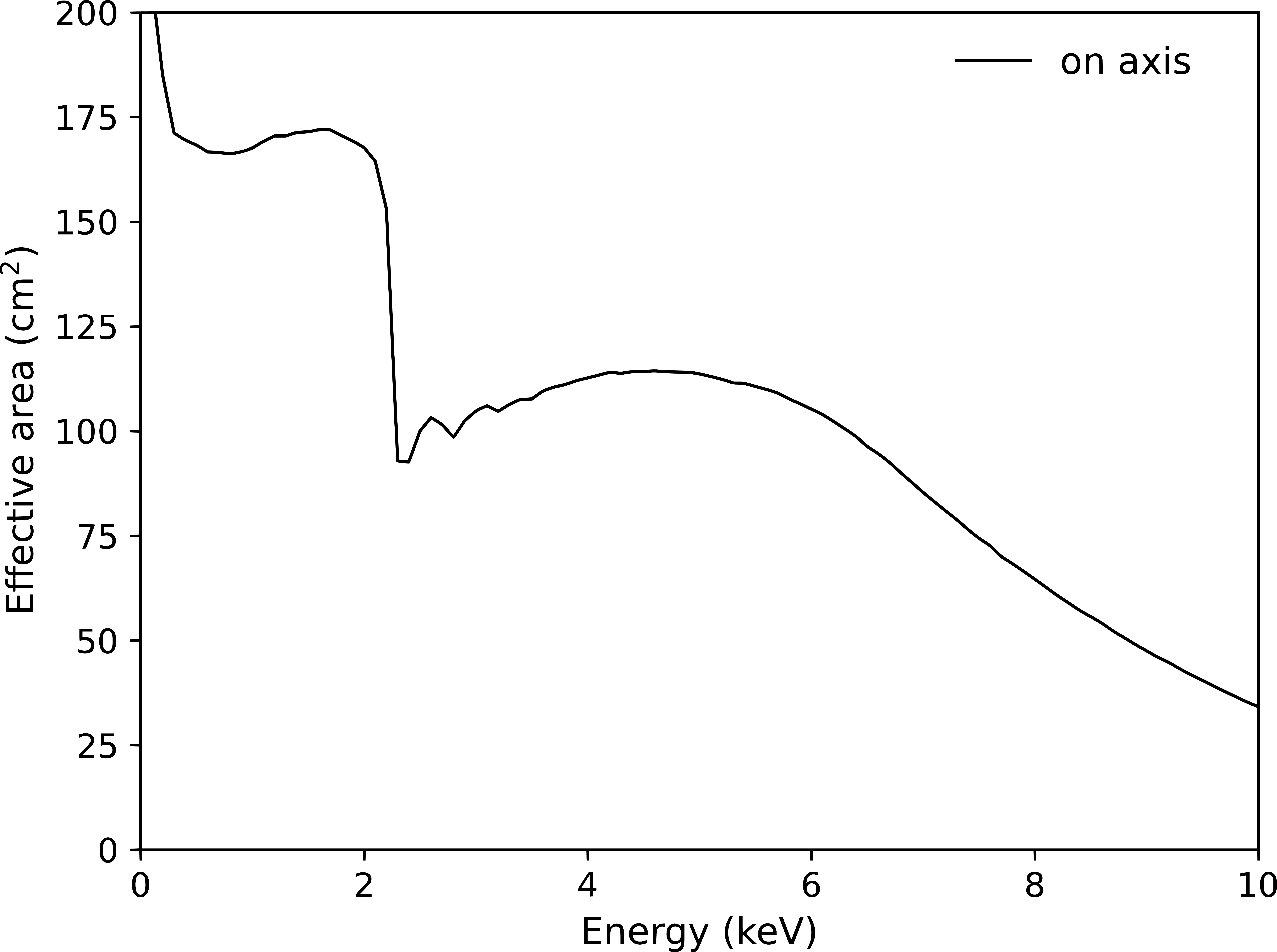} \\
\includegraphics[width=.6\columnwidth]{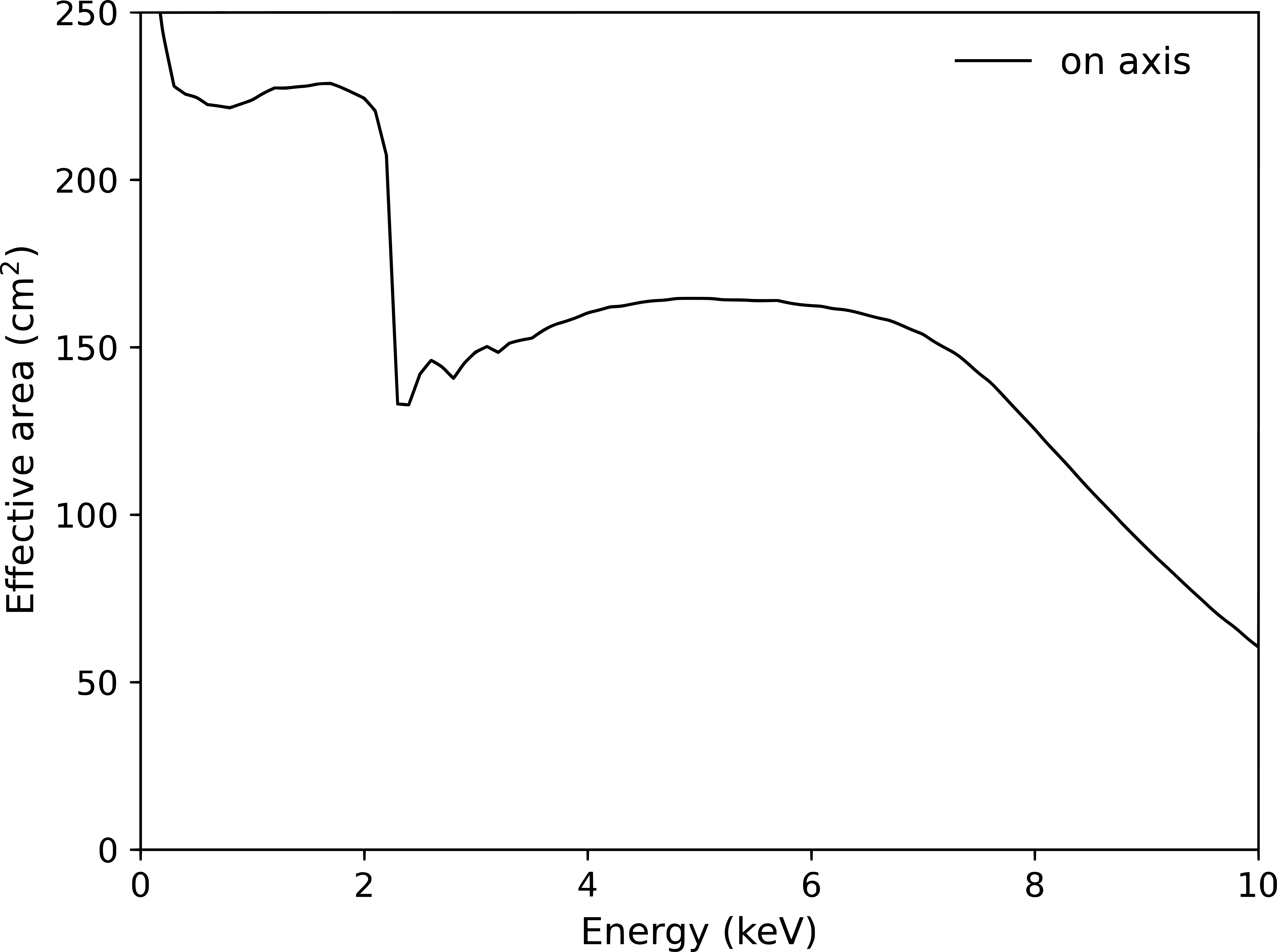} \\
\caption{Top and bottom panels: simulated on-axis effective area curves of 2\,m (type-B} satellites) and 3\,m (type-C satellites) focal lengths, respectively.
\label{FIG:5}
\end{figure}

\subsubsection{Lightweight Wolter-I optics}
Type-B and C satellites will use the lightweight Wolter-I optics to focus incidence X-rays. These Wolter-I optics use metal-ceramic hybrid mirror shells, developed based on the electroformed Ni-Co alloy replication technology and plasma thermal spray processes. In contrast, the conventional Wolter-I optics mounted in some observatories (e.g. \textit{XMM-Newton}, \textit{EP}, \textit{eXTP}) uses Ni shell based on the Electroformed Nickel Replication (ENR) process. As the density of Ni is about 8.9\,${\rm g/cm^{3}}$, the mass of conventional Wolter-I optics would be very heavy. The mass of ENR mirror per 1\,$\rm m^{2}$ sensitive area at 1\,keV is nearly 2000\,kg. This means that it is not suitable for micro-satellites. Thus, the Ni substrate will be changed to Ni-Co alloy with a smaller density. Moreover, to further reduce the mass of the optics system, the thickness of the Ni-Co alloy substrate can be thinner than 70\,$\mu$m, and a layer of Al$_2$O$_3$ will be deposited on the substrate to increase the stiffness of the mirror. The mass of Wolter-I optics with this design can be reduced from nearly 2000\,kg/$\rm m^{2}$ to 500\,kg/$\rm m^{2}$. The specified performance of the Ni-Co alloy lightweight Wolter-I optics can be found in Table ~\ref{tab:types}. The effective area curves of the 2\,m (type-B satellites) and 3\,m (type-C satellites) focal length can be found in Figure~\ref{FIG:5}.

\subsection{Focal plane detector}
\label{sec:det}

As presented in Section~\ref{sec:des}, different types of satellites are dedicated to different types of measurements, including imaging, spectral, timing, and polarization measurements. Accordingly, they will carry different types of focal plane detectors, i.e., 4-pixel SDD array (type-A satellites), multi-pixel SDD array/pn-CCD (type-B satellites), and GMPD/GPD (type-C satellites).

\subsubsection{Type-A satellites}

\begin{figure}[h]
\centering
\includegraphics[width=0.7\columnwidth]{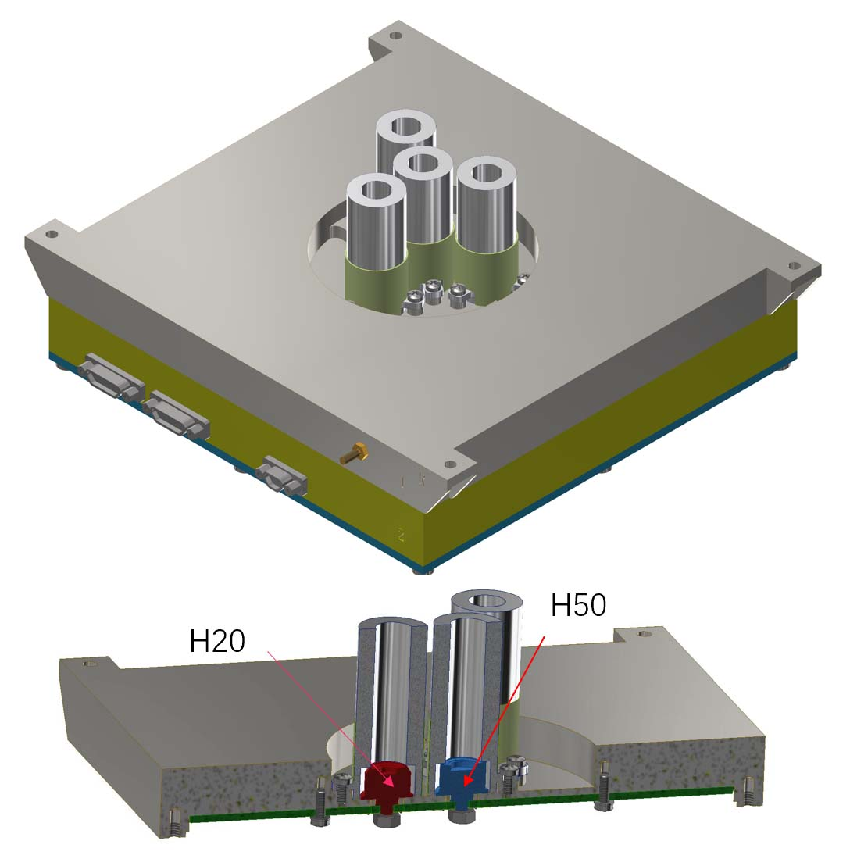} \\
\caption{Diagrams of the detector system (4-pixel SDD array) for type-A satellites. The detector are hidden inside the aluminum baffles partly covered by tantalum, in order to block the in-orbit charged particles. The central detector is H50 SDD, and three H20 SDDs surround the central H50 SDD. 
\label{FIG:6}}
\end{figure}

Type-A satellites are designed for timing observations. In order to study fast transients, we choose SDD as the focal plane detector. The timing resolution of SDD could be better than 1\,$\mu$s. Another advantage of SDD is that the readout anode capacitance is small so that the energy resolution is good, e.g., ideally better than 120\,eV@5.9\,keV. One H50 and three H20 SDDs produced by KETEK are integrated into the 4-pixel SDD array (see Figure~\ref{FIG:6}) on the focal plane. The central H50 SDD is used to detect timing signals from celestial objects, while the surrounding H20 SDDs, including one blind detector blocked by tantalum lid, are used to do background subtraction. In order to reduce the leakage current, which is affected by the radiation damage, the SDDs will be cooled down to $-35^{\circ}$C with Peilter cooler, and will be hidden inside the aluminum baffles partly covered by tantalum. The detailed parameters of the SDD array can be found in the Table~\ref{tab:types}.  

The readout system consists of four identical pre-stage circuits and a signal acquisition circuit. The pre-stage circuit is to supply power to SDD detector, and to amplify and reset the output signal from the preamplifier. We use a digital waveform processing algorithm for data acquisition. The four pre-stage output signal is converted to a digital signal by four 40 MHz 14bits fast ADCs (LTC2247), and the digital signal is processed to obtain the linear 4096-channel energy information. The waveform processing and storage for 4 channels are parallel and independent. The data packets are downloaded via a 20-MHz low-voltage differential signaling (LVDS) interface to the spacecraft. These four ADCs are controlled by a SEU immune Zero FIT Flash FPGA (Microsemi's SmartFusion2 FPGA M2S050TS).

\subsubsection{Type-B} satellites
\begin{figure}[h]
\centering
\includegraphics[width=0.4\columnwidth]{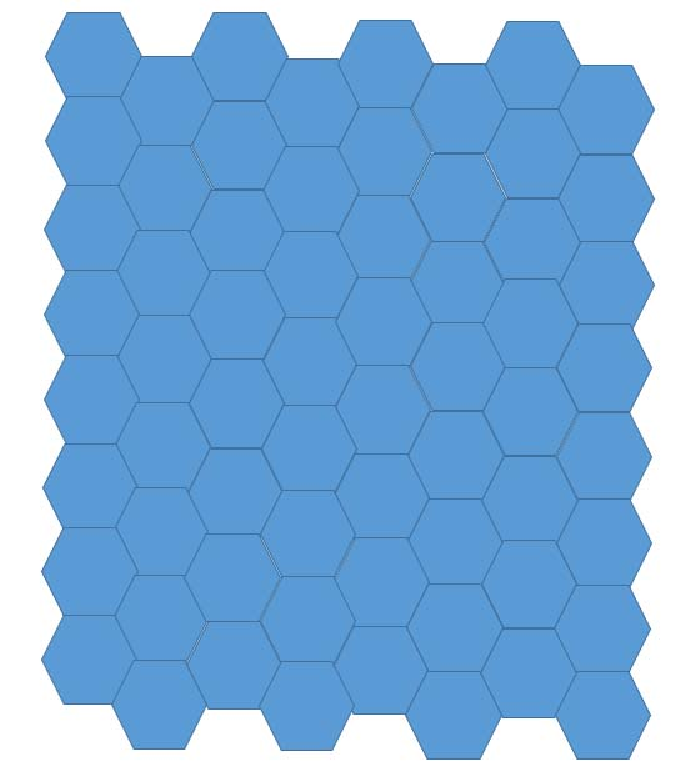} \\
\caption{One SDD module ($8\times8$ pixels SDD array) for type-B} satellites. The detector system is consisted of 4 modules. 
\label{FIG:7}
\end{figure}

Type-B satellites will do imaging, spectral, and timing observations based on the results from Type-A satellites. The multi-pixel SDD array is preferred as the focal plane detector. Compared to the CCD system, the SDD array has excellent timing and energy resolution. In addition, since its readout speed is fast, the pile-up effect is weaker than that of the CCD system. It is more suitable for observations of bright sources. As the focal length of type-C satellites is 2\,m and the angular resolution is about 2\,$^\prime$, the pixel size of the SDD array should be smaller than 1.16\,mm. Considering that the FOV of the telescope is $0.5^{\circ}\times0.5^{\circ}$, the size of the focal plane should be larger than 17.4\,mm$\times$17.4\,mm. Therefore, the focal plane detector consists of 4 SDD detector modules, and each SDD module is composed of a 64-pixel SDD array, as shown in Figure~\ref{FIG:7}. The readout system of the SDD array is ASIC, which is under development and will be integrated with charge sensitive amplifiers, shaping amplifiers, etc. The detailed requirement for the SDD array can be found in Table~\ref{tab:types}. Moreover, the pn-CCD system is also considered as a candidate focal plane detector for type-C satellites, since the energy resolution of pn-CCDs is comparable to that of SDDs, and the pixel size is smaller. But we also note that the displacement damage caused by the charged particles in-orbit will deteriorate the performance of pn-CCDs, and it needs to be cooled down to nearly $-60^\circ$C, which is much lower than that of the SDD. This will increase power consumption, therefore, we prefer the SDD array as the focal plane detector of the type B satellites.

\subsubsection{Type-C} satellites

\begin{figure}[h]
\centering
\includegraphics[width=.8\columnwidth]{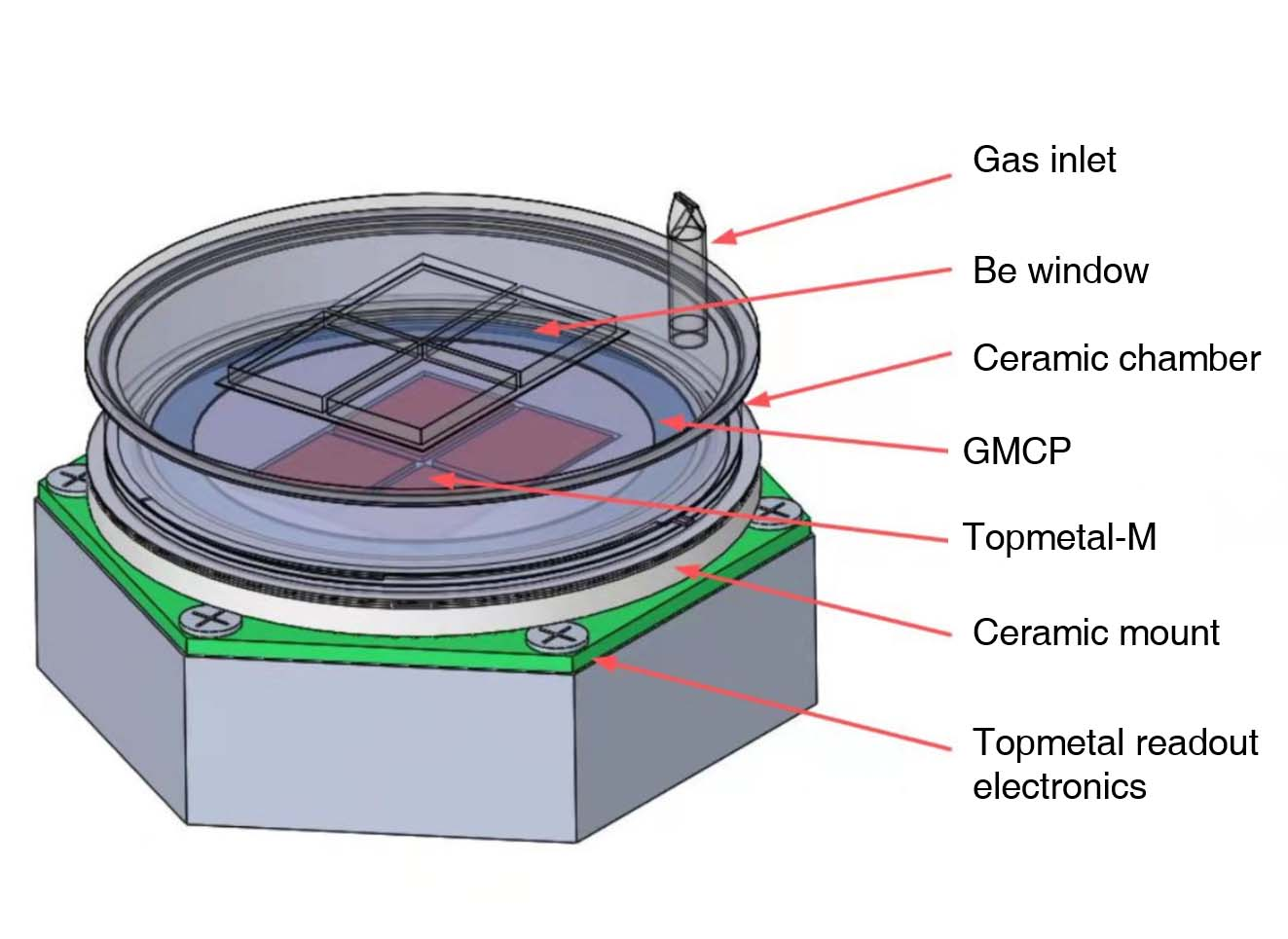} \\
\caption{Preliminary design of the GMPD for type-C satellites. 
\label{FIG:8}}
\end{figure}

Type-C satellites will perform soft X-ray polarization observations. The sensitive energy range is from 2 to 8\,keV. The preferred focal plane detector is GMPD, developed by Guangxi University \cite{Li2021,Huang2021}. As a type of sensitive gas detector, the linearly polarized incidence soft X-rays will generate photoelectrons in the GMPD via the photoelectric effect. The polarizations of X-ray photons determine the directions of emitting photoelectrons, and can be re-constructed by measuring their tracks. The microchannel plate is used as the position sensitive detector to read out the tracking information. The preliminary design of the GMPD is shown in Figure~\ref{FIG:8}, and its specification can be found in Table~\ref{tab:types}. The GPD \cite{Costa2001}, used in some polarization missions, e.g., PolarLight \cite{Feng2019} and \textit{IXPE} \cite{Soffitta2021}, is an option of \textit{CATCH}. The main difference between the GPD and GMPD is that the gas electron multiplier (GEM) is used in the GPD instead of the micro-plate.

\subsection{Orbital configuration of the constellation}
\label{sec:orb}

The high energy particles in the radiation belt can cause radiation damage to semiconductor devices, including Single Event Effects (SEE), Displacement Damage (DD), and Total Ionizing Dose (TID) \cite{Mar2011}. In particular, the former two effects will cause serious degradation of the detector performance. The SEE mainly affect the performance of the digital readout electronics. The high energy particles will generate electron-hole pairs through ionization and lead to abnormal changes in the logic state of semiconductor material, which will be gradually aggravated with time. In addition to the SEE, the high energy particles will also lose energy through non-ionizing interactions with materials. This process, known as the DD, can generate defects in the lattice of the semi-conductor, and cause the increase of leakage current of the semi-conductor detector and the deterioration of spectral resolution. Taking into account these effects, the radiation damage caused by the high-energy in-orbit particles have to be considered carefully to make sure that the detector system can work well during the designed life time of the satellite. The radiation damage suffered by the satellite and the effective observing time depend on the orbital altitudes and inclinations; therefore, it is necessary to evaluate the radiation flux in different orbital configurations, so as to increase the effective observing time and reduce radiation damage.

\subsubsection{In-orbit radiation flux of the charged particles}

As discussed above, the high energy particles (including electrons and protons) in the low-earth orbit from the Van Allen radiation belt will cause significant damage to the semi-conductor detector onboard \textit{CATCH}. The Van Allen radiation belt is a crescent radiation layer formed by a small number of high-energy particles from the solar wind, after breaking into the magnetosphere and being captured by the Earth's magnetic field. There are some kinds of models\footnote{\url{https://www.spenvis.oma.be/help/background/background.html}} to describe the Van Allen radiation belt. For example, the proton flux can be well modeled by the AP series model (i.e. AP-8, AP-9), CRRESPRO and PSB97, and the electron radiation models include the AE series model (AE-8, AE-9), CRRESELE, IGE-2006, etc. The AE-8 (Max)/AP-8 (Min) model and the AE-9 (Mean)/AP-9 (Mean) model will be used in this paper to calculate the in-orbit particle flux.

\begin{figure}[h]
\centering
\includegraphics[width=.7\columnwidth]{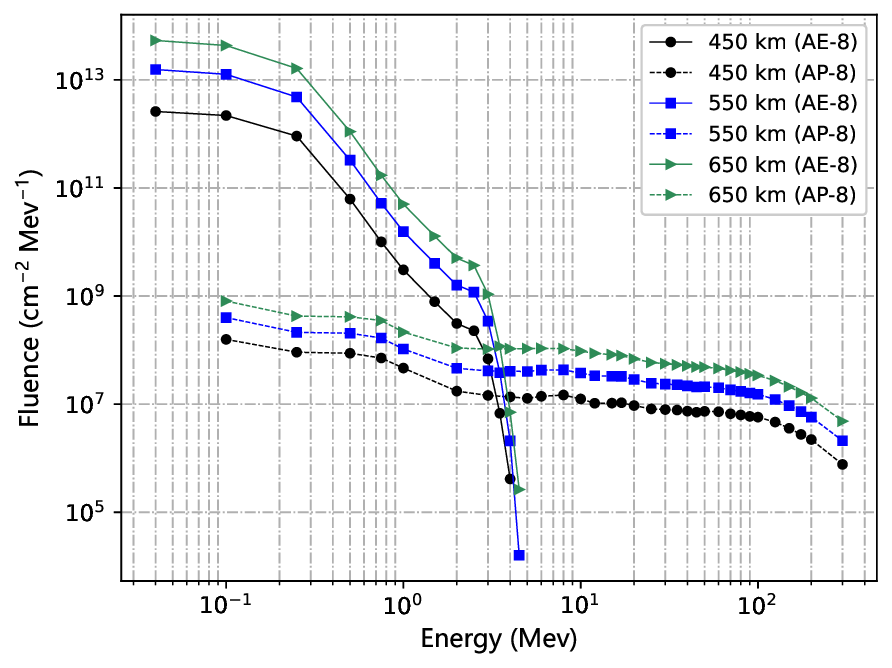}\\
\includegraphics[width=.7\columnwidth]{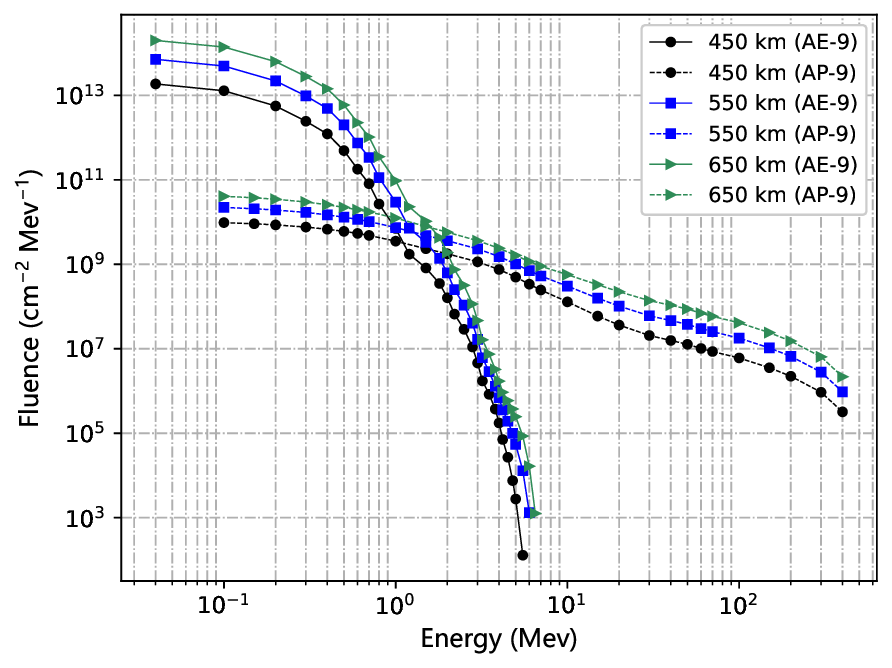}\\
\caption{Two-year differential flux of electrons and protons in an orbit with different altitudes for AE-8/AP-8 (upper panel) and AE-9/AP-9 (lower panel), respectively. The inclination is assumed to be 29$^{\circ}$.  
\label{FIG:9}}
\end{figure}

We calculate the electron and proton differential flux over two years at different altitudes, by assuming an inclination of 29$^\circ$. The energy range of the electron is 0.04--7\,MeV and the proton energy range is 0.1--300\,MeV. As shown in Figure~\ref{FIG:9}, it can be seen that the particle flux becomes larger when the altitude increases. The 650\,km altitude is under the Earth's radiation belt, and so the space environment is relatively harsh. For the 450\,km orbit, although the particle flux is low, the satellite will be affected by atmospheric resistance, leading to a shorter life time. Therefore, the 550\,km orbit is the preferred altitude of \textit{CATCH}. 
\begin{figure}[h]
\centering
\includegraphics[width=.7\columnwidth]{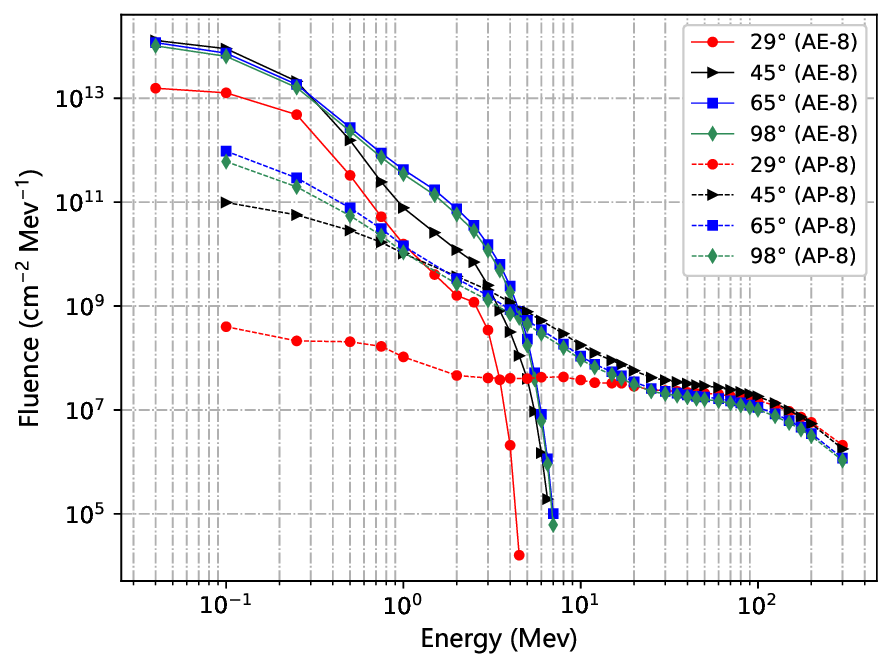} \\
\includegraphics[width=.7\columnwidth]{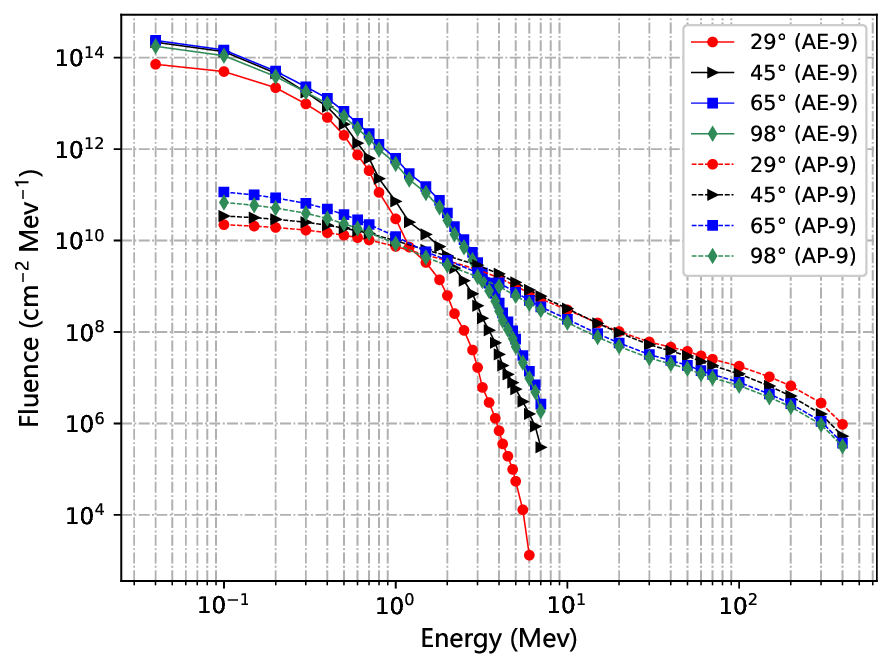} \\
\caption{Two-year differential flux of electron and proton with different inclinations from AE-8/AP-8 (upper panel) and AE-9/AP-9 (lower panel), when assuming the altitude to be 550\,km.
\label{FIG:10}}
\end{figure}
Taking into account an altitude of 550\,km for \textit{CATCH}, the differential flux of electrons and protons at different inclinations are calculated. As shown in Figure~\ref{FIG:10}, the particle flux is low when the inclination is low, and increases with the inclination. But when the inclination increases to $65^{\circ}$, the particle differential flux will not increase anymore. This is consistent with the results of \cite{Sajid2015} and \cite{2020SPIE}. Therefore, a low inclination circular orbit is preferred for \textit{CATCH}, in order to reduce the damage to the semiconductor detector. At present, by considering the launch feasibility and cost, a low inclination of 29$^{\circ}$ is the  optimum choice when launching from China.

\subsubsection{Observing efficiency}
\label{sec:obs_eff}

As a low-earth orbit X-ray constellation, the following effects from the Earth, Sun, Moon, and high particle region will impact the observing efficiency of \textit{CATCH}. 

Earth: the Earth is the main factor to reduce the effective observing time. For a satellite with an orbital altitude of $h$, the probability ($p$) that the entire celestial sphere will be obscured by the earth is

\begin{equation}\label{eq1}
p=\frac{1}{2}-\frac{1}{2}\cos\left[\arcsin{\left(\frac{R_{\rm E}}{R_{\rm E}+h}\right)}\right], 	
\end{equation}
where $R_{\rm E}$ is the radius of the earth. The observation will not be affected by the atmosphere if the angle between the pointing of the telescope and the Earth's edge is larger than the elevation angle (ELV). Moreover, there is another important angle, defined as the Day Earth ELV (DYE\_ELV), which is the minimum permitted observable angle between the side illuminated by the Sun and the pointing direction. The DYE\_ELV should be larger than ELV because the earth’s atmosphere can absorb X-rays and scatter photons from the Sun.

Sun and Moon: The solar chromosphere and corona emit a lot of ultraviolet and X-ray photons, which can saturate the detectors on-board \textit{CATCH}, even permanently damage. Therefore, the angle between the pointing of the telescope and the Sun should be larger than a certain angle to make sure that the detector system can be safe and work well. This angle is defined as the SUN\_ANGLE. Although the moon itself does not emit radiation, it can scatter the solar visible and X-ray light, increasing the background of the telescope. Therefore, the angle between the telescope pointing and the moon should be larger than a certain angle, defined as the MOON\_ANGLE. 

High particle background regions: the magnetic field in the South Atlantic Anomaly (SAA) area is relatively weak and a large number of charged particles will be tracked in this region. The high flux of charged particles will increase the background significantly and can cause damage to the semiconductor detector. In addition to the SAA, some other high-energy particle background regions will also affect the observations of \textit{CATCH}, such as the earth's polar regions\footnote{Although \textit{CATCH} will not be affected by the earth's polar regions due to the inclination of 29$^{\circ}$, some \textit{CATCH} pathfinders may be affected by these regions as they are likely to be deployed in the Sun-synchronous orbit, since this orbit is the most common orbit.}. The charged particles from a certain direction of space will be restricted by the earth's magnetic field, and can only enter the earth when their magnetic rigidity is larger than the geomagnetic cutoff rigidity (COR) of the earth in that direction  \cite{kaspi2017}. The geomagnetic COR gradually decreases with the increase of the geomagnetic altitude. This means that the flux of charged particles from space is larger at higher geomagnetic latitudes. The charged particle background in these regions will increase significantly, and in some situations, the detector system will saturate.
\begin{table}[h]
\begin{center}
\caption{Constraints when calculating the effective observing time} \label{tab:2}
\begin{tabular}{@{}ccccc@{}}
\toprule
ELV & DYE\_ELV  & SUN\_ANGLE & MOON\_ANGLE & COR\\
\midrule
\textgreater{}30$^{\circ}$ & \textgreater{}40$^{\circ}$ & \textgreater{}25$^{\circ}$& \textgreater{}15$^{\circ}$ & \textgreater{}6 \\
\bottomrule
\end{tabular}
\end{center}
\footnotemark{ELV is the earth's elevation angle; DYE\_ELV is the Day Earth ELV; SUN\_ANGLE is the solar avoidance angle; MOON\_ANGLE is the moon avoidance angle; COR is the geomagnetic cutoff rigidity. }
\end{table}


Taking into account these effects, we estimate the observing efficiency of \textit{CATCH} \cite{Zhao2019}. As described above, the candidate orbit of CATCH is the 550\,km low-earth orbit with an inclination of 29$^{\circ}$. Taking the type-B satellites as an example, the focal plane detector is a SDD array with a high timing and spectral resolution. The sensitive energy range of the detector is 0.3--10\,keV, which is similar to the energy range of \textit{NICER} (0.2--12\,keV). Thus, except for the SUN\_ANGLE, we use the observation constraints from \textit{NICER}\footnote{\url{https://heasarc.gsfc.nasa.gov/docs/nicer/mission_guide/}} (see Table~\ref{tab:2}), to calculate the observing efficiency. For the SUN\_ANGLE of \textit{CATCH}, we specifically design it to be 25$^\circ$, in order to increase the observing efficiency (see below). It should be noted that these constraints may be updated after the launch of \textit{CATCH}. 

\begin{figure}[h]
\centering
\includegraphics[width=0.8\textwidth]{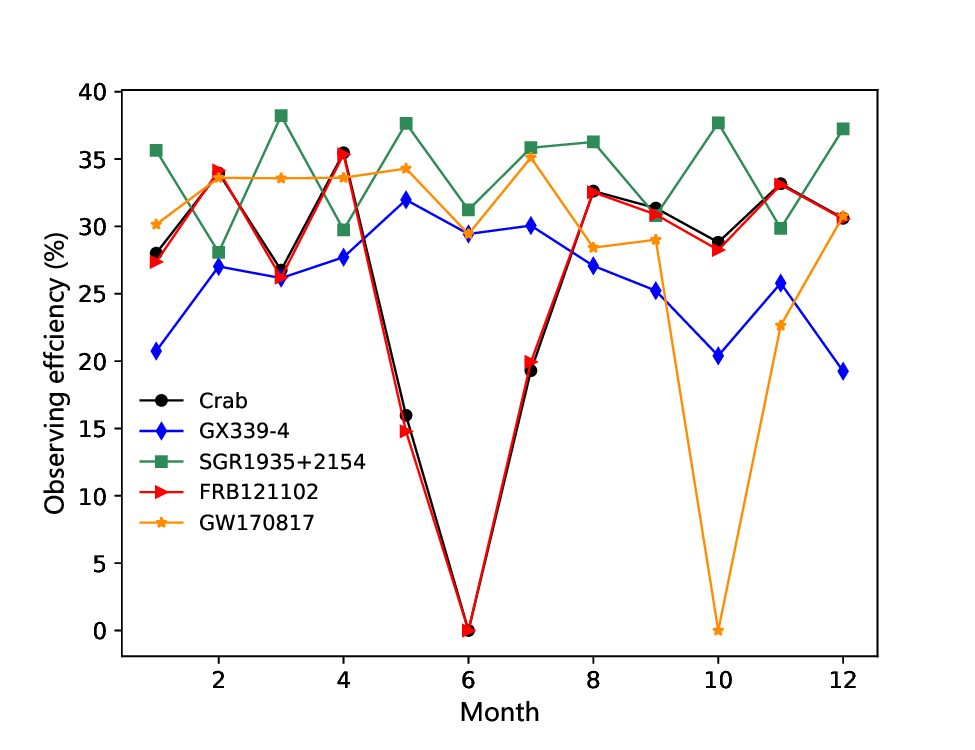} 
\caption{Observing efficiency of five transient and variable sources over one year. 
\label{FIG:11}}
\end{figure}

\begin{figure}[h]
\centering
\includegraphics[width=0.7\textwidth]{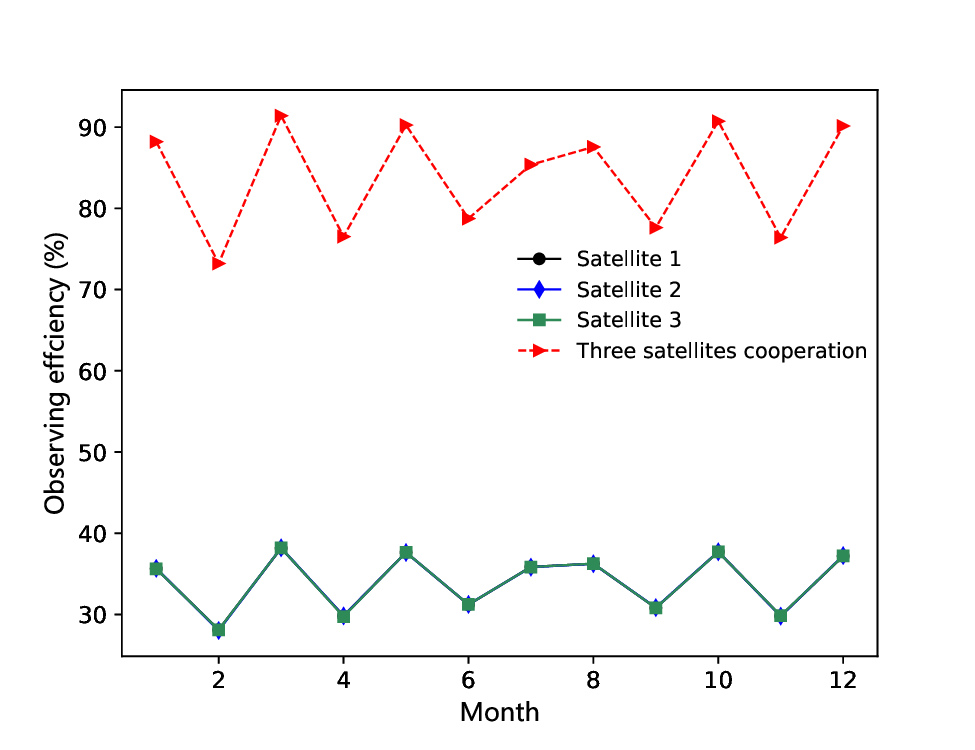} \\
\includegraphics[width=0.7\textwidth]{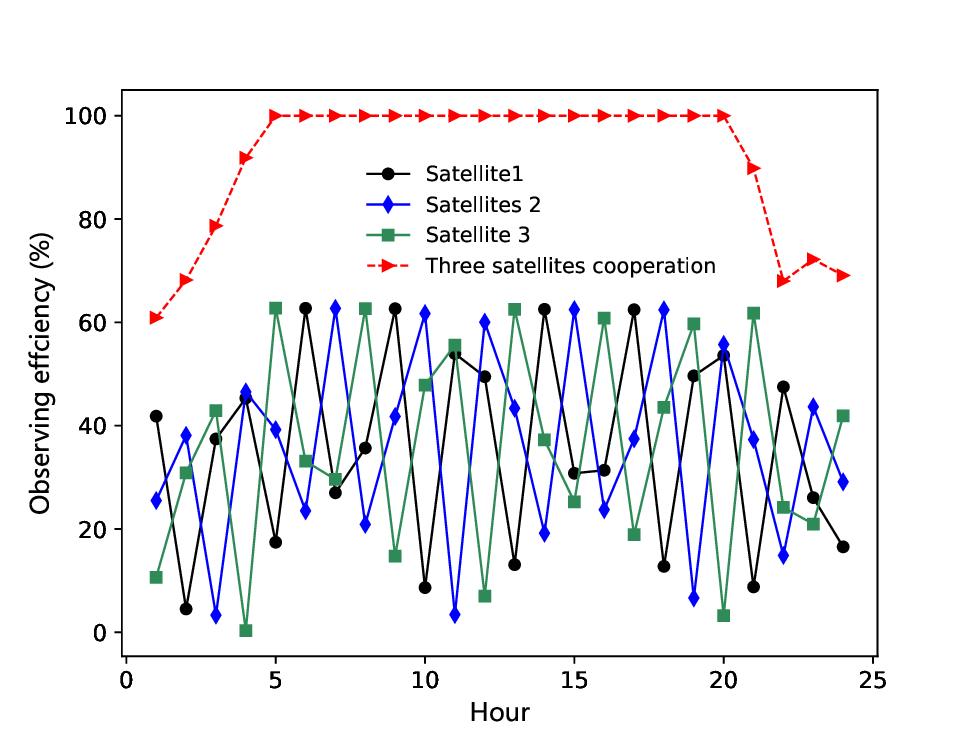} \\ 
\caption{Observing efficiency of the SGR 1935+2154 when deploying three satellites in one orbit with a phase separation of 120$^{\circ}$. Top: annual observing efficiency; Bottom: daily observing efficiency.
\label{FIG:12}}
\end{figure}

Considering one satellite in the \textit{CATCH} constellation, the observing efficiency of five typical transient and variable sources, including the Crab, a black hole binary (GX 339-4), a magnetar (SGR 1935+2154), a repeating fast radio burst source (FRB 121102), and the electromagnetic counterpart of the gravitational-wave event GW170817, are calculated (see Figure~\ref{FIG:11}). The observing efficiency varies throughout the year due to the change of the solar aspect angle. For example, the Crab can be observed with a relatively high efficiency in the spring and winter, but is invisible in June because the declination difference between the Sun and the Crab is only 1$^{\circ}$ on the summer solstice. Moreover, the duration of the invisible period increases with the increase of the SUN\_ANGLE. By extending the solar panels to block the solar radiation (see Figure~\ref{FIG:1}), the \textit{CATCH}'s SUN\_ANGLE of 25$^\circ$ is much smaller than other X-ray astronomy satellites, whose SUN\_ANGLE are typically 70$^\circ$. This design will effectively increase the observation efficiency, and it is very important to observe transients occurring in the sky region close to the Sun. In contrast, most other telescopes can not observe this region due to the large SUN\_ANGLE. 

With the cooperation of different satellites in the constellation, the visibility of one point source will increase significantly. As shown in Figure~\ref{FIG:12}, when deploying three satellites in one orbital plane with a phase separation of $120^{\circ}$, the observing efficiency of SGR 1935+2154 exhibits an obvious increase. The annual average observing efficiency of this source increases from 34.1\% to 84.0\%, and for a visible day (e.g. July 15th, 2023), the observation is almost uninterrupted, except when the satellites enter into the SAA region. If the satellites are deployed in different orbital planes, the efficiency can increase further (nearly 100\%), as in this case, when one satellite enters into the SAA region, we can use another satellite to observe the source. Thus, \textit{CATCH} can perform uninterrupted observations when the source is visible. This is a powerful capability to study fast transients, such as fast radio burst sources. 

\begin{figure}[h]
\centering
\includegraphics[width=\textwidth]{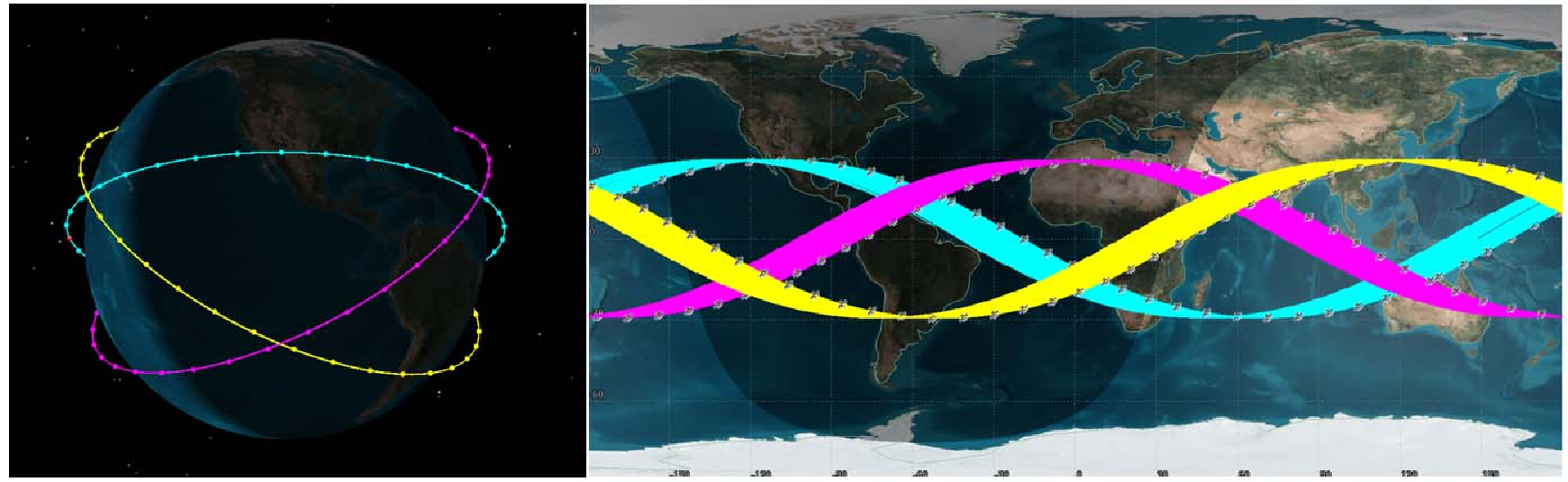} \\

\caption{Preliminary constellation configuration of \textit{CATCH}: 3-dimensional (left panel) and 2-dimensional (left panel) diagrams.} 
\label{FIG:13}
\end{figure}

\subsubsection{Constellation configuration}

By taking the following factors into account, including the current daily average number of transients reported in the TNS\footnote{\url{https://www.wis-tns.org/}}, the improvement of the detection capability ($\sim 1$ order of magnitude better) brought by the operation of next-generation survey telescopes such as the Vera C. Rubin Observatory \cite{Graham2019}, the average exposure for each source, and the proportion of the visible sky region after considering the solar avoidance angle, we roughly estimate that 108 type-A satellites are required for unbiased follow-up observations. The number of satellites is also comparable to the daily average number of X-ray point sources detected in the ROSAT all-sky survey \cite{Boller2016} with a flux meeting the detection capability of type-A satellites.



Since the sources are randomly distributed throughout the sky region and some observing constraints (Section~\ref{sec:obs_eff}) must be considered, we need to deploy the 108 satellites into a higher number of orbital planes, in order to ensure every target can be observed with a long effective exposure time, via the ``relay observation'' through satellites on different orbital planes. However, more orbital planes will substantially increase launch costs. Taking into account observing efficiency and cost, we plan to adopt the WALKER constellation design with 3 orbital planes, as shown in Figure~\ref{FIG:13}. The latitude of each orbit plane is 550\,km, and the orbital inclination is 29$^{\circ}$. Each orbital plane will host 36 type-A satellites with a phase separation of 10$^{\circ}$. Considering the source ratio that the X-ray radiation of multi-wavelength transients can be detected by type-A satellites (taking $\sim5$\% from TDEs as a typical value \cite{Gezari2021,Sazonov2021}), the number of satellites (3 satellites) required for uninterrupted observation of one source (Figure~\ref{FIG:12}), and the uniformity of orbit distribution, 9 type-B and 9 type-C satellites will be deployed in \textit{CATCH}, and one orbit will host 3 satellites of each type with a phase difference of 120$^{\circ}$.

This design ensures that \textit{CATCH} has excellent observing efficiency in order to meet observational requirements. In the future, we will further quantify the coverage of the sky region, the continuity of observations, the effective observation time, the number of valid triggers, the opportunity for joint observations, the launch cost, the operation and maintenance difficulty, the communication efficiency, etc, in order to optimize the constellation configuration.

\begin{figure}[h]
\centering
\includegraphics[width=\textwidth]{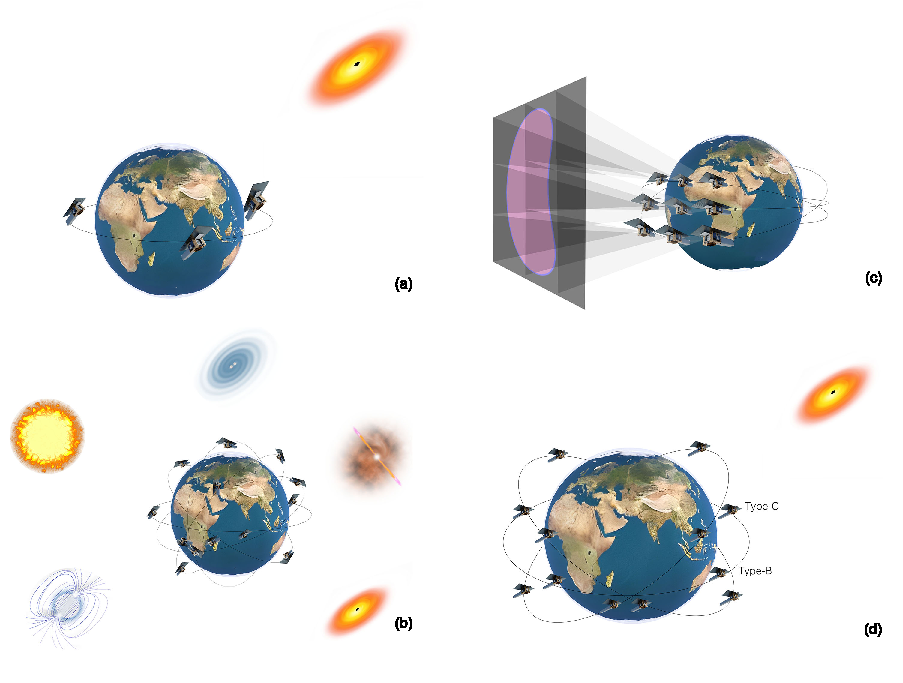} \\
\caption{Four observing modes of \textit{CATCH}: uninterrupted (long-term) monitoring (panel a), all-sky follow-up (panel b), scanning observations executed through a constellation with a flexible FOV (panel c) and multi-dimensional observations (panel d).
\label{FIG:14}}
\end{figure}
\subsection{Triggers and} Observing Mode
\label{sec:obs_mode}

\textit{CATCH} is a constellation to perform follow-up observations for a large number of transients, and study the dynamic universe in multi-dimensional ways. The triggers of multi-wavelength transients and multi-messenger events mainly come from the TNS, the Gamma-ray Coordinates Network (GCN)\footnote{\url{https://gcn.gsfc.nasa.gov/}}, the Astronomer's Telegram (ATel)\footnote{\url{https://www.astronomerstelegram.org/}}, the community alert brokers (e.g. ALeRCE\footnote{\url{https://alerce.science/services/}}, ANTARES\footnote{\url{https://antares.noirlab.edu/}}, Pitt-Google\footnote{\url{https://pitt-broker.readthedocs.io/en/latest/}}, Lasair\footnote{\url{https://lasair-ztf.lsst.ac.uk/}}), and the survey telescopes (such as \textit{GECAM}) having a cooperation with \textit{CATCH}. In future work, we will build a trigger network to quickly collect public trigger information and get triggers from the observatories that cooperate with \textit{CATCH}. The time delay from getting the triggers to the satellites performing the observations, including the time to generate the observation commands via an intelligent control system, the time to upload the commands via the BeiDou Navigation Satellite System, and the time for the satellites to change the pointing, is a few minutes. As the FOV of one type-A satellite is $1^{\circ} \times 1^{\circ}$, the position accuracy of the triggers should be smaller than $1^{\circ}$. For most triggers in radio, optical and X-ray bands, this accuracy can be well satisfied. But for gravitational wave events, the localization accuracy is typically in the order of tens of square degrees, so \textit{CATCH} will perform a scanning mode with multiple satellites to form a larger FOV (see below).

In order to achieve the scientific goals of \textit{CATCH}, the intelligent control system will be developed to control the cooperation between different satellites, and enable them to form four observing modes according to different scientific requirements. As shown in Figure~\ref{FIG:14}, the observing modes include:

\begin{itemize}
\item Uninterrupted (long-term) monitoring: due to the influence of the Earth, Sun, Moon, and high particle regions (see Section~\ref{sec:obs_eff}), the source can not be uninterruptedly monitored by a single satellite. In \textit{CATCH}, the intelligent control system will control the satellites to perform successive coordinated observations. For example, when one satellite enters into the SAA region or is blocked by the Earth, another satellite will take over its job. In this mode, we will not miss out on some important discovery opportunities of fast transients.

\item All-sky follow-up observations: when the transient sources are visible (i.e. the solar aspect angle $>$ SUN\_ANGLE), multiple satellites, controlled by the intelligent control system, can point to the sources in different sky regions to perform all-sky follow-up observations.

\item Scanning mode with a flexible FOV: for some specific targets, such as gravitational wave and high-energy neutrino events, whose localization are large and irregular, multiple satellites will be controlled by the intelligent control system to form a ``large'' telescope, with the FOV similar to the localization region. In comparison to conventional all-sky monitors, this ``large'' telescope can search for signals with better sensitivity and has a greater chance of finding X-ray counterparts.

\item Multi-dimensional observations: for some transients, \textit{CATCH} can simultaneously perform timing, imaging, spectroscopy, and polarization observations, with the cooperation of different types of satellites in the constellation. This multi-dimensional observation allows us to better understand the physical processes of the transients.

\end{itemize}
\begin{figure}[h]
\centering
\includegraphics[width=0.9\columnwidth]{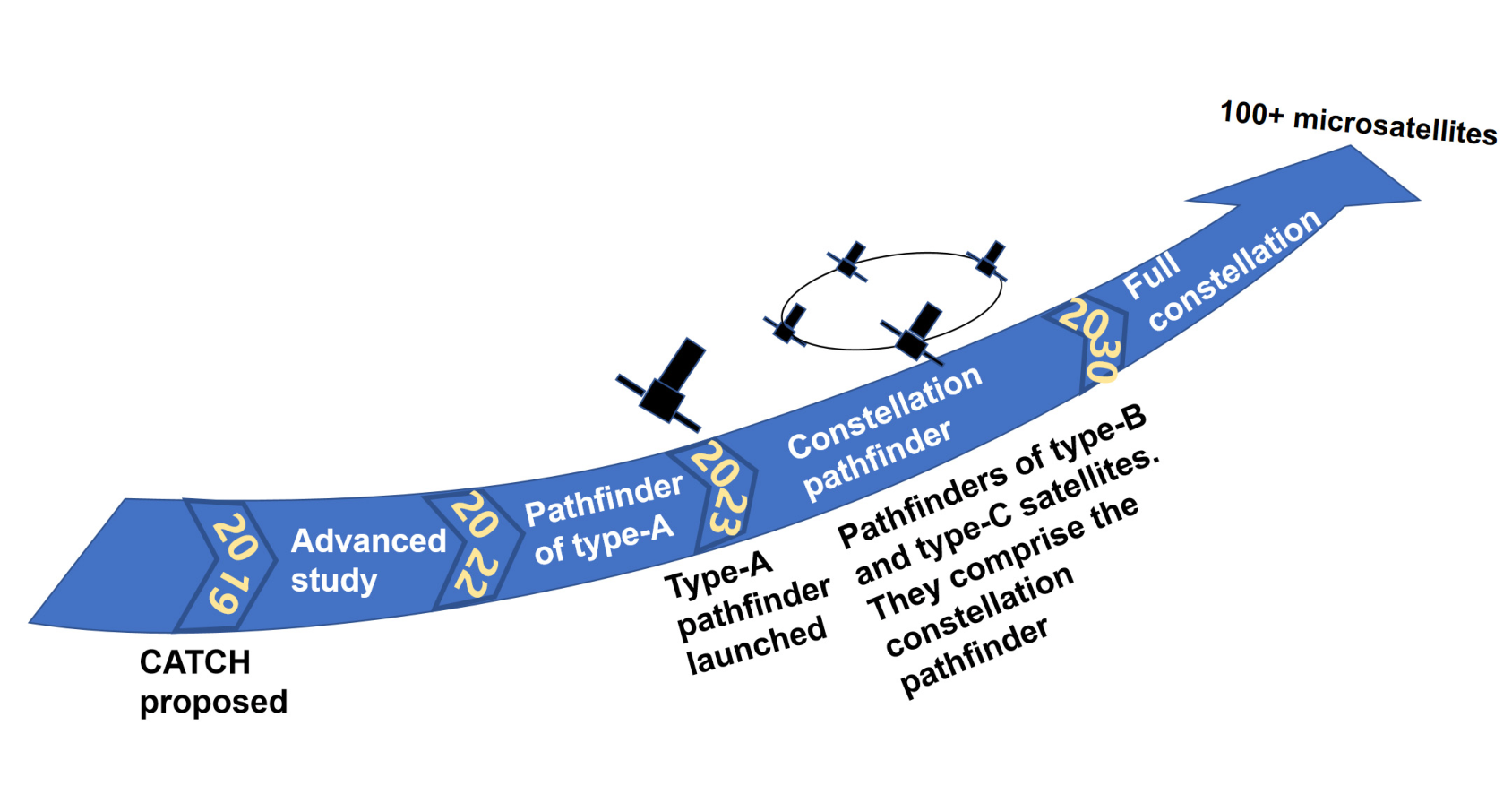} \\
\caption{Blueprint of \textit{CATCH}}. 
\label{FIG:15}
\end{figure}

\subsection{Blueprint of \textit{CATCH} }

The blueprint of \textit{CATCH} is presented in Figure~\ref{FIG:15}. Its concept was proposed in 2019. upported by the Strategic Priority Research Program of the Chinese Academy of Sciences, the National Natural Science Foundation of China (NSFC), and the Scientific and technological innovation project of IHEP, we have finished the preliminary design (means early Phase B) at present, including the preliminary design of the X-ray optics, detector system, deployable mast, and spacecraft, and will launch the first technological and scientific pathfinder of type-A satellite around 2023. This pathfinder is to conduct in-orbit verification of the MPO optics, SDD array detector system, deployable mast and fast pointing capability, and more importantly, it will verify that the X-ray telescope onboard a micro-satellite can still have a sensitive observation capability. After this, the pathfinders of type-B and C satellites will be developed and launched from 2023 to 2030. These pathfinders will do further technological and scientific verification, and comprise a small constellation (i.e. Constellation pathfinder) to test the intelligent control system. According to the in-orbit verification results, the detailed design of the \textit{CATCH} constellation will be updated. This is one significant difference between \textit{CATCH} and conventional missions; for \textit{CATCH}, thanks to its flexible research and development mode brought by satellite miniaturization, some satellites will be launched as early as Phase B in order to realize the iteration of technology and improve the final design. The deployment of the entire constellation is expected to be finished in the early 2030s.

\section{summary}
\label{sec:sum}

In time-domain astronomy, the current-operating and planned survey telescopes will detect a large number of transients, but follow-up instruments are especially scarce. In view of this, we have proposed an intelligent X-ray constellation, \textit{CATCH}, to perform follow-up observations for a substantial amount of transients. \textit{CATCH} consists of 126 satellites in three types (type-A, B, and C), adopting two types of three-axis stabilized small spacecrafts. The BeiDou Navigation Satellite System will enable the spacecrafts to receive triggers from other telescopes and then point to the targets via momentum wheels very quickly. Each satellite is installed with a low cost, high pointing accuracy and high fold ratio deployable mast, in order to increase the focal length of the optics module in a small spacecraft. In current designs, type-A satellites will carry lightweight MPO optics and a 4-pixel SDD array as the focal plane detector. If the cost of MPO pieces reduces in the future, type-A satellites will be upgraded to type-A$^{+}$ satellites with a larger effective area and a higher sensitivity. Type-B will carry lightweight Wolter-I optics and a multi-pixel SDD array. Similar to type-B satellites, type-C satellites will also carry a lightweight Wolter-I optics, but will take a gas polarization detector as the detector system. Type-A satellites are designed to perform timing monitoring for a large number of transients. According to their results, type-B satellites will do further imaging, spectral and timing observations, and type-C satellites will measure the X-ray polarization. By considering the in-orbit radiation flux of the charged particles and the orbital altitude decay, the 550\,km orbit with an inclination of 29$^{\circ}$ is preferred. The constellation configuration is preliminarily designed as a WALKER constellation into three orbital planes, with 36 type-A, 3 type-B and 3 type-C satellites in each orbital plane. With the cooperation of different satellites by the intelligent control system, the observing efficiency will be improved significantly. We have planed four observing modes, including uninterrupted (long-term) monitoring, all-sky follow-up observations, scanning mode with a flexible FOV, and multi-dimensional observations in order to perform follow-up observations for a large number and different types of transients, and study the dynamic universe in multi-dimensional ways.

\section*{Acknowledgements}
We thank the anonymous referee for useful comments that have improved the paper, and thank Tencent Games and Digital Domain for providing the 3D model images of the satellites. We acknowledge funding support from the National Natural Science Foundation of China (NSFC) under grant Nos. 12122306 and 12173056, the Strategic Priority Research Program of the Chinese Academy of Sciences XDA15016400, and the Scientific and technological innovation project of IHEP E15456U2.

\section*{Author’s Contribution}
The manuscript was produced by Panping Li, Qian-Qing Yin, Zhengwei Li, Lian Tao, and Xiangyang Wen. The PI of the CATCH mission is Lian Tao. All authors contributed to the development of the mission concept and/or construction of CATCH.

\section*{Conflict of Interest}
The authors have no competing interests as defined by Springer, or other interests that might be perceived to influence the results and/or discussion reported in this paper.

\section*{Availability of Data and Materials}
If anyone wants to request data, please contact Dr. Tao.

\email{taolian@ihep.ac.cn}

\section*{Funding}
We acknowledge funding support from the National Natural Science Foundation of China (NSFC) under grant Nos. 12122306 and 12173056, the Strategic Priority Research Program of the Chinese Academy of Sciences XDA15016400, and the Scientific and technological innovation project of IHEP E15456U2.
\bibliographystyle{spphys.bst}
\bibliography{ref.bib}

\begin{thebibliography}{10}
\providecommand{\url}[1]{{#1}}
\providecommand{\urlprefix}{URL }
\expandafter\ifx\csname urlstyle\endcsname\relax
  \providecommand{\doi}[1]{DOI \discretionary{}{}{}#1}\else
  \providecommand{\doi}{DOI \discretionary{}{}{}\begingroup
  \urlstyle{rm}\Url}\fi

\bibitem{kantor2014}
J.~{Kantor}, in \emph{The Third Hot-wiring the Transient Universe Workshop},
  ed. by P.R. {Wozniak}, M.J. {Graham}, A.A. {Mahabal}, R.~{Seaman} (2014), pp.
  19--26

\bibitem{lsst2009}
{LSST Science Collaboration}, P.A. {Abell}, J.~{Allison}, S.F. {Anderson}, J.R.
  {Andrew}, J.R.P. {Angel}, L.~{Armus}, D.~{Arnett}, S.J. {Asztalos}, T.S.
  {Axelrod}, S.~{Bailey}, D.R. {Ballantyne}, J.R. {Bankert}, W.A. {Barkhouse},
  J.D. {Barr}, L.F. {Barrientos}, A.J. {Barth}, J.G. {Bartlett}, A.C. {Becker},
  J.~{Becla}, T.C. {Beers}, J.P. {Bernstein}, R.~{Biswas}, M.R. {Blanton}, J.S.
  {Bloom}, J.J. {Bochanski}, P.~{Boeshaar}, K.D. {Borne}, M.~{Bradac}, W.N.
  {Brandt}, C.R. {Bridge}, M.E. {Brown}, R.J. {Brunner}, J.S. {Bullock}, A.J.
  {Burgasser}, J.H. {Burge}, D.L. {Burke}, P.A. {Cargile},
  S.~{Chandrasekharan}, G.~{Chartas}, S.R. {Chesley}, Y.H. {Chu}, D.~{Cinabro},
  M.W. {Claire}, C.F. {Claver}, D.~{Clowe}, A.J. {Connolly}, K.H. {Cook},
  J.~{Cooke}, A.~{Cooray}, K.R. {Covey}, C.S. {Culliton}, R.~{de Jong}, W.H.
  {de Vries}, V.P. {Debattista}, F.~{Delgado}, I.P. {Dell'Antonio},
  S.~{Dhital}, R.~{Di Stefano}, M.~{Dickinson}, B.~{Dilday}, S.G. {Djorgovski},
  G.~{Dobler}, C.~{Donalek}, G.~{Dubois-Felsmann}, J.~{Durech},
  A.~{Eliasdottir}, M.~{Eracleous}, L.~{Eyer}, E.E. {Falco}, X.~{Fan}, C.D.
  {Fassnacht}, H.C. {Ferguson}, Y.R. {Fernandez}, B.D. {Fields},
  D.~{Finkbeiner}, E.E. {Figueroa}, D.B. {Fox}, H.~{Francke}, J.S. {Frank},
  J.~{Frieman}, S.~{Fromenteau}, M.~{Furqan}, G.~{Galaz}, A.~{Gal-Yam},
  P.~{Garnavich}, E.~{Gawiser}, J.~{Geary}, P.~{Gee}, R.R. {Gibson},
  K.~{Gilmore}, E.A. {Grace}, R.F. {Green}, W.J. {Gressler}, C.J. {Grillmair},
  S.~{Habib}, J.S. {Haggerty}, M.~{Hamuy}, A.W. {Harris}, S.L. {Hawley}, A.F.
  {Heavens}, L.~{Hebb}, T.J. {Henry}, E.~{Hileman}, E.J. {Hilton},
  K.~{Hoadley}, J.B. {Holberg}, M.J. {Holman}, S.B. {Howell}, L.~{Infante},
  Z.~{Ivezic}, S.H. {Jacoby}, B.~{Jain}, {R}, {Jedicke}, M.J. {Jee},
  J.~{Garrett Jernigan}, S.W. {Jha}, K.V. {Johnston}, R.L. {Jones}, M.~{Juric},
  M.~{Kaasalainen}, {Styliani}, {Kafka}, S.M. {Kahn}, N.A. {Kaib},
  J.~{Kalirai}, J.~{Kantor}, M.M. {Kasliwal}, C.R. {Keeton}, R.~{Kessler},
  Z.~{Knezevic}, A.~{Kowalski}, V.L. {Krabbendam}, K.S. {Krughoff},
  S.~{Kulkarni}, S.~{Kuhlman}, M.~{Lacy}, S.~{Lepine}, M.~{Liang}, A.~{Lien},
  P.~{Lira}, K.S. {Long}, S.~{Lorenz}, J.M. {Lotz}, R.H. {Lupton}, J.~{Lutz},
  L.M. {Macri}, A.A. {Mahabal}, R.~{Mandelbaum}, P.~{Marshall}, M.~{May}, P.M.
  {McGehee}, B.T. {Meadows}, A.~{Meert}, A.~{Milani}, C.J. {Miller},
  M.~{Miller}, D.~{Mills}, D.~{Minniti}, D.~{Monet}, A.S. {Mukadam},
  E.~{Nakar}, D.R. {Neill}, J.A. {Newman}, S.~{Nikolaev}, M.~{Nordby},
  P.~{O'Connor}, M.~{Oguri}, J.~{Oliver}, S.S. {Olivier}, J.K. {Olsen},
  K.~{Olsen}, E.W. {Olszewski}, H.~{Oluseyi}, N.D. {Padilla}, A.~{Parker},
  J.~{Pepper}, J.R. {Peterson}, C.~{Petry}, P.A. {Pinto}, J.L. {Pizagno},
  B.~{Popescu}, A.~{Prsa}, V.~{Radcka}, M.J. {Raddick}, A.~{Rasmussen},
  A.~{Rau}, J.~{Rho}, J.E. {Rhoads}, G.T. {Richards}, S.T. {Ridgway}, B.E.
  {Robertson}, R.~{Roskar}, A.~{Saha}, A.~{Sarajedini}, E.~{Scannapieco},
  T.~{Schalk}, R.~{Schindler}, S.~{Schmidt}, S.~{Schmidt}, D.P. {Schneider},
  G.~{Schumacher}, R.~{Scranton}, J.~{Sebag}, L.G. {Seppala}, O.~{Shemmer},
  J.D. {Simon}, M.~{Sivertz}, H.A. {Smith}, J.~{Allyn Smith}, N.~{Smith}, A.H.
  {Spitz}, A.~{Stanford}, K.G. {Stassun}, J.~{Strader}, M.A. {Strauss}, C.W.
  {Stubbs}, D.W. {Sweeney}, A.~{Szalay}, P.~{Szkody}, M.~{Takada},
  P.~{Thorman}, D.E. {Trilling}, V.~{Trimble}, A.~{Tyson}, R.~{Van Berg},
  D.~{Vanden Berk}, J.~{VanderPlas}, L.~{Verde}, B.~{Vrsnak}, L.M. {Walkowicz},
  B.D. {Wandelt}, S.~{Wang}, Y.~{Wang}, M.~{Warner}, R.H. {Wechsler}, A.A.
  {West}, O.~{Wiecha}, B.F. {Williams}, B.~{Willman}, D.~{Wittman}, S.C.
  {Wolff}, W.M. {Wood-Vasey}, P.~{Wozniak}, P.~{Young}, A.~{Zentner},
  H.~{Zhan}, arXiv e-prints arXiv:0912.0201 (2009)

\bibitem{fender2015}
R.~{Fender}, A.~{Stewart}, J.P. {Macquart}, I.~{Donnarumma}, T.~{Murphy},
  A.~{Deller}, Z.~{Paragi}, S.~{Chatterjee}, in \emph{Advancing Astrophysics
  with the Square Kilometre Array (AASKA14)} (2015), p.~51

\bibitem{yuan2018}
W.~{Yuan}, C.~{Zhang}, Y.~{Chen}, S.~{Sun}, Y.~{Zhang}, W.~{Cui}, Z.~{Ling},
  M.~{Huang}, D.~{Zhao}, W.~{Wang}, Y.~{Qiu}, Z.~{Liu}, H.~{Pan}, H.~{Cai},
  J.~{Deng}, Z.~{Jia}, C.~{Jin}, H.~{Sun}, H.~{Hu}, F.~{Liu}, M.~{Zhang},
  L.~{Song}, F.~{Lu}, S.~{Jia}, C.~{Li}, H.~{Zhao}, M.~{Ge}, J.~{Zhang},
  W.~{Cui}, Y.~{Wang}, J.~{Wang}, X.~{Sun}, G.~{Jin}, L.~{Li}, F.~{Chen},
  Z.~{Cai}, T.~{Guo}, G.~{Liu}, H.~{Liu}, H.~{Feng}, S.~{Zhang}, B.~{Zhang},
  Z.~{Dai}, X.~{Wu}, L.~{Gou}, Scientia Sinica Physica, Mechanica \&
  Astronomica \textbf{48}(3), 039502 (2018).
\newblock \doi{10.1360/SSPMA2017-00297}

\bibitem{amati2018}
L.~{Amati}, P.~{O'Brien}, D.~{G{\"o}tz}, E.~{Bozzo}, C.~{Tenzer},
  F.~{Frontera}, G.~{Ghirlanda}, C.~{Labanti}, J.P. {Osborne}, G.~{Stratta},
  N.~{Tanvir}, R.~{Willingale}, P.~{Attina}, R.~{Campana}, A.J.
  {Castro-Tirado}, C.~{Contini}, F.~{Fuschino}, A.~{Gomboc}, R.~{Hudec},
  P.~{Orleanski}, E.~{Renotte}, T.~{Rodic}, Z.~{Bagoly}, A.~{Blain},
  P.~{Callanan}, S.~{Covino}, A.~{Ferrara}, E.~{Le Floch}, M.~{Marisaldi},
  S.~{Mereghetti}, P.~{Rosati}, A.~{Vacchi}, P.~{D'Avanzo}, P.~{Giommi},
  S.~{Piranomonte}, L.~{Piro}, V.~{Reglero}, A.~{Rossi}, A.~{Santangelo},
  R.~{Salvaterra}, G.~{Tagliaferri}, S.~{Vergani}, S.~{Vinciguerra},
  M.~{Briggs}, E.~{Campolongo}, R.~{Ciolfi}, V.~{Connaughton}, B.~{Cordier},
  B.~{Morelli}, M.~{Orlandini}, C.~{Adami}, A.~{Argan}, J.L. {Atteia},
  N.~{Auricchio}, L.~{Balazs}, G.~{Baldazzi}, S.~{Basa}, R.~{Basak},
  P.~{Bellutti}, M.G. {Bernardini}, G.~{Bertuccio}, J.~{Braga}, M.~{Branchesi},
  S.~{Brandt}, E.~{Brocato}, C.~{Budtz-Jorgensen}, A.~{Bulgarelli},
  L.~{Burderi}, J.~{Camp}, S.~{Capozziello}, J.~{Caruana}, P.~{Casella},
  B.~{Cenko}, P.~{Chardonnet}, B.~{Ciardi}, S.~{Colafrancesco}, M.G.
  {Dainotti}, V.~{D'Elia}, D.~{De Martino}, M.~{De Pasquale}, E.~{Del Monte},
  M.~{Della Valle}, A.~{Drago}, Y.~{Evangelista}, M.~{Feroci}, F.~{Finelli},
  M.~{Fiorini}, J.~{Fynbo}, A.~{Gal-Yam}, B.~{Gendre}, G.~{Ghisellini},
  A.~{Grado}, C.~{Guidorzi}, M.~{Hafizi}, L.~{Hanlon}, J.~{Hjorth}, L.~{Izzo},
  L.~{Kiss}, P.~{Kumar}, I.~{Kuvvetli}, M.~{Lavagna}, T.~{Li}, F.~{Longo},
  M.~{Lyutikov}, U.~{Maio}, E.~{Maiorano}, P.~{Malcovati}, D.~{Malesani},
  R.~{Margutti}, A.~{Martin-Carrillo}, N.~{Masetti}, S.~{McBreen},
  R.~{Mignani}, G.~{Morgante}, C.~{Mundell}, H.U. {Nargaard-Nielsen},
  L.~{Nicastro}, E.~{Palazzi}, S.~{Paltani}, F.~{Panessa}, G.~{Pareschi},
  A.~{Pe'er}, A.V. {Penacchioni}, E.~{Pian}, E.~{Piedipalumbo}, T.~{Piran},
  G.~{Rauw}, M.~{Razzano}, A.~{Read}, L.~{Rezzolla}, P.~{Romano}, R.~{Ruffini},
  S.~{Savaglio}, V.~{Sguera}, P.~{Schady}, W.~{Skidmore}, L.~{Song},
  E.~{Stanway}, R.~{Starling}, M.~{Topinka}, E.~{Troja}, M.~{van Putten},
  E.~{Vanzella}, S.~{Vercellone}, C.~{Wilson-Hodge}, D.~{Yonetoku}, G.~{Zampa},
  N.~{Zampa}, B.~{Zhang}, B.B. {Zhang}, S.~{Zhang}, S.N. {Zhang},
  A.~{Antonelli}, F.~{Bianco}, S.~{Boci}, M.~{Boer}, M.T. {Botticella},
  O.~{Boulade}, C.~{Butler}, S.~{Campana}, F.~{Capitanio}, A.~{Celotti},
  Y.~{Chen}, M.~{Colpi}, A.~{Comastri}, J.G. {Cuby}, M.~{Dadina}, A.~{De Luca},
  Y.W. {Dong}, S.~{Ettori}, P.~{Gandhi}, E.~{Geza}, J.~{Greiner}, S.~{Guiriec},
  J.~{Harms}, M.~{Hernanz}, A.~{Hornstrup}, I.~{Hutchinson}, G.~{Israel},
  P.~{Jonker}, Y.~{Kaneko}, N.~{Kawai}, K.~{Wiersema}, S.~{Korpela},
  V.~{Lebrun}, F.~{Lu}, A.~{MacFadyen}, G.~{Malaguti}, L.~{Maraschi},
  A.~{Melandri}, M.~{Modjaz}, D.~{Morris}, N.~{Omodei}, A.~{Paizis},
  P.~{P{\'a}ta}, V.~{Petrosian}, A.~{Rachevski}, J.~{Rhoads}, F.~{Ryde},
  L.~{Sabau-Graziati}, N.~{Shigehiro}, M.~{Sims}, J.~{Soomin}, D.~{Sz{\'e}csi},
  Y.~{Urata}, M.~{Uslenghi}, L.~{Valenziano}, G.~{Vianello}, S.~{Vojtech},
  D.~{Watson}, J.~{Zicha}, Advances in Space Research \textbf{62}(1), 191
  (2018).
\newblock \doi{10.1016/j.asr.2018.03.010}

\bibitem{Weisskopf2002}
M.C. {Weisskopf}, B.~{Brinkman}, C.~{Canizares}, G.~{Garmire}, S.~{Murray},
  L.P. {Van Speybroeck}, Publications of the Astronomical Society of the
  Pacific \textbf{114}(791), 1 (2002).
\newblock \doi{10.1086/338108}

\bibitem{Jansen2001}
F.~{Jansen}, D.~{Lumb}, B.~{Altieri}, J.~{Clavel}, M.~{Ehle}, C.~{Erd},
  C.~{Gabriel}, M.~{Guainazzi}, P.~{Gondoin}, R.~{Much}, R.~{Munoz},
  M.~{Santos}, N.~{Schartel}, D.~{Texier}, G.~{Vacanti}, Astronomy \&
  Astrophysics \textbf{365}, L1 (2001).
\newblock \doi{10.1051/0004-6361:20000036}

\bibitem{Gendreau2016}
K.C. {Gendreau}, Z.~{Arzoumanian}, P.W. {Adkins}, C.L. {Albert}, J.F. {Anders},
  A.T. {Aylward}, C.L. {Baker}, E.R. {Balsamo}, W.A. {Bamford}, S.S.
  {Benegalrao}, D.L. {Berry}, S.~{Bhalwani}, J.K. {Black}, C.~{Blaurock}, G.M.
  {Bronke}, G.L. {Brown}, J.G. {Budinoff}, J.D. {Cantwell}, T.~{Cazeau}, P.T.
  {Chen}, T.G. {Clement}, A.T. {Colangelo}, J.S. {Coleman}, J.D. {Coopersmith},
  W.E. {Dehaven}, J.P. {Doty}, M.D. {Egan}, T.~{Enoto}, T.W. {Fan}, D.M.
  {Ferro}, R.~{Foster}, N.M. {Galassi}, L.D. {Gallo}, C.M. {Green}, D.~{Grosh},
  K.Q. {Ha}, M.A. {Hasouneh}, K.B. {Heefner}, P.~{Hestnes}, L.J. {Hoge}, T.M.
  {Jacobs}, J.L. {J{\o}rgensen}, M.A. {Kaiser}, J.W. {Kellogg}, S.J. {Kenyon},
  R.G. {Koenecke}, R.P. {Kozon}, B.~{LaMarr}, M.D. {Lambertson}, A.M. {Larson},
  S.~{Lentine}, J.H. {Lewis}, M.G. {Lilly}, K.A. {Liu}, A.~{Malonis}, S.S.
  {Manthripragada}, C.B. {Markwardt}, B.D. {Matonak}, I.E. {Mcginnis}, R.L.
  {Miller}, A.L. {Mitchell}, J.W. {Mitchell}, J.S. {Mohammed}, C.A. {Monroe},
  K.M. {Montt de Garcia}, P.D. {Mul{\'e}}, L.T. {Nagao}, S.N. {Ngo}, E.D.
  {Norris}, D.A. {Norwood}, J.~{Novotka}, T.~{Okajima}, L.G. {Olsen}, C.O.
  {Onyeachu}, H.Y. {Orosco}, J.R. {Peterson}, K.N. {Pevear}, K.K. {Pham}, S.E.
  {Pollard}, J.S. {Pope}, D.F. {Powers}, C.E. {Powers}, S.R. {Price}, G.Y.
  {Prigozhin}, J.B. {Ramirez}, W.J. {Reid}, R.A. {Remillard}, E.M. {Rogstad},
  G.P. {Rosecrans}, J.N. {Rowe}, J.A. {Sager}, C.A. {Sanders}, B.~{Savadkin},
  M.R. {Saylor}, A.F. {Schaeffer}, N.S. {Schweiss}, S.R. {Semper}, P.J.
  {Serlemitsos}, L.V. {Shackelford}, Y.~{Soong}, J.~{Struebel}, M.L. {Vezie},
  J.S. {Villasenor}, L.B. {Winternitz}, G.I. {Wofford}, M.R. {Wright}, M.Y.
  {Yang}, W.H. {Yu}, in \emph{Space Telescopes and Instrumentation 2016:
  Ultraviolet to Gamma Ray}, \emph{Society of Photo-Optical Instrumentation
  Engineers (SPIE) Conference Series}, vol. 9905, ed. by J.W.A. {den Herder},
  T.~{Takahashi}, M.~{Bautz} (2016), \emph{Society of Photo-Optical
  Instrumentation Engineers (SPIE) Conference Series}, vol. 9905, p. 99051H.
\newblock \doi{10.1117/12.2231304}

\bibitem{Zhang2020}
S.N. {Zhang}, T.~{Li}, F.~{Lu}, L.~{Song}, Y.~{Xu}, C.~{Liu}, Y.~{Chen},
  X.~{Cao}, Q.~{Bu}, Z.~{Chang}, G.~{Chen}, L.~{Chen}, T.~{Chen}, Y.~{Chen},
  Y.~{Chen}, W.~{Cui}, W.~{Cui}, J.~{Deng}, Y.~{Dong}, Y.~{Du}, M.~{Fu},
  G.~{Gao}, H.~{Gao}, M.~{Gao}, M.~{Ge}, Y.~{Gu}, J.~{Guan}, C.~{Gungor},
  C.~{Guo}, D.~{Han}, W.~{Hu}, Y.~{Huang}, J.~{Huo}, S.~{Jia}, L.~{Jiang},
  W.~{Jiang}, J.~{Jin}, Y.~{Jin}, B.~{Li}, C.~{Li}, G.~{Li}, M.~{Li}, W.~{Li},
  X.~{Li}, X.~{Li}, X.~{Li}, Y.~{Li}, Z.~{Li}, Z.~{Li}, X.~{Liang}, J.~{Liao},
  G.~{Liu}, H.~{Liu}, S.~{Liu}, X.~{Liu}, Y.~{Liu}, Y.~{Liu}, B.~{Lu}, X.~{Lu},
  T.~{Luo}, X.~{Ma}, B.~{Meng}, Y.~{Nang}, J.~{Nie}, G.~{Ou}, J.~{Qu},
  N.~{Sai}, R.~{Shang}, G.~{Shen}, L.~{Sun}, Y.~{Tan}, L.~{Tao}, Y.~{Tuo},
  C.~{Wang}, C.~{Wang}, G.~{Wang}, H.~{Wang}, J.~{Wang}, W.~{Wang}, Y.~{Wang},
  X.~{Wen}, B.~{Wu}, B.~{Wu}, M.~{Wu}, G.~{Xiao}, S.~{Xiong}, L.~{Yan},
  J.~{Yang}, S.~{Yang}, Y.~{Yang}, Q.~{Yi}, B.~{Yuan}, A.~{Zhang}, C.~{Zhang},
  C.~{Zhang}, F.~{Zhang}, H.~{Zhang}, J.~{Zhang}, Q.~{Zhang}, S.~{Zhang},
  S.~{Zhang}, T.~{Zhang}, W.~{Zhang}, W.~{Zhang}, W.~{Zhang}, Y.~{Zhang},
  Y.~{Zhang}, Y.~{Zhang}, Y.~{Zhang}, Z.~{Zhang}, Z.~{Zhang}, Z.~{Zhang},
  H.~{Zhao}, X.~{Zhao}, S.~{Zheng}, J.~{Zhou}, Y.~{Zhu}, Y.~{Zhu}, R.~{Zhuang},
  {Insight-HXMT Team}, Science China Physics, Mechanics, and Astronomy
  \textbf{63}(4), 249502 (2020).
\newblock \doi{10.1007/s11433-019-1432-6}

\bibitem{nandra2013}
K.~Nandra, D.~Barret, X.~Barcons, A.~Fabian, J.W.d. Herder, L.~Piro, M.~Watson,
  C.~Adami, J.~Aird, J.M. Afonso, et~al., arXiv preprint arXiv:1306.2307
  (2013)

\bibitem{zhang2019}
S.~{Zhang}, A.~{Santangelo}, M.~{Feroci}, Y.~{Xu}, F.~{Lu}, Y.~{Chen},
  H.~{Feng}, S.~{Zhang}, S.~{Brandt}, M.~{Hernanz}, L.~{Baldini}, E.~{Bozzo},
  R.~{Campana}, A.~{De Rosa}, Y.~{Dong}, Y.~{Evangelista}, V.~{Karas},
  N.~{Meidinger}, A.~{Meuris}, K.~{Nandra}, T.~{Pan}, G.~{Pareschi},
  P.~{Orleanski}, Q.~{Huang}, S.~{Schanne}, G.~{Sironi}, D.~{Spiga},
  J.~{Svoboda}, G.~{Tagliaferri}, C.~{Tenzer}, A.~{Vacchi}, S.~{Zane},
  D.~{Walton}, Z.~{Wang}, B.~{Winter}, X.~{Wu}, J.J.M. {in't Zand},
  M.~{Ahangarianabhari}, G.~{Ambrosi}, F.~{Ambrosino}, M.~{Barbera},
  S.~{Basso}, J.~{Bayer}, R.~{Bellazzini}, P.~{Bellutti}, B.~{Bertucci},
  G.~{Bertuccio}, G.~{Borghi}, X.~{Cao}, F.~{Cadoux}, R.~{Campana},
  F.~{Ceraudo}, T.~{Chen}, Y.~{Chen}, J.~{Chevenez}, M.~{Civitani}, W.~{Cui},
  W.~{Cui}, T.~{Dauser}, E.~{Del Monte}, S.~{Di Cosimo}, S.~{Diebold},
  V.~{Doroshenko}, M.~{Dovciak}, Y.~{Du}, L.~{Ducci}, Q.~{Fan}, Y.~{Favre},
  F.~{Fuschino}, J.L. {G{\'a}lvez}, M.~{Gao}, M.~{Ge}, O.~{Gevin}, M.~{Grassi},
  Q.~{Gu}, Y.~{Gu}, D.~{Han}, B.~{Hong}, W.~{Hu}, L.~{Ji}, S.~{Jia},
  W.~{Jiang}, T.~{Kennedy}, I.~{Kreykenbohm}, I.~{Kuvvetli}, C.~{Labanti},
  L.~{Latronico}, G.~{Li}, M.~{Li}, X.~{Li}, W.~{Li}, Z.~{Li}, O.~{Limousin},
  H.~{Liu}, X.~{Liu}, B.~{Lu}, T.~{Luo}, D.~{Macera}, P.~{Malcovati},
  A.~{Martindale}, M.~{Michalska}, B.~{Meng}, M.~{Minuti}, A.~{Morbidini},
  F.~{Muleri}, S.~{Paltani}, E.~{Perinati}, A.~{Picciotto}, C.~{Piemonte},
  J.~{Qu}, A.~{Rachevski}, I.~{Rashevskaya}, J.~{Rodriguez}, T.~{Schanz},
  Z.~{Shen}, L.~{Sheng}, J.~{Song}, L.~{Song}, C.~{Sgro}, L.~{Sun}, Y.~{Tan},
  P.~{Uttley}, B.~{Wang}, D.~{Wang}, G.~{Wang}, J.~{Wang}, L.~{Wang},
  Y.~{Wang}, A.L. {Watts}, X.~{Wen}, J.~{Wilms}, S.~{Xiong}, J.~{Yang},
  S.~{Yang}, Y.~{Yang}, N.~{Yu}, W.~{Zhang}, G.~{Zampa}, N.~{Zampa}, A.A.
  {Zdziarski}, A.~{Zhang}, C.~{Zhang}, F.~{Zhang}, L.~{Zhang}, T.~{Zhang},
  Y.~{Zhang}, X.~{Zhang}, Z.~{Zhang}, B.~{Zhao}, S.~{Zheng}, Y.~{Zhou},
  N.~{Zorzi}, J.F. {Zwart}, Science China Physics, Mechanics, and Astronomy
  \textbf{62}(2), 29502 (2019).
\newblock \doi{10.1007/s11433-018-9309-2}

\bibitem{Ray2019}
P.S. {Ray}, Z.~{Arzoumanian}, D.~{Ballantyne}, E.~{Bozzo}, S.~{Brandt},
  L.~{Brenneman}, D.~{Chakrabarty}, M.~{Christophersen}, A.~{DeRosa},
  M.~{Feroci}, K.~{Gendreau}, A.~{Goldstein}, D.~{Hartmann}, M.~{Hernanz},
  P.~{Jenke}, E.~{Kara}, T.~{Maccarone}, M.~{McDonald}, M.~{Nowak},
  B.~{Phlips}, R.~{Remillard}, A.~{Stevens}, J.~{Tomsick}, A.~{Watts},
  C.~{Wilson-Hodge}, K.~{Wood}, S.~{Zane}, M.~{Ajello}, W.~{Alston},
  D.~{Altamirano}, V.~{Antoniou}, K.~{Arur}, D.~{Ashton}, K.~{Auchettl},
  T.~{Ayres}, M.~{Bachetti}, M.~{Balokovic}, M.~{Baring}, A.~{Baykal},
  M.~{Begelman}, N.~{Bhat}, S.~{Bogdanov}, M.~{Briggs}, E.~{Bulbul}, P.~{Bult},
  E.~{Burns}, E.~{Cackett}, R.~{Campana}, A.~{Caspi}, Y.~{Cavecchi},
  J.~{Chenevez}, M.~{Cherry}, R.~{Corbet}, M.~{Corcoran}, A.~{Corsi},
  N.~{Degenaar}, J.~{Drake}, S.~{Eikenberry}, T.~{Enoto}, C.~{Fragile},
  F.~{Fuerst}, P.~{Gandhi}, J.~{Garcia}, A.~{Goldstein}, A.~{Gonzalez},
  B.~{Grefenstette}, V.~{Grinberg}, B.~{Grossan}, S.~{Guillot}, T.~{Guver},
  D.~{Haggard}, C.~{Heinke}, S.~{Heinz}, P.~{Hemphill}, J.~{Homan}, M.~{Hui},
  D.~{Huppenkothen}, A.~{Ingram}, J.~{Irwin}, G.~{Jaisawal}, A.~{Jaodand},
  E.~{Kalemci}, D.~{Kaplan}, L.~{Keek}, J.~{Kennea}, M.~{Kerr}, M.~{van der
  Klis}, D.~{Kocevski}, M.~{Koss}, A.~{Kowalski}, D.~{Lai}, F.~{Lamb},
  S.~{Laycock}, J.~{Lazio}, D.~{Lazzati}, D.~{Longcope}, M.~{Loewenstein},
  D.~{Maitra}, W.~{Majid}, W.P. {Maksym}, C.~{Malacaria}, R.~{Margutti},
  A.~{Martindale}, I.~{McHardy}, M.~{Meyer}, M.~{Middleton}, J.~{Miller},
  C.~{Miller}, S.~{Motta}, J.~{Neilsen}, T.~{Nelson}, S.~{Noble}, P.~{O'Brien},
  J.~{Osborne}, R.~{Osten}, F.~{Ozel}, N.~{Palliyaguru}, D.~{Pasham},
  A.~{Patruno}, V.~{Pelassa}, M.~{Petropoulou}, M.~{Pilia}, M.~{Pohl},
  D.~{Pooley}, C.~{Prescod-Weinstein}, D.~{Psaltis}, G.~{Raaijmakers},
  C.~{Reynolds}, T.E. {Riley}, G.~{Salvesen}, A.~{Santangelo}, S.~{Scaringi},
  S.~{Schanne}, J.~{Schnittman}, D.~{Smith}, K.L. {Smith}, B.~{Snios},
  A.~{Steiner}, J.~{Steiner}, L.~{Stella}, T.~{Strohmayer}, M.~{Sun},
  T.~{Tauris}, C.~{Taylor}, A.~{Tohuvavohu}, A.~{Vacchi}, G.~{Vasilopoulos},
  A.~{Veledina}, J.~{Walsh}, N.~{Weinberg}, D.~{Wilkins}, R.~{Willingale},
  J.~{Wilms}, L.~{Winter}, M.~{Wolff}, J.~{in 't Zand}, A.~{Zezas}, B.~{Zhang},
  A.~{Zoghbi}, arXiv e-prints arXiv:1903.03035 (2019)

\bibitem{Gaskin2019}
J.A. {Gaskin}, D.A. {Swartz}, A.~{Vikhlinin}, F.~{{\"O}zel}, K.E. {Gelmis},
  J.W. {Arenberg}, S.R. {Bandler}, M.W. {Bautz}, M.M. {Civitani},
  A.~{Dominguez}, M.E. {Eckart}, A.D. {Falcone}, E.~{Figueroa-Feliciano}, M.D.
  {Freeman}, H.M. {G{\"u}nther}, K.A. {Havey}, R.K. {Heilmann}, K.~{Kilaru},
  R.P. {Kraft}, K.S. {McCarley}, R.L. {McEntaffer}, G.~{Pareschi},
  W.~{Purcell}, P.B. {Reid}, M.L. {Schattenburg}, D.A. {Schwartz}, E.D.
  {Schwartz}, H.D. {Tananbaum}, G.R. {Tremblay}, W.W. {Zhang}, J.A. {Zuhone},
  Journal of Astronomical Telescopes, Instruments, and Systems \textbf{5},
  021001 (2019).
\newblock \doi{10.1117/1.JATIS.5.2.021001}

\bibitem{XRISM2020}
{XRISM Science Team}, arXiv e-prints arXiv:2003.04962 (2020)

\bibitem{XRISM2022}
{XRISM Science Team}, arXiv e-prints arXiv:2202.05399 (2022)

\bibitem{Cui2020}
W.~{Cui}, J.N. {Bregman}, M.P. {Bruijn}, L.B. {Chen}, Y.~{Chen}, C.~{Cui}, T.T.
  {Fang}, B.~{Gao}, H.~{Gao}, J.R. {Gao}, L.~{Gottardi}, K.X. {Gu}, F.L. {Guo},
  J.~{Guo}, C.L. {He}, P.F. {He}, J.W. {den Herder}, Q.S. {Huang}, F.J. {Li},
  J.T. {Li}, J.J. {Li}, L.Y. {Li}, T.P. {Li}, W.B. {Li}, J.T. {Liang}, Y.J.
  {Liang}, G.Y. {Liang}, Y.J. {Liu}, Z.~{Liu}, Z.Y. {Liu}, F.~{Jaeckel},
  L.~{Ji}, W.~{Ji}, H.~{Jin}, X.~{Kang}, Y.X. {Ma}, D.~{McCammon}, H.J. {Mo},
  K.~{Nagayoshi}, K.~{Nelms}, R.~{Qi}, J.~{Quan}, M.L. {Ridder}, Z.X. {Shen},
  A.~{Simionescu}, E.~{Taralli}, Q.D. {Wang}, G.L. {Wang}, J.J. {Wang},
  K.~{Wang}, L.~{Wang}, S.F. {Wang}, S.J. {Wang}, T.G. {Wang}, W.~{Wang}, X.Q.
  {Wang}, Y.L. {Wang}, Y.R. {Wang}, Z.~{Wang}, Z.S. {Wang}, N.Y. {Wen}, M.~{de
  Wit}, S.F. {Wu}, D.~{Xu}, D.D. {Xu}, H.G. {Xu}, X.J. {Xu}, R.X. {Xu}, Y.Q.
  {Xue}, S.Z. {Yi}, J.~{Yu}, L.W. {Yang}, F.~{Yuan}, S.~{Zhang}, W.~{Zhang},
  Z.~{Zhang}, Q.~{Zhong}, Y.~{Zhou}, W.X. {Zhu}, in \emph{Society of
  Photo-Optical Instrumentation Engineers (SPIE) Conference Series},
  \emph{Society of Photo-Optical Instrumentation Engineers (SPIE) Conference
  Series}, vol. 11444 (2020), \emph{Society of Photo-Optical Instrumentation
  Engineers (SPIE) Conference Series}, vol. 11444, p. 114442S.
\newblock \doi{10.1117/12.2560871}

\bibitem{Burrows2005}
D.N. {Burrows}, J.E. {Hill}, J.A. {Nousek}, J.A. {Kennea}, A.~{Wells}, J.P.
  {Osborne}, A.F. {Abbey}, A.~{Beardmore}, K.~{Mukerjee}, A.D.T. {Short},
  G.~{Chincarini}, S.~{Campana}, O.~{Citterio}, A.~{Moretti}, C.~{Pagani},
  G.~{Tagliaferri}, P.~{Giommi}, M.~{Capalbi}, F.~{Tamburelli}, L.~{Angelini},
  G.~{Cusumano}, H.W. {Br{\"a}uninger}, W.~{Burkert}, G.D. {Hartner}, Space
  science reviews \textbf{120}(3-4), 165 (2005).
\newblock \doi{10.1007/s11214-005-5097-2}

\bibitem{Matsuoka2009}
M.~{Matsuoka}, K.~{Kawasaki}, S.~{Ueno}, H.~{Tomida}, M.~{Kohama}, M.~{Suzuki},
  Y.~{Adachi}, M.~{Ishikawa}, T.~{Mihara}, M.~{Sugizaki}, N.~{Isobe},
  Y.~{Nakagawa}, H.~{Tsunemi}, E.~{Miyata}, N.~{Kawai}, J.~{Kataoka},
  M.~{Morii}, A.~{Yoshida}, H.~{Negoro}, M.~{Nakajima}, Y.~{Ueda}, H.~{Chujo},
  K.~{Yamaoka}, O.~{Yamazaki}, S.~{Nakahira}, T.~{You}, R.~{Ishiwata},
  S.~{Miyoshi}, S.~{Eguchi}, K.~{Hiroi}, H.~{Katayama}, K.~{Ebisawa},
  Publications of the Astronomical Society of Japan \textbf{61}, 999 (2009).
\newblock \doi{10.1093/pasj/61.5.999}

\bibitem{Predehl2021}
P.~{Predehl}, R.~{Andritschke}, V.~{Arefiev}, V.~{Babyshkin}, O.~{Batanov},
  W.~{Becker}, H.~{B{\"o}hringer}, A.~{Bogomolov}, T.~{Boller}, K.~{Borm},
  W.~{Bornemann}, H.~{Br{\"a}uninger}, M.~{Br{\"u}ggen}, H.~{Brunner},
  M.~{Brusa}, E.~{Bulbul}, M.~{Buntov}, V.~{Burwitz}, W.~{Burkert}, N.~{Clerc},
  E.~{Churazov}, D.~{Coutinho}, T.~{Dauser}, K.~{Dennerl}, V.~{Doroshenko},
  J.~{Eder}, V.~{Emberger}, T.~{Eraerds}, A.~{Finoguenov}, M.~{Freyberg},
  P.~{Friedrich}, S.~{Friedrich}, M.~{F{\"u}rmetz}, A.~{Georgakakis},
  M.~{Gilfanov}, S.~{Granato}, C.~{Grossberger}, A.~{Gueguen}, P.~{Gureev},
  F.~{Haberl}, O.~{H{\"a}lker}, G.~{Hartner}, G.~{Hasinger}, H.~{Huber},
  L.~{Ji}, A.v. {Kienlin}, W.~{Kink}, F.~{Korotkov}, I.~{Kreykenbohm},
  G.~{Lamer}, I.~{Lomakin}, I.~{Lapshov}, T.~{Liu}, C.~{Maitra},
  N.~{Meidinger}, B.~{Menz}, A.~{Merloni}, T.~{Mernik}, B.~{Mican}, J.~{Mohr},
  S.~{M{\"u}ller}, K.~{Nandra}, V.~{Nazarov}, F.~{Pacaud}, M.~{Pavlinsky},
  E.~{Perinati}, E.~{Pfeffermann}, D.~{Pietschner}, M.E. {Ramos-Ceja},
  A.~{Rau}, J.~{Reiffers}, T.H. {Reiprich}, J.~{Robrade}, M.~{Salvato},
  J.~{Sanders}, A.~{Santangelo}, M.~{Sasaki}, H.~{Scheuerle}, C.~{Schmid},
  J.~{Schmitt}, A.~{Schwope}, A.~{Shirshakov}, M.~{Steinmetz}, I.~{Stewart},
  L.~{Str{\"u}der}, R.~{Sunyaev}, C.~{Tenzer}, L.~{Tiedemann},
  J.~{Tr{\"u}mper}, V.~{Voron}, P.~{Weber}, J.~{Wilms}, V.~{Yaroshenko},
  Astronomy \& Astrophysics \textbf{647}, A1 (2021).
\newblock \doi{10.1051/0004-6361/202039313}

\bibitem{Camp2019}
J.~{Camp}, J.~{Abel}, S.~{Barthelmy}, M.~{Bautz}, E.~{Behar}, E.~{Berger},
  S.~{Spolaor}, S.B. {Cenko}, N.~{Cornish}, T.~{Dal Canton}, C.~{Fryer},
  S.~{Gezari}, P.~{Gorenstein}, S.~{Guiriec}, D.~{Hartmann}, V.~{Kalogera},
  C.~{Kouveliotou}, J.~{Kruk}, A.~{Kutyrev}, R.~{Margutti}, F.~{Marshall},
  B.~{Metzger}, C.~{Miller}, S.~{Noble}, J.~{Perkins}, A.~{Ptak}, B.~{Purcell},
  J.~{Racusin}, J.~{Schlieder}, J.~{Schnittman}, A.~{Sesana}, P.~{Shawhan},
  L.~{Singer}, A.~{van der Horst}, R.~{Willingale}, K.~{Wood}, W.~{Zhang}, in
  \emph{Bulletin of the American Astronomical Society}, vol.~51 (2019),
  vol.~51, p.~85

\bibitem{Harrison2013}
F.A. {Harrison}, W.W. {Craig}, F.E. {Christensen}, C.J. {Hailey}, W.W. {Zhang},
  S.E. {Boggs}, D.~{Stern}, W.R. {Cook}, K.~{Forster}, P.~{Giommi}, B.W.
  {Grefenstette}, Y.~{Kim}, T.~{Kitaguchi}, J.E. {Koglin}, K.K. {Madsen}, P.H.
  {Mao}, H.~{Miyasaka}, K.~{Mori}, M.~{Perri}, M.J. {Pivovaroff},
  S.~{Puccetti}, V.R. {Rana}, N.J. {Westergaard}, J.~{Willis}, A.~{Zoglauer},
  H.~{An}, M.~{Bachetti}, N.M. {Barri{\`e}re}, E.C. {Bellm}, V.~{Bhalerao},
  N.F. {Brejnholt}, F.~{Fuerst}, C.C. {Liebe}, C.B. {Markwardt}, M.~{Nynka},
  J.K. {Vogel}, D.J. {Walton}, D.R. {Wik}, D.M. {Alexander}, L.R. {Cominsky},
  A.E. {Hornschemeier}, A.~{Hornstrup}, V.M. {Kaspi}, G.M. {Madejski},
  G.~{Matt}, S.~{Molendi}, D.M. {Smith}, J.A. {Tomsick}, M.~{Ajello}, D.R.
  {Ballantyne}, M.~{Balokovi{\'c}}, D.~{Barret}, F.E. {Bauer}, R.D.
  {Blandford}, W.N. {Brandt}, L.W. {Brenneman}, J.~{Chiang}, D.~{Chakrabarty},
  J.~{Chenevez}, A.~{Comastri}, F.~{Dufour}, M.~{Elvis}, A.C. {Fabian},
  D.~{Farrah}, C.L. {Fryer}, E.V. {Gotthelf}, J.E. {Grindlay}, D.J. {Helfand},
  R.~{Krivonos}, D.L. {Meier}, J.M. {Miller}, L.~{Natalucci}, P.~{Ogle}, E.O.
  {Ofek}, A.~{Ptak}, S.P. {Reynolds}, J.R. {Rigby}, G.~{Tagliaferri}, S.E.
  {Thorsett}, E.~{Treister}, C.M. {Urry}, The Astrophysical Journal
  \textbf{770}(2), 103 (2013).
\newblock \doi{10.1088/0004-637X/770/2/103}

\bibitem{Singh2014}
K.P. {Singh}, S.N. {Tandon}, P.C. {Agrawal}, H.M. {Antia}, R.K. {Manchanda},
  J.S. {Yadav}, S.~{Seetha}, M.C. {Ramadevi}, A.R. {Rao}, D.~{Bhattacharya},
  B.~{Paul}, P.~{Sreekumar}, S.~{Bhattacharyya}, G.C. {Stewart},
  J.~{Hutchings}, S.A. {Annapurni}, S.K. {Ghosh}, J.~{Murthy}, A.~{Pati}, N.K.
  {Rao}, C.S. {Stalin}, V.~{Girish}, K.~{Sankarasubramanian}, S.~{Vadawale},
  V.B. {Bhalerao}, G.C. {Dewangan}, D.K. {Dedhia}, M.K. {Hingar}, T.B.
  {Katoch}, A.T. {Kothare}, I.~{Mirza}, K.~{Mukerjee}, H.~{Shah}, P.~{Shah},
  R.~{Mohan}, A.K. {Sangal}, S.~{Nagabhusana}, S.~{Sriram}, J.P. {Malkar},
  S.~{Sreekumar}, A.F. {Abbey}, G.M. {Hansford}, A.P. {Beardmore}, M.R.
  {Sharma}, S.~{Murthy}, R.~{Kulkarni}, G.~{Meena}, V.C. {Babu}, J.~{Postma},
  in \emph{Space Telescopes and Instrumentation 2014: Ultraviolet to Gamma
  Ray}, \emph{Society of Photo-Optical Instrumentation Engineers (SPIE)
  Conference Series}, vol. 9144, ed. by T.~{Takahashi}, J.W.A. {den Herder},
  M.~{Bautz} (2014), \emph{Society of Photo-Optical Instrumentation Engineers
  (SPIE) Conference Series}, vol. 9144, p. 91441S.
\newblock \doi{10.1117/12.2062667}

\bibitem{Weisskopf2016}
M.C. {Weisskopf}, B.~{Ramsey}, S.~{O'Dell}, A.~{Tennant}, R.~{Elsner},
  P.~{Soffitta}, R.~{Bellazzini}, E.~{Costa}, J.~{Kolodziejczak}, V.~{Kaspi},
  F.~{Muleri}, H.~{Marshall}, G.~{Matt}, R.~{Romani}, in \emph{Space Telescopes
  and Instrumentation 2016: Ultraviolet to Gamma Ray}, \emph{Society of
  Photo-Optical Instrumentation Engineers (SPIE) Conference Series}, vol. 9905,
  ed. by J.W.A. {den Herder}, T.~{Takahashi}, M.~{Bautz} (2016), \emph{Society
  of Photo-Optical Instrumentation Engineers (SPIE) Conference Series}, vol.
  9905, p. 990517.
\newblock \doi{10.1117/12.2235240}

\bibitem{Bernardini2021}
M.G. {Bernardini}, B.~{Cordier}, J.~{Wei}, Galaxies \textbf{9}(4), 113 (2021).
\newblock \doi{10.3390/galaxies9040113}

\bibitem{Barret2018}
D.~{Barret}, T.~{Lam Trong}, J.W. {den Herder}, L.~{Piro}, M.~{Cappi},
  J.~{Houvelin}, R.~{Kelley}, J.M. {Mas-Hesse}, K.~{Mitsuda}, S.~{Paltani},
  G.~{Rauw}, A.~{Rozanska}, J.~{Wilms}, S.~{Bandler}, M.~{Barbera},
  X.~{Barcons}, E.~{Bozzo}, M.T. {Ceballos}, I.~{Charles}, E.~{Costantini},
  A.~{Decourchelle}, R.~{den Hartog}, L.~{Duband}, J.M. {Duval}, F.~{Fiore},
  F.~{Gatti}, A.~{Goldwurm}, B.~{Jackson}, P.~{Jonker}, C.~{Kilbourne},
  C.~{Macculi}, M.~{Mendez}, S.~{Molendi}, P.~{Orleanski}, F.~{Pajot},
  E.~{Pointecouteau}, F.~{Porter}, G.W. {Pratt}, D.~{Pr{\^e}le}, L.~{Ravera},
  K.~{Sato}, J.~{Schaye}, K.~{Shinozaki}, T.~{Thibert}, L.~{Valenziano},
  V.~{Valette}, J.~{Vink}, N.~{Webb}, M.~{Wise}, N.~{Yamasaki}, F.~{Douchin},
  J.M. {Mesnager}, B.~{Pontet}, A.~{Pradines}, G.~{Branduardi-Raymont},
  E.~{Bulbul}, M.~{Dadina}, S.~{Ettori}, A.~{Finoguenov}, Y.~{Fukazawa},
  A.~{Janiuk}, J.~{Kaastra}, P.~{Mazzotta}, J.~{Miller}, G.~{Miniutti},
  Y.~{Naze}, F.~{Nicastro}, S.~{Scioritino}, A.~{Simonescu}, J.M. {Torrejon},
  B.~{Frezouls}, H.~{Geoffray}, P.~{Peille}, C.~{Aicardi}, J.~{Andr{\'e}},
  C.~{Daniel}, A.~{Cl{\'e}net}, C.~{Etcheverry}, E.~{Gloaguen}, G.~{Hervet},
  A.~{Jolly}, A.~{Ledot}, I.~{Paillet}, R.~{Schmisser}, B.~{Vella}, J.C.
  {Damery}, K.~{Boyce}, M.~{Dipirro}, S.~{Lotti}, D.~{Schwander}, S.~{Smith},
  B.J. {Van Leeuwen}, H.~{van Weers}, N.~{Clerc}, B.~{Cobo}, T.~{Dauser},
  C.~{Kirsch}, E.~{Cucchetti}, M.~{Eckart}, P.~{Ferrando}, L.~{Natalucci}, in
  \emph{Space Telescopes and Instrumentation 2018: Ultraviolet to Gamma Ray},
  \emph{Society of Photo-Optical Instrumentation Engineers (SPIE) Conference
  Series}, vol. 10699, ed. by J.W.A. {den Herder}, S.~{Nikzad}, K.~{Nakazawa}
  (2018), \emph{Society of Photo-Optical Instrumentation Engineers (SPIE)
  Conference Series}, vol. 10699, p. 106991G.
\newblock \doi{10.1117/12.2312409}

\bibitem{Espinoza2011}
C.M. {Espinoza}, A.G. {Lyne}, B.W. {Stappers}, M.~{Kramer}, Monthly Notices of
  the Royal Astronomical Society \textbf{414}(2), 1679 (2011).
\newblock \doi{10.1111/j.1365-2966.2011.18503.x}

\bibitem{kaspi2017}
V.M. {Kaspi}, A.M. {Beloborodov}, Annual Review of Astronomy and Astrophysics
  \textbf{55}(1), 261 (2017).
\newblock \doi{10.1146/annurev-astro-081915-023329}

\bibitem{Abbott2017}
B.P. {Abbott}, R.~{Abbott}, T.D. {Abbott}, F.~{Acernese}, K.~{Ackley},
  C.~{Adams}, T.~{Adams}, P.~{Addesso}, R.X. {Adhikari}, V.B. {Adya},
  C.~{Affeldt}, M.~{Afrough}, B.~{Agarwal}, M.~{Agathos}, K.~{Agatsuma},
  N.~{Aggarwal}, O.D. {Aguiar}, L.~{Aiello}, A.~{Ain}, P.~{Ajith}, B.~{Allen},
  G.~{Allen}, A.~{Allocca}, P.A. {Altin}, A.~{Amato}, A.~{Ananyeva}, S.B.
  {Anderson}, W.G. {Anderson}, S.V. {Angelova}, S.~{Antier}, S.~{Appert},
  K.~{Arai}, M.C. {Araya}, J.S. {Areeda}, N.~{Arnaud}, K.G. {Arun},
  S.~{Ascenzi}, G.~{Ashton}, M.~{Ast}, S.M. {Aston}, P.~{Astone}, D.V.
  {Atallah}, P.~{Aufmuth}, C.~{Aulbert}, K.~{AultONeal}, C.~{Austin},
  A.~{Avila-Alvarez}, S.~{Babak}, P.~{Bacon}, M.K.M. {Bader}, S.~{Bae}, P.T.
  {Baker}, F.~{Baldaccini}, G.~{Ballardin}, S.W. {Ballmer}, S.~{Banagiri}, J.C.
  {Barayoga}, S.E. {Barclay}, B.C. {Barish}, D.~{Barker}, K.~{Barkett},
  F.~{Barone}, B.~{Barr}, L.~{Barsotti}, M.~{Barsuglia}, D.~{Barta}, S.D.
  {Barthelmy}, J.~{Bartlett}, I.~{Bartos}, R.~{Bassiri}, A.~{Basti}, J.C.
  {Batch}, M.~{Bawaj}, J.C. {Bayley}, M.~{Bazzan}, B.~{B{\'e}csy}, C.~{Beer},
  M.~{Bejger}, I.~{Belahcene}, A.S. {Bell}, B.K. {Berger}, G.~{Bergmann}, J.J.
  {Bero}, C.P.L. {Berry}, D.~{Bersanetti}, A.~{Bertolini}, J.~{Betzwieser},
  S.~{Bhagwat}, R.~{Bhandare}, I.A. {Bilenko}, G.~{Billingsley}, C.R.
  {Billman}, J.~{Birch}, R.~{Birney}, O.~{Birnholtz}, S.~{Biscans},
  S.~{Biscoveanu}, A.~{Bisht}, M.~{Bitossi}, C.~{Biwer}, M.A. {Bizouard}, J.K.
  {Blackburn}, J.~{Blackman}, C.D. {Blair}, D.G. {Blair}, R.M. {Blair},
  S.~{Bloemen}, O.~{Bock}, N.~{Bode}, M.~{Boer}, G.~{Bogaert}, A.~{Bohe},
  F.~{Bondu}, E.~{Bonilla}, R.~{Bonnand}, B.A. {Boom}, R.~{Bork}, V.~{Boschi},
  S.~{Bose}, K.~{Bossie}, Y.~{Bouffanais}, A.~{Bozzi}, C.~{Bradaschia}, P.R.
  {Brady}, M.~{Branchesi}, J.E. {Brau}, T.~{Briant}, A.~{Brillet},
  M.~{Brinkmann}, V.~{Brisson}, P.~{Brockill}, J.E. {Broida}, A.F. {Brooks},
  D.A. {Brown}, D.D. {Brown}, S.~{Brunett}, C.C. {Buchanan}, A.~{Buikema},
  T.~{Bulik}, H.J. {Bulten}, A.~{Buonanno}, D.~{Buskulic}, C.~{Buy}, R.L.
  {Byer}, M.~{Cabero}, L.~{Cadonati}, G.~{Cagnoli}, C.~{Cahillane},
  J.~{Calder{\'o}n Bustillo}, T.A. {Callister}, E.~{Calloni}, J.B. {Camp},
  M.~{Canepa}, P.~{Canizares}, K.C. {Cannon}, H.~{Cao}, J.~{Cao}, C.D.
  {Capano}, E.~{Capocasa}, F.~{Carbognani}, S.~{Caride}, M.F. {Carney},
  J.~{Casanueva Diaz}, C.~{Casentini}, S.~{Caudill}, M.~{Cavagli{\`a}},
  F.~{Cavalier}, R.~{Cavalieri}, G.~{Cella}, C.B. {Cepeda},
  P.~{Cerd{\'a}-Dur{\'a}n}, G.~{Cerretani}, E.~{Cesarini}, S.J. {Chamberlin},
  M.~{Chan}, S.~{Chao}, P.~{Charlton}, E.~{Chase}, E.~{Chassande-Mottin},
  D.~{Chatterjee}, K.~{Chatziioannou}, B.D. {Cheeseboro}, H.Y. {Chen},
  X.~{Chen}, Y.~{Chen}, H.P. {Cheng}, H.~{Chia}, A.~{Chincarini}, A.~{Chiummo},
  T.~{Chmiel}, H.S. {Cho}, M.~{Cho}, J.H. {Chow}, N.~{Christensen}, Q.~{Chu},
  A.J.K. {Chua}, S.~{Chua}, A.K.W. {Chung}, S.~{Chung}, G.~{Ciani},
  R.~{Ciolfi}, C.E. {Cirelli}, A.~{Cirone}, F.~{Clara}, J.A. {Clark},
  P.~{Clearwater}, F.~{Cleva}, C.~{Cocchieri}, E.~{Coccia}, P.F. {Cohadon},
  D.~{Cohen}, A.~{Colla}, C.G. {Collette}, L.R. {Cominsky}, J.~{Constancio},
  M., L.~{Conti}, S.J. {Cooper}, P.~{Corban}, T.R. {Corbitt},
  I.~{Cordero-Carri{\'o}n}, K.R. {Corley}, N.~{Cornish}, A.~{Corsi},
  S.~{Cortese}, C.A. {Costa}, M.W. {Coughlin}, S.B. {Coughlin}, J.P. {Coulon},
  S.T. {Countryman}, P.~{Couvares}, P.B. {Covas}, E.E. {Cowan}, D.M. {Coward},
  M.J. {Cowart}, D.C. {Coyne}, R.~{Coyne}, J.D.E. {Creighton}, T.D.
  {Creighton}, J.~{Cripe}, S.G. {Crowder}, T.J. {Cullen}, A.~{Cumming},
  L.~{Cunningham}, E.~{Cuoco}, T.~{Dal Canton}, G.~{D{\'a}lya}, S.L.
  {Danilishin}, S.~{D'Antonio}, K.~{Danzmann}, A.~{Dasgupta}, C.F. {Da Silva
  Costa}, V.~{Dattilo}, I.~{Dave}, M.~{Davier}, D.~{Davis}, E.J. {Daw},
  B.~{Day}, S.~{De}, D.~{DeBra}, J.~{Degallaix}, M.~{De Laurentis},
  S.~{Del{\'e}glise}, W.~{Del Pozzo}, N.~{Demos}, T.~{Denker}, T.~{Dent},
  R.~{De Pietri}, V.~{Dergachev}, R.~{De Rosa}, R.T. {DeRosa}, C.~{De Rossi},
  R.~{DeSalvo}, O.~{de Varona}, J.~{Devenson}, S.~{Dhurandhar}, M.C.
  {D{\'\i}az}, L.~{Di Fiore}, M.~{Di Giovanni}, T.~{Di Girolamo}, A.~{Di
  Lieto}, S.~{Di Pace}, I.~{Di Palma}, F.~{Di Renzo}, Z.~{Doctor},
  V.~{Dolique}, F.~{Donovan}, K.L. {Dooley}, S.~{Doravari}, I.~{Dorrington},
  R.~{Douglas}, M.~{Dovale {\'A}lvarez}, T.P. {Downes}, M.~{Drago},
  C.~{Dreissigacker}, J.C. {Driggers}, Z.~{Du}, M.~{Ducrot}, P.~{Dupej}, S.E.
  {Dwyer}, T.B. {Edo}, M.C. {Edwards}, A.~{Effler}, P.~{Ehrens}, J.~{Eichholz},
  S.S. {Eikenberry}, R.A. {Eisenstein}, R.C. {Essick}, D.~{Estevez}, Z.B.
  {Etienne}, T.~{Etzel}, M.~{Evans}, T.M. {Evans}, M.~{Factourovich},
  V.~{Fafone}, H.~{Fair}, S.~{Fairhurst}, X.~{Fan}, S.~{Farinon}, B.~{Farr},
  W.M. {Farr}, E.J. {Fauchon-Jones}, M.~{Favata}, M.~{Fays}, C.~{Fee},
  H.~{Fehrmann}, J.~{Feicht}, M.M. {Fejer}, A.~{Fernandez-Galiana},
  I.~{Ferrante}, E.C. {Ferreira}, F.~{Ferrini}, F.~{Fidecaro}, D.~{Finstad},
  I.~{Fiori}, D.~{Fiorucci}, M.~{Fishbach}, R.P. {Fisher}, M.~{Fitz-Axen},
  R.~{Flaminio}, M.~{Fletcher}, H.~{Fong}, J.A. {Font}, P.W.F. {Forsyth}, S.S.
  {Forsyth}, J.D. {Fournier}, S.~{Frasca}, F.~{Frasconi}, Z.~{Frei},
  A.~{Freise}, R.~{Frey}, V.~{Frey}, E.M. {Fries}, P.~{Fritschel}, V.V.
  {Frolov}, P.~{Fulda}, M.~{Fyffe}, H.~{Gabbard}, B.U. {Gadre}, S.M. {Gaebel},
  J.R. {Gair}, L.~{Gammaitoni}, M.R. {Ganija}, S.G. {Gaonkar},
  C.~{Garcia-Quiros}, F.~{Garufi}, B.~{Gateley}, S.~{Gaudio}, G.~{Gaur},
  V.~{Gayathri}, N.~{Gehrels}, G.~{Gemme}, E.~{Genin}, A.~{Gennai},
  D.~{George}, J.~{George}, L.~{Gergely}, V.~{Germain}, S.~{Ghonge},
  A.~{Ghosh}, A.~{Ghosh}, S.~{Ghosh}, J.A. {Giaime}, K.D. {Giardina},
  A.~{Giazotto}, K.~{Gill}, L.~{Glover}, E.~{Goetz}, R.~{Goetz}, S.~{Gomes},
  B.~{Goncharov}, G.~{Gonz{\'a}lez}, J.M. {Gonzalez Castro}, A.~{Gopakumar},
  M.L. {Gorodetsky}, S.E. {Gossan}, M.~{Gosselin}, R.~{Gouaty}, A.~{Grado},
  C.~{Graef}, M.~{Granata}, A.~{Grant}, S.~{Gras}, C.~{Gray}, G.~{Greco}, A.C.
  {Green}, E.M. {Gretarsson}, B.~{Griswold}, P.~{Groot}, H.~{Grote},
  S.~{Grunewald}, P.~{Gruning}, G.M. {Guidi}, X.~{Guo}, A.~{Gupta}, M.K.
  {Gupta}, K.E. {Gushwa}, E.K. {Gustafson}, R.~{Gustafson}, O.~{Halim}, B.R.
  {Hall}, E.D. {Hall}, E.Z. {Hamilton}, G.~{Hammond}, M.~{Haney}, M.M. {Hanke},
  J.~{Hanks}, C.~{Hanna}, M.D. {Hannam}, O.A. {Hannuksela}, J.~{Hanson},
  T.~{Hardwick}, J.~{Harms}, G.M. {Harry}, I.W. {Harry}, M.J. {Hart}, C.J.
  {Haster}, K.~{Haughian}, J.~{Healy}, A.~{Heidmann}, M.C. {Heintze},
  H.~{Heitmann}, P.~{Hello}, G.~{Hemming}, M.~{Hendry}, I.S. {Heng},
  J.~{Hennig}, A.W. {Heptonstall}, M.~{Heurs}, S.~{Hild}, T.~{Hinderer},
  D.~{Hoak}, D.~{Hofman}, K.~{Holt}, D.E. {Holz}, P.~{Hopkins}, C.~{Horst},
  J.~{Hough}, E.A. {Houston}, E.J. {Howell}, A.~{Hreibi}, Y.M. {Hu}, E.A.
  {Huerta}, D.~{Huet}, B.~{Hughey}, S.~{Husa}, S.H. {Huttner}, T.~{Huynh-Dinh},
  N.~{Indik}, R.~{Inta}, G.~{Intini}, H.N. {Isa}, J.M. {Isac}, M.~{Isi}, B.R.
  {Iyer}, K.~{Izumi}, T.~{Jacqmin}, K.~{Jani}, P.~{Jaranowski}, S.~{Jawahar},
  F.~{Jim{\'e}nez-Forteza}, W.W. {Johnson}, D.I. {Jones}, R.~{Jones}, R.J.G.
  {Jonker}, L.~{Ju}, J.~{Junker}, C.V. {Kalaghatgi}, V.~{Kalogera}, B.~{Kamai},
  S.~{Kandhasamy}, G.~{Kang}, J.B. {Kanner}, S.J. {Kapadia}, S.~{Karki}, K.S.
  {Karvinen}, M.~{Kasprzack}, M.~{Katolik}, E.~{Katsavounidis}, W.~{Katzman},
  S.~{Kaufer}, K.~{Kawabe}, F.~{K{\'e}f{\'e}lian}, D.~{Keitel}, A.J. {Kemball},
  R.~{Kennedy}, C.~{Kent}, J.S. {Key}, F.Y. {Khalili}, I.~{Khan}, S.~{Khan},
  Z.~{Khan}, E.A. {Khazanov}, N.~{Kijbunchoo}, C.~{Kim}, J.C. {Kim}, K.~{Kim},
  W.~{Kim}, W.S. {Kim}, Y.M. {Kim}, S.J. {Kimbrell}, E.J. {King}, P.J. {King},
  M.~{Kinley-Hanlon}, R.~{Kirchhoff}, J.S. {Kissel}, L.~{Kleybolte},
  S.~{Klimenko}, T.D. {Knowles}, P.~{Koch}, S.M. {Koehlenbeck}, S.~{Koley},
  V.~{Kondrashov}, A.~{Kontos}, M.~{Korobko}, W.Z. {Korth}, I.~{Kowalska}, D.B.
  {Kozak}, C.~{Kr{\"a}mer}, V.~{Kringel}, B.~{Krishnan}, A.~{Kr{\'o}lak},
  G.~{Kuehn}, P.~{Kumar}, R.~{Kumar}, S.~{Kumar}, L.~{Kuo}, A.~{Kutynia},
  S.~{Kwang}, B.D. {Lackey}, K.H. {Lai}, M.~{Landry}, R.N. {Lang}, J.~{Lange},
  B.~{Lantz}, R.K. {Lanza}, S.L. {Larson}, A.~{Lartaux-Vollard}, P.D. {Lasky},
  M.~{Laxen}, A.~{Lazzarini}, C.~{Lazzaro}, P.~{Leaci}, S.~{Leavey}, C.H.
  {Lee}, H.K. {Lee}, H.M. {Lee}, H.W. {Lee}, K.~{Lee}, J.~{Lehmann},
  A.~{Lenon}, M.~{Leonardi}, N.~{Leroy}, N.~{Letendre}, Y.~{Levin}, T.G.F.
  {Li}, S.D. {Linker}, T.B. {Littenberg}, J.~{Liu}, R.K.L. {Lo}, N.A.
  {Lockerbie}, L.T. {London}, J.E. {Lord}, M.~{Lorenzini}, V.~{Loriette},
  M.~{Lormand}, G.~{Losurdo}, J.D. {Lough}, C.O. {Lousto}, G.~{Lovelace},
  H.~{L{\"u}ck}, D.~{Lumaca}, A.P. {Lundgren}, R.~{Lynch}, Y.~{Ma}, R.~{Macas},
  S.~{Macfoy}, B.~{Machenschalk}, M.~{MacInnis}, D.M. {Macleod}, I.~{Maga{\~n}a
  Hernandez}, F.~{Maga{\~n}a-Sandoval}, L.~{Maga{\~n}a Zertuche}, R.M. {Magee},
  E.~{Majorana}, I.~{Maksimovic}, N.~{Man}, V.~{Mandic}, V.~{Mangano}, G.L.
  {Mansell}, M.~{Manske}, M.~{Mantovani}, F.~{Marchesoni}, F.~{Marion},
  S.~{M{\'a}rka}, Z.~{M{\'a}rka}, C.~{Markakis}, A.S. {Markosyan},
  A.~{Markowitz}, E.~{Maros}, A.~{Marquina}, P.~{Marsh}, F.~{Martelli},
  L.~{Martellini}, I.W. {Martin}, R.M. {Martin}, D.V. {Martynov}, K.~{Mason},
  E.~{Massera}, A.~{Masserot}, T.J. {Massinger}, M.~{Masso-Reid},
  S.~{Mastrogiovanni}, A.~{Matas}, F.~{Matichard}, L.~{Matone}, N.~{Mavalvala},
  N.~{Mazumder}, R.~{McCarthy}, D.E. {McClelland}, S.~{McCormick},
  L.~{McCuller}, S.C. {McGuire}, G.~{McIntyre}, J.~{McIver}, D.J. {McManus},
  L.~{McNeill}, T.~{McRae}, S.T. {McWilliams}, D.~{Meacher}, G.D. {Meadors},
  M.~{Mehmet}, J.~{Meidam}, E.~{Mejuto-Villa}, A.~{Melatos}, G.~{Mendell}, R.A.
  {Mercer}, E.L. {Merilh}, M.~{Merzougui}, S.~{Meshkov}, C.~{Messenger},
  C.~{Messick}, R.~{Metzdorff}, P.M. {Meyers}, H.~{Miao}, C.~{Michel},
  H.~{Middleton}, E.E. {Mikhailov}, L.~{Milano}, A.L. {Miller}, B.B. {Miller},
  J.~{Miller}, M.~{Millhouse}, M.C. {Milovich-Goff}, O.~{Minazzoli},
  Y.~{Minenkov}, J.~{Ming}, C.~{Mishra}, S.~{Mitra}, V.P. {Mitrofanov},
  G.~{Mitselmakher}, R.~{Mittleman}, D.~{Moffa}, A.~{Moggi}, K.~{Mogushi},
  M.~{Mohan}, S.R.P. {Mohapatra}, M.~{Montani}, C.J. {Moore}, D.~{Moraru},
  G.~{Moreno}, S.R. {Morriss}, B.~{Mours}, C.M. {Mow-Lowry}, G.~{Mueller}, A.W.
  {Muir}, A.~{Mukherjee}, D.~{Mukherjee}, S.~{Mukherjee}, N.~{Mukund},
  A.~{Mullavey}, J.~{Munch}, E.A. {Mu{\~n}iz}, M.~{Muratore}, P.G. {Murray},
  K.~{Napier}, I.~{Nardecchia}, L.~{Naticchioni}, R.K. {Nayak}, J.~{Neilson},
  G.~{Nelemans}, T.J.N. {Nelson}, M.~{Nery}, A.~{Neunzert}, L.~{Nevin}, J.M.
  {Newport}, G.~{Newton}, K.K.Y. {Ng}, P.~{Nguyen}, T.T. {Nguyen},
  D.~{Nichols}, A.B. {Nielsen}, S.~{Nissanke}, A.~{Nitz}, A.~{Noack},
  F.~{Nocera}, D.~{Nolting}, C.~{North}, L.K. {Nuttall}, J.~{Oberling}, G.D.
  {O'Dea}, G.H. {Ogin}, J.J. {Oh}, S.H. {Oh}, F.~{Ohme}, M.A. {Okada},
  M.~{Oliver}, P.~{Oppermann}, R.J. {Oram}, B.~{O'Reilly}, R.~{Ormiston}, L.F.
  {Ortega}, R.~{O'Shaughnessy}, S.~{Ossokine}, D.J. {Ottaway}, H.~{Overmier},
  B.J. {Owen}, A.E. {Pace}, J.~{Page}, M.A. {Page}, A.~{Pai}, S.A. {Pai}, J.R.
  {Palamos}, O.~{Palashov}, C.~{Palomba}, A.~{Pal-Singh}, H.~{Pan}, H.W. {Pan},
  B.~{Pang}, P.T.H. {Pang}, C.~{Pankow}, F.~{Pannarale}, B.C. {Pant},
  F.~{Paoletti}, A.~{Paoli}, M.A. {Papa}, A.~{Parida}, W.~{Parker},
  D.~{Pascucci}, A.~{Pasqualetti}, R.~{Passaquieti}, D.~{Passuello},
  M.~{Patil}, B.~{Patricelli}, B.L. {Pearlstone}, M.~{Pedraza}, R.~{Pedurand},
  L.~{Pekowsky}, A.~{Pele}, S.~{Penn}, C.J. {Perez}, A.~{Perreca}, L.M.
  {Perri}, H.P. {Pfeiffer}, M.~{Phelps}, O.J. {Piccinni}, M.~{Pichot},
  F.~{Piergiovanni}, V.~{Pierro}, G.~{Pillant}, L.~{Pinard}, I.M. {Pinto},
  M.~{Pirello}, M.~{Pitkin}, M.~{Poe}, R.~{Poggiani}, P.~{Popolizio}, E.K.
  {Porter}, A.~{Post}, J.~{Powell}, J.~{Prasad}, J.W.W. {Pratt}, G.~{Pratten},
  V.~{Predoi}, T.~{Prestegard}, L.R. {Price}, M.~{Prijatelj}, M.~{Principe},
  S.~{Privitera}, G.A. {Prodi}, L.G. {Prokhorov}, O.~{Puncken}, M.~{Punturo},
  P.~{Puppo}, M.~{P{\"u}rrer}, H.~{Qi}, V.~{Quetschke}, E.A. {Quintero},
  R.~{Quitzow-James}, F.J. {Raab}, D.S. {Rabeling}, H.~{Radkins}, P.~{Raffai},
  S.~{Raja}, C.~{Rajan}, B.~{Rajbhandari}, M.~{Rakhmanov}, K.E. {Ramirez},
  A.~{Ramos-Buades}, P.~{Rapagnani}, V.~{Raymond}, M.~{Razzano}, J.~{Read},
  T.~{Regimbau}, L.~{Rei}, S.~{Reid}, D.H. {Reitze}, W.~{Ren}, S.D. {Reyes},
  F.~{Ricci}, P.M. {Ricker}, S.~{Rieger}, K.~{Riles}, M.~{Rizzo}, N.A.
  {Robertson}, R.~{Robie}, F.~{Robinet}, A.~{Rocchi}, L.~{Rolland}, J.G.
  {Rollins}, V.J. {Roma}, R.~{Romano}, C.L. {Romel}, J.H. {Romie},
  D.~{Rosi{\'n}ska}, M.P. {Ross}, S.~{Rowan}, A.~{R{\"u}diger}, P.~{Ruggi},
  G.~{Rutins}, K.~{Ryan}, S.~{Sachdev}, T.~{Sadecki}, L.~{Sadeghian},
  M.~{Sakellariadou}, L.~{Salconi}, M.~{Saleem}, F.~{Salemi}, A.~{Samajdar},
  L.~{Sammut}, L.M. {Sampson}, E.J. {Sanchez}, L.E. {Sanchez},
  N.~{Sanchis-Gual}, V.~{Sandberg}, J.R. {Sanders}, B.~{Sassolas}, B.S.
  {Sathyaprakash}, P.R. {Saulson}, O.~{Sauter}, R.L. {Savage}, A.~{Sawadsky},
  P.~{Schale}, M.~{Scheel}, J.~{Scheuer}, J.~{Schmidt}, P.~{Schmidt},
  R.~{Schnabel}, R.M.S. {Schofield}, A.~{Sch{\"o}nbeck}, E.~{Schreiber},
  D.~{Schuette}, B.W. {Schulte}, B.F. {Schutz}, S.G. {Schwalbe}, J.~{Scott},
  S.M. {Scott}, E.~{Seidel}, D.~{Sellers}, A.S. {Sengupta}, D.~{Sentenac},
  V.~{Sequino}, A.~{Sergeev}, D.A. {Shaddock}, T.J. {Shaffer}, A.A. {Shah},
  M.S. {Shahriar}, M.B. {Shaner}, L.~{Shao}, B.~{Shapiro}, P.~{Shawhan},
  A.~{Sheperd}, D.H. {Shoemaker}, D.M. {Shoemaker}, K.~{Siellez}, X.~{Siemens},
  M.~{Sieniawska}, D.~{Sigg}, A.D. {Silva}, L.P. {Singer}, A.~{Singh},
  A.~{Singhal}, A.M. {Sintes}, B.J.J. {Slagmolen}, B.~{Smith}, J.R. {Smith},
  R.J.E. {Smith}, S.~{Somala}, E.J. {Son}, J.A. {Sonnenberg}, B.~{Sorazu},
  F.~{Sorrentino}, T.~{Souradeep}, A.P. {Spencer}, A.K. {Srivastava},
  K.~{Staats}, A.~{Staley}, M.~{Steinke}, J.~{Steinlechner}, S.~{Steinlechner},
  D.~{Steinmeyer}, S.P. {Stevenson}, R.~{Stone}, D.J. {Stops}, K.A. {Strain},
  G.~{Stratta}, S.E. {Strigin}, A.~{Strunk}, R.~{Sturani}, A.L. {Stuver}, T.Z.
  {Summerscales}, L.~{Sun}, S.~{Sunil}, J.~{Suresh}, P.J. {Sutton}, B.L.
  {Swinkels}, M.J. {Szczepa{\'n}czyk}, M.~{Tacca}, S.C. {Tait}, C.~{Talbot},
  D.~{Talukder}, D.B. {Tanner}, M.~{T{\'a}pai}, A.~{Taracchini}, J.D. {Tasson},
  J.A. {Taylor}, R.~{Taylor}, S.V. {Tewari}, T.~{Theeg}, F.~{Thies}, E.G.
  {Thomas}, M.~{Thomas}, P.~{Thomas}, K.A. {Thorne}, K.S. {Thorne},
  E.~{Thrane}, S.~{Tiwari}, V.~{Tiwari}, K.V. {Tokmakov}, K.~{Toland},
  M.~{Tonelli}, Z.~{Tornasi}, A.~{Torres-Forn{\'e}}, C.I. {Torrie},
  D.~{T{\"o}yr{\"a}}, F.~{Travasso}, G.~{Traylor}, J.~{Trinastic}, M.C.
  {Tringali}, L.~{Trozzo}, K.W. {Tsang}, M.~{Tse}, R.~{Tso}, L.~{Tsukada},
  D.~{Tsuna}, D.~{Tuyenbayev}, K.~{Ueno}, D.~{Ugolini}, C.S. {Unnikrishnan},
  A.L. {Urban}, S.A. {Usman}, H.~{Vahlbruch}, G.~{Vajente}, G.~{Valdes},
  N.~{van Bakel}, M.~{van Beuzekom}, J.F.J. {van den Brand}, C.~{Van Den
  Broeck}, D.C. {Vander-Hyde}, L.~{van der Schaaf}, J.V. {van Heijningen}, A.A.
  {van Veggel}, M.~{Vardaro}, V.~{Varma}, S.~{Vass}, M.~{Vas{\'u}th},
  A.~{Vecchio}, G.~{Vedovato}, J.~{Veitch}, P.J. {Veitch}, K.~{Venkateswara},
  G.~{Venugopalan}, D.~{Verkindt}, F.~{Vetrano}, A.~{Vicer{\'e}}, A.D. {Viets},
  S.~{Vinciguerra}, D.J. {Vine}, J.Y. {Vinet}, S.~{Vitale}, T.~{Vo},
  H.~{Vocca}, C.~{Vorvick}, S.P. {Vyatchanin}, A.R. {Wade}, L.E. {Wade},
  M.~{Wade}, R.~{Walet}, M.~{Walker}, L.~{Wallace}, S.~{Walsh}, G.~{Wang},
  H.~{Wang}, J.Z. {Wang}, W.H. {Wang}, Y.F. {Wang}, R.L. {Ward}, J.~{Warner},
  M.~{Was}, J.~{Watchi}, B.~{Weaver}, L.W. {Wei}, M.~{Weinert}, A.J.
  {Weinstein}, R.~{Weiss}, L.~{Wen}, E.K. {Wessel}, P.~{Wessels},
  J.~{Westerweck}, T.~{Westphal}, K.~{Wette}, J.T. {Whelan}, S.E. {Whitcomb},
  B.F. {Whiting}, C.~{Whittle}, D.~{Wilken}, D.~{Williams}, R.D. {Williams},
  A.R. {Williamson}, J.L. {Willis}, B.~{Willke}, M.H. {Wimmer}, W.~{Winkler},
  C.C. {Wipf}, H.~{Wittel}, G.~{Woan}, J.~{Woehler}, J.~{Wofford}, K.W.K.
  {Wong}, J.~{Worden}, J.L. {Wright}, D.S. {Wu}, D.M. {Wysocki}, S.~{Xiao},
  H.~{Yamamoto}, C.C. {Yancey}, L.~{Yang}, M.J. {Yap}, M.~{Yazback}, H.~{Yu},
  H.~{Yu}, M.~{Yvert}, A.~{Zadro{\.z}ny}, M.~{Zanolin}, T.~{Zelenova}, J.P.
  {Zendri}, M.~{Zevin}, L.~{Zhang}, M.~{Zhang}, T.~{Zhang}, Y.H. {Zhang},
  C.~{Zhao}, M.~{Zhou}, Z.~{Zhou}, S.J. {Zhu}, X.J. {Zhu}, A.B. {Zimmerman},
  M.E. {Zucker}, J.~{Zweizig}, {LIGO Scientific Collaboration}, {Virgo
  Collaboration}, C.A. {Wilson-Hodge}, E.~{Bissaldi}, L.~{Blackburn}, M.S.
  {Briggs}, E.~{Burns}, W.H. {Cleveland}, V.~{Connaughton}, M.H. {Gibby}, M.M.
  {Giles}, A.~{Goldstein}, R.~{Hamburg}, P.~{Jenke}, C.M. {Hui}, R.M. {Kippen},
  D.~{Kocevski}, S.~{McBreen}, C.A. {Meegan}, W.S. {Paciesas}, S.~{Poolakkil},
  R.D. {Preece}, J.~{Racusin}, O.J. {Roberts}, M.~{Stanbro}, P.~{Veres},
  A.~{von Kienlin}, F.~{GBM}, V.~{Savchenko}, C.~{Ferrigno}, E.~{Kuulkers},
  A.~{Bazzano}, E.~{Bozzo}, S.~{Brandt}, J.~{Chenevez}, T.J.L. {Courvoisier},
  R.~{Diehl}, A.~{Domingo}, L.~{Hanlon}, E.~{Jourdain}, P.~{Laurent},
  F.~{Lebrun}, A.~{Lutovinov}, A.~{Martin-Carrillo}, S.~{Mereghetti},
  L.~{Natalucci}, J.~{Rodi}, J.P. {Roques}, R.~{Sunyaev}, P.~{Ubertini},
  {INTEGRAL}, M.G. {Aartsen}, M.~{Ackermann}, J.~{Adams}, J.A. {Aguilar},
  M.~{Ahlers}, M.~{Ahrens}, I.A. {Samarai}, D.~{Altmann}, K.~{Andeen},
  T.~{Anderson}, I.~{Ansseau}, G.~{Anton}, C.~{Arg{\"u}elles}, J.~{Auffenberg},
  S.~{Axani}, H.~{Bagherpour}, X.~{Bai}, J.P. {Barron}, S.W. {Barwick},
  V.~{Baum}, R.~{Bay}, J.J. {Beatty}, J.~{Becker Tjus}, E.~{Bernardini}, D.Z.
  {Besson}, G.~{Binder}, D.~{Bindig}, E.~{Blaufuss}, S.~{Blot}, C.~{Bohm},
  M.~{B{\"o}rner}, F.~{Bos}, D.~{Bose}, S.~{B{\"o}ser}, O.~{Botner},
  E.~{Bourbeau}, J.~{Bourbeau}, F.~{Bradascio}, J.~{Braun}, L.~{Brayeur},
  M.~{Brenzke}, H.P. {Bretz}, S.~{Bron}, J.~{Brostean-Kaiser}, A.~{Burgman},
  T.~{Carver}, J.~{Casey}, M.~{Casier}, E.~{Cheung}, D.~{Chirkin},
  A.~{Christov}, K.~{Clark}, L.~{Classen}, S.~{Coenders}, G.H. {Collin}, J.M.
  {Conrad}, D.F. {Cowen}, R.~{Cross}, M.~{Day}, J.P.A.M. {de Andr{\'e}}, C.~{De
  Clercq}, J.J. {DeLaunay}, H.~{Dembinski}, S.~{De Ridder}, P.~{Desiati}, K.D.
  {de Vries}, G.~{de Wasseige}, M.~{de With}, T.~{DeYoung}, J.C.
  {D{\'\i}az-V{\'e}lez}, V.~{di Lorenzo}, H.~{Dujmovic}, J.P. {Dumm},
  M.~{Dunkman}, E.~{Dvorak}, B.~{Eberhardt}, T.~{Ehrhardt}, B.~{Eichmann},
  P.~{Eller}, P.A. {Evenson}, S.~{Fahey}, A.R. {Fazely}, J.~{Felde},
  K.~{Filimonov}, C.~{Finley}, S.~{Flis}, A.~{Franckowiak}, E.~{Friedman},
  T.~{Fuchs}, T.K. {Gaisser}, J.~{Gallagher}, L.~{Gerhardt}, K.~{Ghorbani},
  W.~{Giang}, T.~{Glauch}, T.~{Gl{\"u}senkamp}, A.~{Goldschmidt}, J.G.
  {Gonzalez}, D.~{Grant}, Z.~{Griffith}, C.~{Haack}, A.~{Hallgren},
  F.~{Halzen}, K.~{Hanson}, D.~{Hebecker}, D.~{Heereman}, K.~{Helbing},
  R.~{Hellauer}, S.~{Hickford}, J.~{Hignight}, G.C. {Hill}, K.D. {Hoffman},
  R.~{Hoffmann}, B.~{Hokanson-Fasig}, K.~{Hoshina}, F.~{Huang}, M.~{Huber},
  K.~{Hultqvist}, M.~{H{\"u}nnefeld}, S.~{In}, A.~{Ishihara}, E.~{Jacobi}, G.S.
  {Japaridze}, M.~{Jeong}, K.~{Jero}, B.J.P. {Jones}, P.~{Kalaczynski},
  W.~{Kang}, A.~{Kappes}, T.~{Karg}, A.~{Karle}, M.~{Kauer}, A.~{Keivani}, J.L.
  {Kelley}, A.~{Kheirandish}, J.~{Kim}, M.~{Kim}, T.~{Kintscher}, J.~{Kiryluk},
  T.~{Kittler}, S.R. {Klein}, G.~{Kohnen}, R.~{Koirala}, H.~{Kolanoski},
  L.~{K{\"o}pke}, C.~{Kopper}, S.~{Kopper}, J.P. {Koschinsky}, D.J. {Koskinen},
  M.~{Kowalski}, K.~{Krings}, M.~{Kroll}, G.~{Kr{\"u}ckl}, J.~{Kunnen},
  S.~{Kunwar}, N.~{Kurahashi}, T.~{Kuwabara}, A.~{Kyriacou}, M.~{Labare}, J.L.
  {Lanfranchi}, M.J. {Larson}, F.~{Lauber}, M.~{Lesiak-Bzdak}, M.~{Leuermann},
  Q.R. {Liu}, L.~{Lu}, J.~{L{\"u}nemann}, W.~{Luszczak}, J.~{Madsen},
  G.~{Maggi}, K.B.M. {Mahn}, S.~{Mancina}, R.~{Maruyama}, K.~{Mase},
  R.~{Maunu}, F.~{McNally}, K.~{Meagher}, M.~{Medici}, M.~{Meier}, T.~{Menne},
  G.~{Merino}, T.~{Meures}, S.~{Miarecki}, J.~{Micallef}, G.~{Moment{\'e}},
  T.~{Montaruli}, R.W. {Moore}, M.~{Moulai}, R.~{Nahnhauer}, P.~{Nakarmi},
  U.~{Naumann}, G.~{Neer}, H.~{Niederhausen}, S.C. {Nowicki}, D.R. {Nygren},
  A.~{Obertacke Pollmann}, A.~{Olivas}, A.~{O'Murchadha}, T.~{Palczewski},
  H.~{Pandya}, D.V. {Pankova}, P.~{Peiffer}, J.A. {Pepper}, C.~{P{\'e}rez de
  los Heros}, D.~{Pieloth}, E.~{Pinat}, P.B. {Price}, G.T. {Przybylski},
  C.~{Raab}, L.~{R{\"a}del}, M.~{Rameez}, K.~{Rawlins}, I.C. {Rea},
  R.~{Reimann}, B.~{Relethford}, M.~{Relich}, E.~{Resconi}, W.~{Rhode},
  M.~{Richman}, S.~{Robertson}, M.~{Rongen}, C.~{Rott}, T.~{Ruhe},
  D.~{Ryckbosch}, D.~{Rysewyk}, T.~{S{\"a}lzer}, S.E. {Sanchez Herrera},
  A.~{Sandrock}, J.~{Sandroos}, M.~{Santander}, S.~{Sarkar}, S.~{Sarkar},
  K.~{Satalecka}, P.~{Schlunder}, T.~{Schmidt}, A.~{Schneider}, S.~{Schoenen},
  S.~{Sch{\"o}neberg}, L.~{Schumacher}, D.~{Seckel}, S.~{Seunarine},
  J.~{Soedingrekso}, D.~{Soldin}, M.~{Song}, G.M. {Spiczak}, C.~{Spiering},
  J.~{Stachurska}, M.~{Stamatikos}, T.~{Stanev}, A.~{Stasik}, J.~{Stettner},
  A.~{Steuer}, T.~{Stezelberger}, R.G. {Stokstad}, A.~{St{\"o}ssl}, N.L.
  {Strotjohann}, T.~{Stuttard}, G.W. {Sullivan}, M.~{Sutherland}, I.~{Taboada},
  J.~{Tatar}, F.~{Tenholt}, S.~{Ter-Antonyan}, A.~{Terliuk},
  G.~{Te{\v{s}}i{\'c}}, S.~{Tilav}, P.A. {Toale}, M.N. {Tobin}, S.~{Toscano},
  D.~{Tosi}, M.~{Tselengidou}, C.F. {Tung}, A.~{Turcati}, C.F. {Turley},
  B.~{Ty}, E.~{Unger}, M.~{Usner}, J.~{Vandenbroucke}, W.~{Van Driessche},
  N.~{van Eijndhoven}, S.~{Vanheule}, J.~{van Santen}, M.~{Vehring},
  E.~{Vogel}, M.~{Vraeghe}, C.~{Walck}, A.~{Wallace}, M.~{Wallraff}, F.D.
  {Wandler}, N.~{Wandkowsky}, A.~{Waza}, C.~{Weaver}, M.J. {Weiss}, C.~{Wendt},
  J.~{Werthebach}, B.J. {Whelan}, K.~{Wiebe}, C.H. {Wiebusch}, L.~{Wille}, D.R.
  {Williams}, L.~{Wills}, M.~{Wolf}, T.R. {Wood}, E.~{Woolsey}, K.~{Woschnagg},
  D.L. {Xu}, X.W. {Xu}, Y.~{Xu}, J.P. {Yanez}, G.~{Yodh}, S.~{Yoshida},
  T.~{Yuan}, M.~{Zoll}, {IceCube Collaboration}, A.~{Balasubramanian},
  S.~{Mate}, V.~{Bhalerao}, D.~{Bhattacharya}, A.~{Vibhute}, G.C. {Dewangan},
  A.R. {Rao}, S.V. {Vadawale}, {AstroSat Cadmium Zinc Telluride Imager Team},
  D.S. {Svinkin}, K.~{Hurley}, R.L. {Aptekar}, D.D. {Frederiks}, S.V.
  {Golenetskii}, A.V. {Kozlova}, A.L. {Lysenko}, P.P. {Oleynik}, A.E.
  {Tsvetkova}, M.V. {Ulanov}, T.~{Cline}, {IPN Collaboration}, T.P. {Li}, S.L.
  {Xiong}, S.N. {Zhang}, F.J. {Lu}, L.M. {Song}, X.L. {Cao}, Z.~{Chang},
  G.~{Chen}, L.~{Chen}, T.X. {Chen}, Y.~{Chen}, Y.B. {Chen}, Y.P. {Chen},
  W.~{Cui}, W.W. {Cui}, J.K. {Deng}, Y.W. {Dong}, Y.Y. {Du}, M.X. {Fu}, G.H.
  {Gao}, H.~{Gao}, M.~{Gao}, M.Y. {Ge}, Y.D. {Gu}, J.~{Guan}, C.C. {Guo}, D.W.
  {Han}, W.~{Hu}, Y.~{Huang}, J.~{Huo}, S.M. {Jia}, L.H. {Jiang}, W.C. {Jiang},
  J.~{Jin}, Y.J. {Jin}, B.~{Li}, C.K. {Li}, G.~{Li}, M.S. {Li}, W.~{Li},
  X.~{Li}, X.B. {Li}, X.F. {Li}, Y.G. {Li}, Z.J. {Li}, Z.W. {Li}, X.H. {Liang},
  J.Y. {Liao}, C.Z. {Liu}, G.Q. {Liu}, H.W. {Liu}, S.Z. {Liu}, X.J. {Liu},
  Y.~{Liu}, Y.N. {Liu}, B.~{Lu}, X.F. {Lu}, T.~{Luo}, X.~{Ma}, B.~{Meng},
  Y.~{Nang}, J.Y. {Nie}, G.~{Ou}, J.L. {Qu}, N.~{Sai}, L.~{Sun}, Y.~{Tan},
  L.~{Tao}, W.H. {Tao}, Y.L. {Tuo}, G.F. {Wang}, H.Y. {Wang}, J.~{Wang}, W.S.
  {Wang}, Y.S. {Wang}, X.Y. {Wen}, B.B. {Wu}, M.~{Wu}, G.C. {Xiao}, H.~{Xu},
  Y.P. {Xu}, L.L. {Yan}, J.W. {Yang}, S.~{Yang}, Y.J. {Yang}, A.M. {Zhang},
  C.L. {Zhang}, C.M. {Zhang}, F.~{Zhang}, H.M. {Zhang}, J.~{Zhang}, Q.~{Zhang},
  S.~{Zhang}, T.~{Zhang}, W.~{Zhang}, W.C. {Zhang}, W.Z. {Zhang}, Y.~{Zhang},
  Y.~{Zhang}, Y.F. {Zhang}, Y.J. {Zhang}, Z.~{Zhang}, Z.L. {Zhang}, H.S.
  {Zhao}, J.L. {Zhao}, X.F. {Zhao}, S.J. {Zheng}, Y.~{Zhu}, Y.X. {Zhu}, C.L.
  {Zou}, {Insight-HXMT Collaboration}, A.~{Albert}, M.~{Andr{\'e}},
  M.~{Anghinolfi}, M.~{Ardid}, J.J. {Aubert}, J.~{Aublin}, T.~{Avgitas},
  B.~{Baret}, J.~{Barrios-Mart{\'\i}}, S.~{Basa}, B.~{Belhorma}, V.~{Bertin},
  S.~{Biagi}, R.~{Bormuth}, S.~{Bourret}, M.C. {Bouwhuis},
  H.~{Br{\^a}nza{\c{s}}}, R.~{Bruijn}, J.~{Brunner}, J.~{Busto}, A.~{Capone},
  L.~{Caramete}, J.~{Carr}, S.~{Celli}, R.~{Cherkaoui El Moursli},
  T.~{Chiarusi}, M.~{Circella}, J.A.B. {Coelho}, A.~{Coleiro}, R.~{Coniglione},
  H.~{Costantini}, P.~{Coyle}, A.~{Creusot}, A.F. {D{\'\i}az}, A.~{Deschamps},
  G.~{De Bonis}, C.~{Distefano}, I.~{Di Palma}, A.~{Domi}, C.~{Donzaud},
  D.~{Dornic}, D.~{Drouhin}, T.~{Eberl}, I.~{El Bojaddaini}, N.~{El Khayati},
  D.~{Els{\"a}sser}, A.~{Enzenh{\"o}fer}, A.~{Ettahiri}, F.~{Fassi},
  I.~{Felis}, L.A. {Fusco}, P.~{Gay}, V.~{Giordano}, H.~{Glotin},
  T.~{Gr{\'e}goire}, R.G. {Ruiz}, K.~{Graf}, S.~{Hallmann}, H.~{van Haren},
  A.J. {Heijboer}, Y.~{Hello}, J.J. {Hern{\'a}ndez-Rey}, J.~{H{\"o}ssl},
  J.~{Hofest{\"a}dt}, C.~{Hugon}, G.~{Illuminati}, C.W. {James}, M.~{de Jong},
  M.~{Jongen}, M.~{Kadler}, O.~{Kalekin}, U.~{Katz}, D.~{Kiessling},
  A.~{Kouchner}, M.~{Kreter}, I.~{Kreykenbohm}, V.~{Kulikovskiy}, C.~{Lachaud},
  R.~{Lahmann}, D.~{Lef{\`e}vre}, E.~{Leonora}, M.~{Lotze}, S.~{Loucatos},
  M.~{Marcelin}, A.~{Margiotta}, A.~{Marinelli}, J.A. {Mart{\'\i}nez-Mora},
  R.~{Mele}, K.~{Melis}, T.~{Michael}, P.~{Migliozzi}, A.~{Moussa}, S.~{Navas},
  E.~{Nezri}, M.~{Organokov}, G.E. {P{\u{a}}v{\u{a}}la{\c{s}}},
  C.~{Pellegrino}, C.~{Perrina}, P.~{Piattelli}, V.~{Popa}, T.~{Pradier},
  L.~{Quinn}, C.~{Racca}, G.~{Riccobene}, A.~{S{\'a}nchez-Losa},
  M.~{Salda{\~n}a}, I.~{Salvadori}, D.F.E. {Samtleben}, M.~{Sanguineti},
  P.~{Sapienza}, C.~{Sieger}, M.~{Spurio}, T.~{Stolarczyk}, M.~{Taiuti},
  Y.~{Tayalati}, A.~{Trovato}, D.~{Turpin}, C.~{T{\"o}nnis}, B.~{Vallage},
  V.~{Van Elewyck}, F.~{Versari}, D.~{Vivolo}, A.~{Vizzoca}, J.~{Wilms}, J.D.
  {Zornoza}, J.~{Z{\'u}{\~n}iga}, {ANTARES Collaboration}, A.P. {Beardmore},
  A.A. {Breeveld}, D.N. {Burrows}, S.B. {Cenko}, G.~{Cusumano}, A.~{D'A{\`\i}},
  M.~{de Pasquale}, S.W.K. {Emery}, P.A. {Evans}, P.~{Giommi}, C.~{Gronwall},
  J.A. {Kennea}, H.A. {Krimm}, N.P.M. {Kuin}, A.~{Lien}, F.E. {Marshall},
  A.~{Melandri}, J.A. {Nousek}, S.R. {Oates}, J.P. {Osborne}, C.~{Pagani}, K.L.
  {Page}, D.M. {Palmer}, M.~{Perri}, M.H. {Siegel}, B.~{Sbarufatti},
  G.~{Tagliaferri}, A.~{Tohuvavohu}, {Swift Collaboration}, M.~{Tavani},
  F.~{Verrecchia}, A.~{Bulgarelli}, Y.~{Evangelista}, L.~{Pacciani},
  M.~{Feroci}, C.~{Pittori}, A.~{Giuliani}, E.~{Del Monte}, I.~{Donnarumma},
  A.~{Argan}, A.~{Trois}, A.~{Ursi}, M.~{Cardillo}, G.~{Piano}, F.~{Longo},
  F.~{Lucarelli}, P.~{Munar-Adrover}, F.~{Fuschino}, C.~{Labanti},
  M.~{Marisaldi}, G.~{Minervini}, V.~{Fioretti}, N.~{Parmiggiani},
  F.~{Gianotti}, M.~{Trifoglio}, G.~{Di Persio}, L.A. {Antonelli},
  G.~{Barbiellini}, P.~{Caraveo}, P.W. {Cattaneo}, E.~{Costa},
  S.~{Colafrancesco}, F.~{D'Amico}, A.~{Ferrari}, A.~{Morselli}, F.~{Paoletti},
  P.~{Picozza}, M.~{Pilia}, A.~{Rappoldi}, P.~{Soffitta}, S.~{Vercellone},
  {AGILE Team}, R.J. {Foley}, D.A. {Coulter}, C.D. {Kilpatrick}, M.R. {Drout},
  A.L. {Piro}, B.J. {Shappee}, M.R. {Siebert}, J.D. {Simon}, N.~{Ulloa},
  D.~{Kasen}, B.F. {Madore}, A.~{Murguia-Berthier}, Y.C. {Pan}, J.X.
  {Prochaska}, E.~{Ramirez-Ruiz}, A.~{Rest}, C.~{Rojas-Bravo}, {1M2H Team},
  E.~{Berger}, M.~{Soares-Santos}, J.~{Annis}, K.D. {Alexander}, S.~{Allam},
  E.~{Balbinot}, P.~{Blanchard}, D.~{Brout}, R.E. {Butler}, R.~{Chornock}, E.R.
  {Cook}, P.~{Cowperthwaite}, H.T. {Diehl}, A.~{Drlica-Wagner}, M.R. {Drout},
  F.~{Durret}, T.~{Eftekhari}, D.A. {Finley}, W.~{Fong}, J.A. {Frieman}, C.L.
  {Fryer}, J.~{Garc{\'\i}a-Bellido}, R.A. {Gruendl}, W.~{Hartley}, K.~{Herner},
  R.~{Kessler}, H.~{Lin}, P.A.A. {Lopes}, A.C.C. {Louren{\c{c}}o},
  R.~{Margutti}, J.L. {Marshall}, T.~{Matheson}, G.E. {Medina}, B.D. {Metzger},
  R.R. {Mu{\~n}oz}, J.~{Muir}, M.~{Nicholl}, P.~{Nugent}, A.~{Palmese},
  F.~{Paz-Chinch{\'o}n}, E.~{Quataert}, M.~{Sako}, M.~{Sauseda}, D.J.
  {Schlegel}, D.~{Scolnic}, L.F. {Secco}, N.~{Smith}, F.~{Sobreira}, V.A.
  {Villar}, A.K. {Vivas}, W.~{Wester}, P.K.G. {Williams}, B.~{Yanny},
  A.~{Zenteno}, Y.~{Zhang}, T.M.C. {Abbott}, M.~{Banerji}, K.~{Bechtol},
  A.~{Benoit-L{\'e}vy}, E.~{Bertin}, D.~{Brooks}, E.~{Buckley-Geer}, D.L.
  {Burke}, D.~{Capozzi}, A.~{Carnero Rosell}, M.~{Carrasco Kind}, F.J.
  {Castander}, M.~{Crocce}, C.E. {Cunha}, C.B. {D'Andrea}, L.N. {da Costa},
  C.~{Davis}, D.L. {DePoy}, S.~{Desai}, J.P. {Dietrich}, T.F. {Eifler},
  E.~{Fernandez}, B.~{Flaugher}, P.~{Fosalba}, E.~{Gaztanaga}, D.W. {Gerdes},
  T.~{Giannantonio}, D.A. {Goldstein}, D.~{Gruen}, J.~{Gschwend},
  G.~{Gutierrez}, K.~{Honscheid}, D.J. {James}, T.~{Jeltema}, M.W.G. {Johnson},
  M.D. {Johnson}, S.~{Kent}, E.~{Krause}, R.~{Kron}, K.~{Kuehn}, O.~{Lahav},
  M.~{Lima}, M.A.G. {Maia}, M.~{March}, P.~{Martini}, R.G. {McMahon},
  F.~{Menanteau}, C.J. {Miller}, R.~{Miquel}, J.J. {Mohr}, R.C. {Nichol},
  R.L.C. {Ogando}, A.A. {Plazas}, A.K. {Romer}, A.~{Roodman}, E.S. {Rykoff},
  E.~{Sanchez}, V.~{Scarpine}, R.~{Schindler}, M.~{Schubnell},
  I.~{Sevilla-Noarbe}, E.~{Sheldon}, M.~{Smith}, R.C. {Smith}, A.~{Stebbins},
  E.~{Suchyta}, M.E.C. {Swanson}, G.~{Tarle}, R.C. {Thomas}, M.A. {Troxel},
  D.L. {Tucker}, V.~{Vikram}, A.R. {Walker}, R.H. {Wechsler}, J.~{Weller}, J.L.
  {Carlin}, M.S.S. {Gill}, T.S. {Li}, J.~{Marriner}, E.~{Neilsen}, {Dark Energy
  Camera GW-EM Collaboration}, {DES Collaboration}, J.B. {Haislip}, V.V.
  {Kouprianov}, D.E. {Reichart}, D.J. {Sand}, L.~{Tartaglia}, S.~{Valenti},
  S.~{Yang}, {DLT40 Collaboration}, S.~{Benetti}, E.~{Brocato}, S.~{Campana},
  E.~{Cappellaro}, S.~{Covino}, P.~{D'Avanzo}, V.~{D'Elia}, F.~{Getman},
  G.~{Ghirlanda}, G.~{Ghisellini}, L.~{Limatola}, L.~{Nicastro}, E.~{Palazzi},
  E.~{Pian}, S.~{Piranomonte}, A.~{Possenti}, A.~{Rossi}, O.S. {Salafia},
  L.~{Tomasella}, L.~{Amati}, L.A. {Antonelli}, M.G. {Bernardini}, F.~{Bufano},
  M.~{Capaccioli}, P.~{Casella}, M.~{Dadina}, G.~{De Cesare}, A.~{Di Paola},
  G.~{Giuffrida}, A.~{Giunta}, G.L. {Israel}, M.~{Lisi}, E.~{Maiorano},
  M.~{Mapelli}, N.~{Masetti}, A.~{Pescalli}, L.~{Pulone}, R.~{Salvaterra},
  P.~{Schipani}, M.~{Spera}, A.~{Stamerra}, L.~{Stella}, V.~{Testa},
  M.~{Turatto}, D.~{Vergani}, G.~{Aresu}, M.~{Bachetti}, F.~{Buffa},
  M.~{Burgay}, M.~{Buttu}, T.~{Caria}, E.~{Carretti}, V.~{Casasola},
  P.~{Castangia}, G.~{Carboni}, S.~{Casu}, R.~{Concu}, A.~{Corongiu}, G.L.
  {Deiana}, E.~{Egron}, A.~{Fara}, F.~{Gaudiomonte}, V.~{Gusai}, A.~{Ladu},
  S.~{Loru}, S.~{Leurini}, L.~{Marongiu}, A.~{Melis}, G.~{Melis}, C.~{Migoni},
  S.~{Milia}, A.~{Navarrini}, A.~{Orlati}, P.~{Ortu}, S.~{Palmas},
  A.~{Pellizzoni}, D.~{Perrodin}, T.~{Pisanu}, S.~{Poppi}, S.~{Righini},
  A.~{Saba}, G.~{Serra}, M.~{Serrau}, M.~{Stagni}, G.~{Surcis}, V.~{Vacca},
  G.P. {Vargiu}, L.K. {Hunt}, Z.P. {Jin}, S.~{Klose}, C.~{Kouveliotou}, P.A.
  {Mazzali}, P.~{M{\o}ller}, L.~{Nava}, T.~{Piran}, J.~{Selsing}, S.D.
  {Vergani}, K.~{Wiersema}, K.~{Toma}, A.B. {Higgins}, C.G. {Mundell}, S.~{di
  Serego Alighieri}, D.~{G{\'o}tz}, W.~{Gao}, A.~{Gomboc}, L.~{Kaper},
  S.~{Kobayashi}, D.~{Kopac}, J.~{Mao}, R.L.C. {Starling}, I.~{Steele}, A.J.
  {van der Horst}, {GRAWITA: GRAvitational Wave Inaf TeAm}, F.~{Acero}, W.B.
  {Atwood}, L.~{Baldini}, G.~{Barbiellini}, D.~{Bastieri}, B.~{Berenji},
  R.~{Bellazzini}, E.~{Bissaldi}, R.D. {Blandford}, E.D. {Bloom}, R.~{Bonino},
  E.~{Bottacini}, J.~{Bregeon}, R.~{Buehler}, S.~{Buson}, R.A. {Cameron},
  R.~{Caputo}, P.A. {Caraveo}, E.~{Cavazzuti}, A.~{Chekhtman}, C.C. {Cheung},
  J.~{Chiang}, S.~{Ciprini}, J.~{Cohen-Tanugi}, L.R. {Cominsky},
  D.~{Costantin}, A.~{Cuoco}, F.~{D'Ammando}, F.~{de Palma}, S.W. {Digel},
  N.~{Di Lalla}, M.~{Di Mauro}, L.~{Di Venere}, R.~{Dubois}, S.J. {Fegan}, W.B.
  {Focke}, A.~{Franckowiak}, Y.~{Fukazawa}, S.~{Funk}, P.~{Fusco},
  F.~{Gargano}, D.~{Gasparrini}, N.~{Giglietto}, F.~{Giordano}, M.~{Giroletti},
  T.~{Glanzman}, D.~{Green}, M.H. {Grondin}, L.~{Guillemot}, S.~{Guiriec}, A.K.
  {Harding}, D.~{Horan}, G.~{J{\'o}hannesson}, T.~{Kamae}, S.~{Kensei},
  M.~{Kuss}, G.~{La Mura}, L.~{Latronico}, M.~{Lemoine-Goumard}, F.~{Longo},
  F.~{Loparco}, M.N. {Lovellette}, P.~{Lubrano}, J.D. {Magill}, S.~{Maldera},
  A.~{Manfreda}, M.N. {Mazziotta}, J.E. {McEnery}, M.~{Meyer}, P.F.
  {Michelson}, N.~{Mirabal}, M.E. {Monzani}, E.~{Moretti}, A.~{Morselli}, I.V.
  {Moskalenko}, M.~{Negro}, E.~{Nuss}, R.~{Ojha}, N.~{Omodei}, M.~{Orienti},
  E.~{Orlando}, M.~{Palatiello}, V.S. {Paliya}, D.~{Paneque},
  M.~{Pesce-Rollins}, F.~{Piron}, T.A. {Porter}, G.~{Principe}, S.~{Rain{\`o}},
  R.~{Rando}, M.~{Razzano}, S.~{Razzaque}, A.~{Reimer}, O.~{Reimer},
  T.~{Reposeur}, L.S. {Rochester}, P.M. {Saz Parkinson}, C.~{Sgr{\`o}}, E.J.
  {Siskind}, F.~{Spada}, G.~{Spandre}, D.J. {Suson}, M.~{Takahashi},
  Y.~{Tanaka}, J.G. {Thayer}, J.B. {Thayer}, D.J. {Thompson}, L.~{Tibaldo},
  D.F. {Torres}, E.~{Torresi}, E.~{Troja}, T.M. {Venters}, G.~{Vianello},
  G.~{Zaharijas}, {Fermi Large Area Telescope Collaboration}, J.R. {Allison},
  K.W. {Bannister}, D.~{Dobie}, D.L. {Kaplan}, E.~{Lenc}, C.~{Lynch},
  T.~{Murphy}, E.M. {Sadler}, A.~{Australia Telescope Compact Array},
  A.~{Hotan}, C.W. {James}, S.~{Oslowski}, W.~{Raja}, R.M. {Shannon},
  M.~{Whiting}, A.~{Australian SKA Pathfinder}, I.~{Arcavi}, D.A. {Howell},
  C.~{McCully}, G.~{Hosseinzadeh}, D.~{Hiramatsu}, D.~{Poznanski}, J.~{Barnes},
  M.~{Zaltzman}, S.~{Vasylyev}, D.~{Maoz}, {Las Cumbres Observatory Group},
  J.~{Cooke}, M.~{Bailes}, C.~{Wolf}, A.T. {Deller}, C.~{Lidman}, L.~{Wang},
  B.~{Gendre}, I.~{Andreoni}, K.~{Ackley}, T.A. {Pritchard}, M.S. {Bessell},
  S.W. {Chang}, A.~{M{\"o}ller}, C.A. {Onken}, R.A. {Scalzo},
  R.~{Ridden-Harper}, R.G. {Sharp}, B.E. {Tucker}, T.J. {Farrell}, E.~{Elmer},
  S.~{Johnston}, V.~{Venkatraman Krishnan}, E.F. {Keane}, J.A. {Green},
  A.~{Jameson}, L.~{Hu}, B.~{Ma}, T.~{Sun}, X.~{Wu}, X.~{Wang}, Z.~{Shang},
  Y.~{Hu}, M.C.B. {Ashley}, X.~{Yuan}, X.~{Li}, C.~{Tao}, Z.~{Zhu}, H.~{Zhang},
  N.B. {Suntzeff}, J.~{Zhou}, J.~{Yang}, B.~{Orange}, D.~{Morris},
  A.~{Cucchiara}, T.~{Giblin}, A.~{Klotz}, J.~{Staff}, P.~{Thierry}, B.P.
  {Schmidt}, {OzGrav}, D.~{(Deeper}, {Wider}, F.~{program}, {AST3}, {CAASTRO
  Collaborations}, N.R. {Tanvir}, A.J. {Levan}, Z.~{Cano}, A.~{de
  Ugarte-Postigo}, C.~{Gonz{\'a}lez-Fern{\'a}ndez}, J.~{Greiner}, J.~{Hjorth},
  M.~{Irwin}, T.~{Kr{\"u}hler}, I.~{Mandel}, B.~{Milvang-Jensen}, P.~{O'Brien},
  E.~{Rol}, S.~{Rosetti}, S.~{Rosswog}, A.~{Rowlinson}, D.T.H. {Steeghs}, C.C.
  {Th{\"o}ne}, K.~{Ulaczyk}, D.~{Watson}, S.H. {Bruun}, R.~{Cutter},
  R.~{Figuera Jaimes}, Y.I. {Fujii}, A.S. {Fruchter}, B.~{Gompertz},
  P.~{Jakobsson}, G.~{Hodosan}, U.G. {J{\`e}rgensen}, T.~{Kangas}, D.A. {Kann},
  M.~{Rabus}, S.L. {Schr{\o}der}, E.R. {Stanway}, R.A.M.J. {Wijers}, {VINROUGE
  Collaboration}, V.M. {Lipunov}, E.S. {Gorbovskoy}, V.G. {Kornilov}, N.V.
  {Tyurina}, P.V. {Balanutsa}, A.S. {Kuznetsov}, D.M. {Vlasenko}, R.C.
  {Podesta}, C.~{Lopez}, F.~{Podesta}, H.O. {Levato}, C.~{Saffe}, C.C.
  {Mallamaci}, N.M. {Budnev}, O.A. {Gress}, D.A. {Kuvshinov}, I.A. {Gorbunov},
  V.V. {Vladimirov}, D.S. {Zimnukhov}, A.V. {Gabovich}, V.V. {Yurkov}, Y.P.
  {Sergienko}, R.~{Rebolo}, M.~{Serra-Ricart}, A.G. {Tlatov}, Y.V.
  {Ishmuhametova}, {MASTER Collaboration}, F.~{Abe}, K.~{Aoki}, W.~{Aoki},
  Y.~{Asakura}, S.~{Baar}, S.~{Barway}, I.A. {Bond}, M.~{Doi}, F.~{Finet},
  T.~{Fujiyoshi}, H.~{Furusawa}, S.~{Honda}, R.~{Itoh}, N.~{Kanda}, K.S.
  {Kawabata}, M.~{Kawabata}, J.H. {Kim}, S.~{Koshida}, D.~{Kuroda}, C.H. {Lee},
  W.~{Liu}, K.~{Matsubayashi}, S.~{Miyazaki}, K.~{Morihana}, T.~{Morokuma},
  K.~{Motohara}, K.L. {Murata}, H.~{Nagai}, H.~{Nagashima}, T.~{Nagayama},
  T.~{Nakaoka}, F.~{Nakata}, R.~{Ohsawa}, T.~{Ohshima}, K.~{Ohta}, H.~{Okita},
  T.~{Saito}, Y.~{Saito}, S.~{Sako}, Y.~{Sekiguchi}, T.~{Sumi}, A.~{Tajitsu},
  J.~{Takahashi}, M.~{Takayama}, Y.~{Tamura}, I.~{Tanaka}, M.~{Tanaka},
  T.~{Terai}, N.~{Tominaga}, P.J. {Tristram}, M.~{Uemura}, Y.~{Utsumi}, M.S.
  {Yamaguchi}, N.~{Yasuda}, M.~{Yoshida}, T.~{Zenko}, {J-GEM}, S.M. {Adams},
  G.C. {Anupama}, J.~{Bally}, S.~{Barway}, E.~{Bellm}, N.~{Blagorodnova},
  C.~{Cannella}, P.~{Chandra}, D.~{Chatterjee}, T.E. {Clarke}, B.E. {Cobb},
  D.O. {Cook}, C.~{Copperwheat}, K.~{De}, S.W.K. {Emery}, U.~{Feindt},
  K.~{Foster}, O.D. {Fox}, D.A. {Frail}, C.~{Fremling}, C.~{Frohmaier}, J.A.
  {Garcia}, S.~{Ghosh}, S.~{Giacintucci}, A.~{Goobar}, O.~{Gottlieb}, B.W.
  {Grefenstette}, G.~{Hallinan}, F.~{Harrison}, M.~{Heida}, G.~{Helou}, A.Y.Q.
  {Ho}, A.~{Horesh}, K.~{Hotokezaka}, W.H. {Ip}, R.~{Itoh}, B.~{Jacobs}, J.E.
  {Jencson}, D.~{Kasen}, M.M. {Kasliwal}, N.E. {Kassim}, H.~{Kim}, B.S.
  {Kiran}, N.P.M. {Kuin}, S.R. {Kulkarni}, T.~{Kupfer}, R.M. {Lau},
  K.~{Madsen}, P.A. {Mazzali}, A.A. {Miller}, H.~{Miyasaka}, K.~{Mooley}, S.T.
  {Myers}, E.~{Nakar}, C.C. {Ngeow}, P.~{Nugent}, E.O. {Ofek},
  N.~{Palliyaguru}, M.~{Pavana}, D.A. {Perley}, W.M. {Peters}, S.~{Pike},
  T.~{Piran}, H.~{Qi}, R.M. {Quimby}, J.~{Rana}, S.~{Rosswog}, F.~{Rusu}, E.M.
  {Sadler}, A.~{Van Sistine}, J.~{Sollerman}, Y.~{Xu}, L.~{Yan}, Y.~{Yatsu},
  P.C. {Yu}, C.~{Zhang}, W.~{Zhao}, {GROWTH}, {JAGWAR}, {Caltech-NRAO},
  {TTU-NRAO}, {NuSTAR Collaborations}, K.C. {Chambers}, M.E. {Huber}, A.S.B.
  {Schultz}, J.~{Bulger}, H.~{Flewelling}, E.A. {Magnier}, T.B. {Lowe}, R.J.
  {Wainscoat}, C.~{Waters}, M.~{Willman}, {Pan-STARRS}, K.~{Ebisawa},
  C.~{Hanyu}, S.~{Harita}, T.~{Hashimoto}, K.~{Hidaka}, T.~{Hori},
  M.~{Ishikawa}, N.~{Isobe}, W.~{Iwakiri}, H.~{Kawai}, N.~{Kawai},
  T.~{Kawamuro}, T.~{Kawase}, Y.~{Kitaoka}, K.~{Makishima}, M.~{Matsuoka},
  T.~{Mihara}, T.~{Morita}, K.~{Morita}, S.~{Nakahira}, M.~{Nakajima},
  Y.~{Nakamura}, H.~{Negoro}, S.~{Oda}, A.~{Sakamaki}, R.~{Sasaki},
  M.~{Serino}, M.~{Shidatsu}, R.~{Shimomukai}, Y.~{Sugawara}, S.~{Sugita},
  M.~{Sugizaki}, Y.~{Tachibana}, Y.~{Takao}, A.~{Tanimoto}, H.~{Tomida},
  Y.~{Tsuboi}, H.~{Tsunemi}, Y.~{Ueda}, S.~{Ueno}, S.~{Yamada}, K.~{Yamaoka},
  M.~{Yamauchi}, F.~{Yatabe}, T.~{Yoneyama}, T.~{Yoshii}, {MAXI Team}, D.M.
  {Coward}, H.~{Crisp}, D.~{Macpherson}, I.~{Andreoni}, R.~{Laugier},
  K.~{Noysena}, A.~{Klotz}, B.~{Gendre}, P.~{Thierry}, D.~{Turpin},
  T.~{Consortium}, M.~{Im}, C.~{Choi}, J.~{Kim}, Y.~{Yoon}, G.~{Lim}, S.K.
  {Lee}, C.U. {Lee}, S.L. {Kim}, S.W. {Ko}, J.~{Joe}, M.K. {Kwon}, P.J. {Kim},
  S.K. {Lim}, J.S. {Choi}, {KU Collaboration}, J.P.U. {Fynbo}, D.~{Malesani},
  D.~{Xu}, N.~{Optical Telescope}, S.J. {Smartt}, A.~{Jerkstrand},
  E.~{Kankare}, S.A. {Sim}, M.~{Fraser}, C.~{Inserra}, K.~{Maguire},
  G.~{Leloudas}, M.~{Magee}, L.J. {Shingles}, K.W. {Smith}, D.R. {Young},
  R.~{Kotak}, A.~{Gal-Yam}, J.D. {Lyman}, D.S. {Homan}, C.~{Agliozzo}, J.P.
  {Anderson}, C.R. {Angus}, C.~{Ashall}, C.~{Barbarino}, F.E. {Bauer},
  M.~{Berton}, M.T. {Botticella}, M.~{Bulla}, G.~{Cannizzaro}, R.~{Cartier},
  A.~{Cikota}, P.~{Clark}, A.~{De Cia}, M.~{Della Valle}, M.~{Dennefeld},
  L.~{Dessart}, G.~{Dimitriadis}, N.~{Elias-Rosa}, R.E. {Firth},
  A.~{Fl{\"o}rs}, C.~{Frohmaier}, L.~{Galbany}, S.~{Gonz{\'a}lez-Gait{\'a}n},
  M.~{Gromadzki}, C.P. {Guti{\'e}rrez}, A.~{Hamanowicz}, J.~{Harmanen}, K.E.
  {Heintz}, M.S. {Hernandez}, S.T. {Hodgkin}, I.M. {Hook}, L.~{Izzo}, P.A.
  {James}, P.G. {Jonker}, W.E. {Kerzendorf}, Z.~{Kostrzewa-Rutkowska},
  M.~{Kromer}, H.~{Kuncarayakti}, A.~{Lawrence}, I.~{Manulis}, S.~{Mattila},
  O.~{McBrien}, A.~{M{\"u}ller}, J.~{Nordin}, D.~{O'Neill}, F.~{Onori}, J.T.
  {Palmerio}, A.~{Pastorello}, F.~{Patat}, G.~{Pignata}, P.~{Podsiadlowski},
  A.~{Razza}, T.~{Reynolds}, R.~{Roy}, A.J. {Ruiter}, K.A. {Rybicki},
  L.~{Salmon}, M.L. {Pumo}, S.J. {Prentice}, I.R. {Seitenzahl}, M.~{Smith},
  J.~{Sollerman}, M.~{Sullivan}, H.~{Szegedi}, F.~{Taddia}, S.~{Taubenberger},
  G.~{Terreran}, B.~{Van Soelen}, J.~{Vos}, N.A. {Walton}, D.E. {Wright},
  {\L}.~{Wyrzykowski}, O.~{Yaron}, a.~{pre=''(''>ePESSTO}, T.W. {Chen},
  T.~{Kr{\"u}hler}, P.~{Schady}, P.~{Wiseman}, J.~{Greiner}, A.~{Rau},
  T.~{Schweyer}, S.~{Klose}, A.~{Nicuesa Guelbenzu}, {GROND}, N.T.
  {Palliyaguru}, T.~{Tech University}, M.M. {Shara}, T.~{Williams},
  P.~{Vaisanen}, S.B. {Potter}, E.~{Romero Colmenero}, S.~{Crawford}, D.A.H.
  {Buckley}, J.~{Mao}, {SALT Group}, M.C. {D{\'\i}az}, L.M. {Macri},
  D.~{Garc{\'\i}a Lambas}, C.~{Mendes de Oliveira}, J.L. {Nilo Castell{\'o}n},
  T.~{Ribeiro}, B.~{S{\'a}nchez}, W.~{Schoenell}, L.R. {Abramo}, S.~{Akras},
  J.S. {Alcaniz}, R.~{Artola}, M.~{Beroiz}, S.~{Bonoli}, J.~{Cabral},
  R.~{Camuccio}, V.~{Chavushyan}, P.~{Coelho}, C.~{Colazo}, M.V.
  {Costa-Duarte}, H.~{Cuevas Larenas}, M.~{Dom{\'\i}nguez Romero},
  D.~{Dultzin}, D.~{Fern{\'a}ndez}, J.~{Garc{\'\i}a}, C.~{Girardini}, D.R.
  {Gon{\c{c}}alves}, T.S. {Gon{\c{c}}alves}, S.~{Gurovich},
  Y.~{Jim{\'e}nez-Teja}, A.~{Kanaan}, M.~{Lares}, R.~{Lopes de Oliveira},
  O.~{L{\'o}pez-Cruz}, R.~{Melia}, A.~{Molino}, N.~{Padilla}, T.~{Pe{\~n}uela},
  V.M. {Placco}, C.~{Qui{\~n}ones}, A.~{Ram{\'\i}rez Rivera}, V.~{Renzi},
  L.~{Riguccini}, E.~{R{\'\i}os-L{\'o}pez}, H.~{Rodriguez}, L.~{Sampedro},
  M.~{Schneiter}, L.~{Sodr{\'e}}, M.~{Starck}, S.~{Torres-Flores},
  M.~{Tornatore}, A.~{Zadro{\.z}ny}, M.~{Castillo}, {TOROS: Transient Robotic
  Observatory of South Collaboration}, A.J. {Castro-Tirado}, J.C. {Tello}, Y.D.
  {Hu}, B.B. {Zhang}, R.~{Cunniffe}, A.~{Castell{\'o}n}, D.~{Hiriart}, M.D.
  {Caballero-Garc{\'\i}a}, M.~{Jel{\'\i}nek}, P.~{Kub{\'a}nek}, C.~{P{\'e}rez
  del Pulgar}, I.H. {Park}, S.~{Jeong}, J.M. {Castro Cer{\'o}n}, S.B. {Pandey},
  P.C. {Yock}, R.~{Querel}, Y.~{Fan}, C.~{Wang}, {BOOTES Collaboration},
  A.~{Beardsley}, I.S. {Brown}, B.~{Crosse}, D.~{Emrich}, T.~{Franzen}, B.M.
  {Gaensler}, L.~{Horsley}, M.~{Johnston-Hollitt}, D.~{Kenney}, M.F. {Morales},
  D.~{Pallot}, M.~{Sokolowski}, K.~{Steele}, S.J. {Tingay}, C.M. {Trott},
  M.~{Walker}, R.~{Wayth}, A.~{Williams}, C.~{Wu}, M.~{Murchison Widefield
  Array}, A.~{Yoshida}, T.~{Sakamoto}, Y.~{Kawakubo}, K.~{Yamaoka},
  I.~{Takahashi}, Y.~{Asaoka}, S.~{Ozawa}, S.~{Torii}, Y.~{Shimizu},
  T.~{Tamura}, W.~{Ishizaki}, M.L. {Cherry}, S.~{Ricciarini}, A.V.
  {Penacchioni}, P.S. {Marrocchesi}, {CALET Collaboration}, A.S. {Pozanenko},
  A.A. {Volnova}, E.D. {Mazaeva}, P.Y. {Minaev}, M.A. {Krugov}, A.V. {Kusakin},
  I.V. {Reva}, A.S. {Moskvitin}, V.V. {Rumyantsev}, R.~{Inasaridze}, E.V.
  {Klunko}, N.~{Tungalag}, S.E. {Schmalz}, O.~{Burhonov}, {IKI-GW Follow-up
  Collaboration}, H.~{Abdalla}, A.~{Abramowski}, F.~{Aharonian}, F.~{Ait
  Benkhali}, E.O. {Ang{\"u}ner}, M.~{Arakawa}, M.~{Arrieta}, P.~{Aubert},
  M.~{Backes}, A.~{Balzer}, M.~{Barnard}, Y.~{Becherini}, J.~{Becker Tjus},
  D.~{Berge}, S.~{Bernhard}, K.~{Bernl{\"o}hr}, R.~{Blackwell},
  M.~{B{\"o}ttcher}, C.~{Boisson}, J.~{Bolmont}, S.~{Bonnefoy}, P.~{Bordas},
  J.~{Bregeon}, F.~{Brun}, P.~{Brun}, M.~{Bryan}, M.~{B{\"u}chele}, T.~{Bulik},
  M.~{Capasso}, S.~{Caroff}, A.~{Carosi}, S.~{Casanova}, M.~{Cerruti},
  N.~{Chakraborty}, R.C.G. {Chaves}, A.~{Chen}, J.~{Chevalier},
  S.~{Colafrancesco}, B.~{Condon}, J.~{Conrad}, I.D. {Davids}, J.~{Decock},
  C.~{Deil}, J.~{Devin}, P.~{deWilt}, L.~{Dirson}, A.~{Djannati-Ata{\"\i}},
  A.~{Donath}, L.~{O'C. Drury}, K.~{Dutson}, J.~{Dyks}, T.~{Edwards},
  K.~{Egberts}, G.~{Emery}, J.P. {Ernenwein}, S.~{Eschbach}, C.~{Farnier},
  S.~{Fegan}, M.V. {Fernandes}, A.~{Fiasson}, G.~{Fontaine}, S.~{Funk},
  M.~{F{\"u}ssling}, S.~{Gabici}, Y.A. {Gallant}, T.~{Garrigoux},
  F.~{Gat{\'e}}, G.~{Giavitto}, B.~{Giebels}, D.~{Glawion}, J.F. {Glicenstein},
  D.~{Gottschall}, M.H. {Grondin}, J.~{Hahn}, M.~{Haupt}, J.~{Hawkes},
  G.~{Heinzelmann}, G.~{Henri}, G.~{Hermann}, J.A. {Hinton}, W.~{Hofmann},
  C.~{Hoischen}, T.L. {Holch}, M.~{Holler}, D.~{Horns}, A.~{Ivascenko},
  H.~{Iwasaki}, A.~{Jacholkowska}, M.~{Jamrozy}, D.~{Jankowsky},
  F.~{Jankowsky}, M.~{Jingo}, L.~{Jouvin}, I.~{Jung-Richardt}, M.A.
  {Kastendieck}, K.~{Katarzy{\'n}ski}, M.~{Katsuragawa}, D.~{Kerszberg},
  D.~{Khangulyan}, B.~{Kh{\'e}lifi}, J.~{King}, S.~{Klepser}, D.~{Klochkov},
  W.~{Klu{\'z}niak}, N.~{Komin}, K.~{Kosack}, S.~{Krakau}, M.~{Kraus}, P.P.
  {Kr{\"u}ger}, H.~{Laffon}, G.~{Lamanna}, J.~{Lau}, J.P. {Lees},
  J.~{Lefaucheur}, A.~{Lemi{\`e}re}, M.~{Lemoine-Goumard}, J.P. {Lenain},
  E.~{Leser}, T.~{Lohse}, M.~{Lorentz}, R.~{Liu}, I.~{Lypova}, D.~{Malyshev},
  V.~{Marandon}, A.~{Marcowith}, C.~{Mariaud}, R.~{Marx}, G.~{Maurin},
  N.~{Maxted}, M.~{Mayer}, P.J. {Meintjes}, M.~{Meyer}, A.M.W. {Mitchell},
  R.~{Moderski}, M.~{Mohamed}, L.~{Mohrmann}, K.~{Mor{\r{a}}}, E.~{Moulin},
  T.~{Murach}, S.~{Nakashima}, M.~{de Naurois}, H.~{Ndiyavala},
  F.~{Niederwanger}, J.~{Niemiec}, L.~{Oakes}, P.~{O'Brien}, H.~{Odaka},
  S.~{Ohm}, M.~{Ostrowski}, I.~{Oya}, M.~{Padovani}, M.~{Panter}, R.D.
  {Parsons}, N.W. {Pekeur}, G.~{Pelletier}, C.~{Perennes}, P.O. {Petrucci},
  B.~{Peyaud}, Q.~{Piel}, S.~{Pita}, V.~{Poireau}, H.~{Poon}, D.~{Prokhorov},
  H.~{Prokoph}, G.~{P{\"u}hlhofer}, M.~{Punch}, A.~{Quirrenbach}, S.~{Raab},
  R.~{Rauth}, A.~{Reimer}, O.~{Reimer}, M.~{Renaud}, R.~{de los Reyes},
  F.~{Rieger}, L.~{Rinchiuso}, C.~{Romoli}, G.~{Rowell}, B.~{Rudak}, C.B.
  {Rulten}, V.~{Sahakian}, S.~{Saito}, D.A. {Sanchez}, A.~{Santangelo},
  M.~{Sasaki}, R.~{Schlickeiser}, F.~{Sch{\"u}ssler}, A.~{Schulz},
  U.~{Schwanke}, S.~{Schwemmer}, M.~{Seglar-Arroyo}, M.~{Settimo}, A.S.
  {Seyffert}, N.~{Shafi}, I.~{Shilon}, K.~{Shiningayamwe}, R.~{Simoni},
  H.~{Sol}, F.~{Spanier}, M.~{Spir-Jacob}, {\L}.~{Stawarz}, R.~{Steenkamp},
  C.~{Stegmann}, C.~{Steppa}, I.~{Sushch}, T.~{Takahashi}, J.P. {Tavernet},
  T.~{Tavernier}, A.M. {Taylor}, R.~{Terrier}, L.~{Tibaldo}, D.~{Tiziani},
  M.~{Tluczykont}, C.~{Trichard}, M.~{Tsirou}, N.~{Tsuji}, R.~{Tuffs},
  Y.~{Uchiyama}, D.J. {van der Walt}, C.~{van Eldik}, C.~{van Rensburg},
  B.~{van Soelen}, G.~{Vasileiadis}, J.~{Veh}, C.~{Venter}, A.~{Viana},
  P.~{Vincent}, J.~{Vink}, F.~{Voisin}, H.J. {V{\"o}lk}, T.~{Vuillaume},
  Z.~{Wadiasingh}, S.J. {Wagner}, P.~{Wagner}, R.M. {Wagner}, R.~{White},
  A.~{Wierzcholska}, P.~{Willmann}, A.~{W{\"o}rnlein}, D.~{Wouters}, R.~{Yang},
  D.~{Zaborov}, M.~{Zacharias}, R.~{Zanin}, A.A. {Zdziarski}, A.~{Zech},
  F.~{Zefi}, A.~{Ziegler}, J.~{Zorn}, N.~{{\.Z}ywucka}, {H.~E.~S.~S.
  Collaboration}, R.P. {Fender}, J.W. {Broderick}, A.~{Rowlinson}, R.A.M.J.
  {Wijers}, A.J. {Stewart}, S.~{ter Veen}, A.~{Shulevski}, {LOFAR
  Collaboration}, M.~{Kavic}, J.H. {Simonetti}, C.~{League}, J.~{Tsai}, K.S.
  {Obenberger}, K.~{Nathaniel}, G.B. {Taylor}, J.D. {Dowell}, S.L. {Liebling},
  J.A. {Estes}, M.~{Lippert}, I.~{Sharma}, P.~{Vincent}, B.~{Farella}, L.L.
  {Wavelength Array}, A.U. {Abeysekara}, A.~{Albert}, R.~{Alfaro},
  C.~{Alvarez}, R.~{Arceo}, J.C. {Arteaga-Vel{\'a}zquez}, D.~{Avila Rojas},
  H.A. {Ayala Solares}, A.S. {Barber}, J.~{Becerra Gonzalez}, A.~{Becerril},
  E.~{Belmont-Moreno}, S.Y. {BenZvi}, D.~{Berley}, A.~{Bernal}, J.~{Braun},
  C.~{Brisbois}, K.S. {Caballero-Mora}, T.~{Capistr{\'a}n},
  A.~{Carrami{\~n}ana}, S.~{Casanova}, M.~{Castillo}, U.~{Cotti}, J.~{Cotzomi},
  S.~{Couti{\~n}o de Le{\'o}n}, C.~{De Le{\'o}n}, E.~{De la Fuente}, R.~{Diaz
  Hernandez}, S.~{Dichiara}, B.L. {Dingus}, M.A. {DuVernois}, J.C.
  {D{\'\i}az-V{\'e}lez}, R.W. {Ellsworth}, K.~{Engel},
  O.~{Enr{\'\i}quez-Rivera}, D.W. {Fiorino}, H.~{Fleischhack}, N.~{Fraija},
  J.A. {Garc{\'\i}a-Gonz{\'a}lez}, F.~{Garfias}, M.~{Gerhardt},
  A.~{Gonz{\~o}lez Mu{\~n}oz}, M.M. {Gonz{\'a}lez}, J.A. {Goodman},
  Z.~{Hampel-Arias}, J.P. {Harding}, S.~{Hernandez}, A.~{Hernandez-Almada},
  B.~{Hona}, P.~{H{\"u}ntemeyer}, A.~{Iriarte}, A.~{Jardin-Blicq}, V.~{Joshi},
  S.~{Kaufmann}, D.~{Kieda}, A.~{Lara}, R.J. {Lauer}, D.~{Lennarz},
  H.~{Le{\'o}n Vargas}, J.T. {Linnemann}, A.L. {Longinotti}, G.L. {Raya},
  R.~{Luna-Garc{\'\i}a}, R.~{L{\'o}pez-Coto}, K.~{Malone}, S.S. {Marinelli},
  O.~{Martinez}, I.~{Martinez-Castellanos}, J.~{Mart{\'\i}nez-Castro},
  H.~{Mart{\'\i}nez-Huerta}, J.A. {Matthews}, P.~{Miranda-Romagnoli},
  E.~{Moreno}, M.~{Mostaf{\'a}}, L.~{Nellen}, M.~{Newbold}, M.U. {Nisa},
  R.~{Noriega-Papaqui}, R.~{Pelayo}, J.~{Pretz}, E.G. {P{\'e}rez-P{\'e}rez},
  Z.~{Ren}, C.D. {Rho}, C.~{Rivi{\`e}re}, D.~{Rosa-Gonz{\'a}lez},
  M.~{Rosenberg}, E.~{Ruiz-Velasco}, H.~{Salazar}, F.~{Salesa Greus},
  A.~{Sandoval}, M.~{Schneider}, H.~{Schoorlemmer}, G.~{Sinnis}, A.J. {Smith},
  R.W. {Springer}, P.~{Surajbali}, O.~{Tibolla}, K.~{Tollefson}, I.~{Torres},
  T.N. {Ukwatta}, T.~{Weisgarber}, S.~{Westerhoff}, I.G. {Wisher}, J.~{Wood},
  T.~{Yapici}, G.B. {Yodh}, P.W. {Younk}, H.~{Zhou}, J.D. {{\'A}lvarez}, {HAWC
  Collaboration}, A.~{Aab}, P.~{Abreu}, M.~{Aglietta}, I.F.M. {Albuquerque},
  J.M. {Albury}, I.~{Allekotte}, A.~{Almela}, J.~{Alvarez Castillo},
  J.~{Alvarez-Mu{\~n}iz}, G.A. {Anastasi}, L.~{Anchordoqui}, B.~{Andrada},
  S.~{Andringa}, C.~{Aramo}, N.~{Arsene}, H.~{Asorey}, P.~{Assis}, G.~{Avila},
  A.M. {Badescu}, A.~{Balaceanu}, F.~{Barbato}, R.J. {Barreira Luz}, K.H.
  {Becker}, J.A. {Bellido}, C.~{Berat}, M.E. {Bertaina}, X.~{Bertou}, P.L.
  {Biermann}, J.~{Biteau}, S.G. {Blaess}, A.~{Blanco}, J.~{Blazek}, C.~{Bleve},
  M.~{Boh{\'a}{\v{c}}ov{\'a}}, C.~{Bonifazi}, N.~{Borodai}, A.M. {Botti},
  J.~{Brack}, I.~{Brancus}, T.~{Bretz}, A.~{Bridgeman}, F.L. {Briechle},
  P.~{Buchholz}, A.~{Bueno}, S.~{Buitink}, M.~{Buscemi}, K.S. {Caballero-Mora},
  L.~{Caccianiga}, A.~{Cancio}, F.~{Canfora}, R.~{Caruso}, A.~{Castellina},
  F.~{Catalani}, G.~{Cataldi}, L.~{Cazon}, A.G. {Chavez}, J.A. {Chinellato},
  J.~{Chudoba}, R.W. {Clay}, A.C. {Cobos Cerutti}, R.~{Colalillo},
  A.~{Coleman}, L.~{Collica}, M.R. {Coluccia}, R.~{Concei{\c{c}}{\~a}o},
  G.~{Consolati}, F.~{Contreras}, M.J. {Cooper}, S.~{Coutu}, C.E. {Covault},
  J.~{Cronin}, S.~{D'Amico}, B.~{Daniel}, S.~{Dasso}, K.~{Daumiller}, B.R.
  {Dawson}, J.A. {Day}, R.M. {de Almeida}, S.J. {de Jong}, G.~{De Mauro},
  J.R.T. {de Mello Neto}, I.~{De Mitri}, J.~{de Oliveira}, V.~{de Souza},
  J.~{Debatin}, O.~{Deligny}, M.L. {D{\'\i}az Castro}, F.~{Diogo},
  C.~{Dobrigkeit}, J.C. {D'Olivo}, Q.~{Dorosti}, R.C. {Dos Anjos}, M.T. {Dova},
  A.~{Dundovic}, J.~{Ebr}, R.~{Engel}, M.~{Erdmann}, M.~{Erfani}, C.O.
  {Escobar}, J.~{Espadanal}, A.~{Etchegoyen}, H.~{Falcke}, J.~{Farmer},
  G.~{Farrar}, A.C. {Fauth}, N.~{Fazzini}, F.~{Feldbusch}, F.~{Fenu},
  B.~{Fick}, J.M. {Figueira}, A.~{Filip{\v{c}}i{\v{c}}}, M.M. {Freire},
  T.~{Fujii}, A.~{Fuster}, R.~{Ga{\"\i}or}, B.~{Garc{\'\i}a}, F.~{Gat{\'e}},
  H.~{Gemmeke}, A.~{Gherghel-Lascu}, P.L. {Ghia}, U.~{Giaccari},
  M.~{Giammarchi}, M.~{Giller}, D.~{G{\l}as}, C.~{Glaser}, G.~{Golup},
  M.~{G{\'o}mez Berisso}, P.F. {G{\'o}mez Vitale}, N.~{Gonz{\'a}lez},
  A.~{Gorgi}, M.~{Gottowik}, A.F. {Grillo}, T.D. {Grubb}, F.~{Guarino}, G.P.
  {Guedes}, R.~{Halliday}, M.R. {Hampel}, P.~{Hansen}, D.~{Harari}, T.A.
  {Harrison}, V.M. {Harvey}, A.~{Haungs}, T.~{Hebbeker}, D.~{Heck},
  P.~{Heimann}, A.E. {Herve}, G.C. {Hill}, C.~{Hojvat}, E.~{Holt}, P.~{Homola},
  J.R. {H{\"o}randel}, P.~{Horvath}, M.~{Hrabovsk{\'y}}, T.~{Huege},
  J.~{Hulsman}, A.~{Insolia}, P.G. {Isar}, I.~{Jandt}, J.A. {Johnsen},
  M.~{Josebachuili}, J.~{Jurysek}, A.~{K{\"a}{\"a}p{\"a}}, K.H. {Kampert},
  B.~{Keilhauer}, N.~{Kemmerich}, J.~{Kemp}, R.M. {Kieckhafer}, H.O. {Klages},
  M.~{Kleifges}, J.~{Kleinfeller}, R.~{Krause}, N.~{Krohm}, D.~{Kuempel},
  G.~{Kukec Mezek}, N.~{Kunka}, A.~{Kuotb Awad}, B.L. {Lago}, D.~{LaHurd}, R.G.
  {Lang}, M.~{Lauscher}, R.~{Legumina}, M.A. {Leigui de Oliveira},
  A.~{Letessier-Selvon}, I.~{Lhenry-Yvon}, K.~{Link}, D.~{Lo Presti},
  L.~{Lopes}, R.~{L{\'o}pez}, A.~{L{\'o}pez Casado}, R.~{Lorek}, Q.~{Luce},
  A.~{Lucero}, M.~{Malacari}, M.~{Mallamaci}, D.~{Mandat}, P.~{Mantsch}, A.G.
  {Mariazzi}, I.C. {Maris}, G.~{Marsella}, D.~{Martello}, H.~{Martinez},
  O.~{Mart{\'\i}nez Bravo}, J.J. {Mas{\'\i}as Meza}, H.J. {Mathes},
  S.~{Mathys}, J.~{Matthews}, G.~{Matthiae}, E.~{Mayotte}, P.O. {Mazur},
  C.~{Medina}, G.~{Medina-Tanco}, D.~{Melo}, A.~{Menshikov}, K.D. {Merenda},
  S.~{Michal}, M.I. {Micheletti}, L.~{Middendorf}, L.~{Miramonti},
  B.~{Mitrica}, D.~{Mockler}, S.~{Mollerach}, F.~{Montanet}, C.~{Morello},
  G.~{Morlino}, A.L. {M{\"u}ller}, G.~{M{\"u}ller}, M.A. {Muller},
  S.~{M{\"u}ller}, R.~{Mussa}, I.~{Naranjo}, P.H. {Nguyen},
  M.~{Niculescu-Oglinzanu}, M.~{Niechciol}, L.~{Niemietz}, T.~{Niggemann},
  D.~{Nitz}, D.~{Nosek}, V.~{Novotny}, L.~{No{\v{z}}ka}, L.A. {N{\'u}{\~n}ez},
  F.~{Oikonomou}, A.~{Olinto}, M.~{Palatka}, J.~{Pallotta}, P.~{Papenbreer},
  G.~{Parente}, A.~{Parra}, T.~{Paul}, M.~{Pech}, F.~{Pedreira},
  J.~{P{\c{e}}kala}, J.~{Pe{\~n}a-Rodriguez}, L.A.S. {Pereira}, M.~{Perlin},
  L.~{Perrone}, C.~{Peters}, S.~{Petrera}, J.~{Phuntsok}, T.~{Pierog},
  M.~{Pimenta}, V.~{Pirronello}, M.~{Platino}, M.~{Plum}, J.~{Poh},
  C.~{Porowski}, R.R. {Prado}, P.~{Privitera}, M.~{Prouza}, E.J. {Quel},
  S.~{Querchfeld}, S.~{Quinn}, R.~{Ramos-Pollan}, J.~{Rautenberg},
  D.~{Ravignani}, J.~{Ridky}, F.~{Riehn}, M.~{Risse}, P.~{Ristori}, V.~{Rizi},
  W.~{Rodrigues de Carvalho}, G.~{Rodriguez Fernandez}, J.~{Rodriguez Rojo},
  M.J. {Roncoroni}, M.~{Roth}, E.~{Roulet}, A.C. {Rovero}, P.~{Ruehl}, S.J.
  {Saffi}, A.~{Saftoiu}, F.~{Salamida}, H.~{Salazar}, A.~{Saleh}, G.~{Salina},
  F.~{S{\'a}nchez}, P.~{Sanchez-Lucas}, E.M. {Santos}, E.~{Santos},
  F.~{Sarazin}, R.~{Sarmento}, C.~{Sarmiento-Cano}, R.~{Sato}, M.~{Schauer},
  V.~{Scherini}, H.~{Schieler}, M.~{Schimp}, D.~{Schmidt}, O.~{Scholten},
  P.~{Schov{\'a}nek}, F.G. {Schr{\"o}der}, S.~{Schr{\"o}der}, A.~{Schulz},
  J.~{Schumacher}, S.J. {Sciutto}, A.~{Segreto}, A.~{Shadkam}, R.C. {Shellard},
  G.~{Sigl}, G.~{Silli}, R.~{{\v{S}}m{\'\i}da}, G.R. {Snow}, P.~{Sommers},
  S.~{Sonntag}, J.F. {Soriano}, R.~{Squartini}, D.~{Stanca}, S.~{Stani{\v{c}}},
  J.~{Stasielak}, P.~{Stassi}, M.~{Stolpovskiy}, F.~{Strafella}, A.~{Streich},
  F.~{Suarez}, M.~{Suarez-Dur{\'a}n}, T.~{Sudholz}, T.~{Suomij{\"a}rvi}, A.D.
  {Supanitsky}, J.~{{\v{S}}up{\'\i}k}, J.~{Swain}, Z.~{Szadkowski},
  A.~{Taboada}, O.A. {Taborda}, C.~{Timmermans}, C.J. {Todero Peixoto},
  L.~{Tomankova}, B.~{Tom{\'e}}, G.~{Torralba Elipe}, P.~{Travnicek},
  M.~{Trini}, M.~{Tueros}, R.~{Ulrich}, M.~{Unger}, M.~{Urban}, J.F.
  {Vald{\'e}s Galicia}, I.~{Vali{\~n}o}, L.~{Valore}, G.~{van Aar}, P.~{van
  Bodegom}, A.M. {van den Berg}, A.~{van Vliet}, E.~{Varela}, B.~{Vargas
  C{\'a}rdenas}, R.A. {V{\'a}zquez}, D.~{Veberi{\v{c}}}, C.~{Ventura}, I.D.
  {Vergara Quispe}, V.~{Verzi}, J.~{Vicha}, L.~{Villase{\~n}or}, S.~{Vorobiov},
  H.~{Wahlberg}, O.~{Wainberg}, D.~{Walz}, A.A. {Watson}, M.~{Weber},
  A.~{Weindl}, M.~{Wiede{\'n}ski}, L.~{Wiencke}, H.~{Wilczy{\'n}ski},
  M.~{Wirtz}, D.~{Wittkowski}, B.~{Wundheiler}, L.~{Yang}, A.~{Yushkov},
  E.~{Zas}, D.~{Zavrtanik}, M.~{Zavrtanik}, A.~{Zepeda}, B.~{Zimmermann},
  M.~{Ziolkowski}, Z.~{Zong}, F.~{Zuccarello}, {Pierre Auger Collaboration},
  S.~{Kim}, S.~{Schulze}, F.E. {Bauer}, J.M. {Corral-Santana}, I.~{de
  Gregorio-Monsalvo}, J.~{Gonz{\'a}lez-L{\'o}pez}, D.H. {Hartmann}, C.H.
  {Ishwara-Chandra}, S.~{Mart{\'\i}n}, A.~{Mehner}, K.~{Misra}, M.J.
  {Micha{\l}owski}, L.~{Resmi}, {ALMA Collaboration}, Z.~{Paragi}, I.~{Agudo},
  T.~{An}, R.~{Beswick}, C.~{Casadio}, S.~{Frey}, P.~{Jonker}, M.~{Kettenis},
  B.~{Marcote}, J.~{Moldon}, A.~{Szomoru}, H.J. {van Langevelde}, J.~{Yang},
  {Euro VLBI Team}, A.~{Cwiek}, M.~{Cwiok}, H.~{Czyrkowski}, R.~{Dabrowski},
  G.~{Kasprowicz}, L.~{Mankiewicz}, K.~{Nawrocki}, R.~{Opiela}, L.W.
  {Piotrowski}, G.~{Wrochna}, M.~{Zaremba}, A.F. {{\.Z}arnecki}, {Pi of Sky
  Collaboration}, D.~{Haggard}, M.~{Nynka}, J.J. {Ruan}, {Chandra Team at
  McGill University}, P.A. {Bland}, T.~{Booler}, H.A.R. {Devillepoix}, J.S. {de
  Gois}, P.J. {Hancock}, R.M. {Howie}, J.~{Paxman}, E.K. {Sansom}, M.C.
  {Towner}, D.~{Desert Fireball Network}, J.~{Tonry}, M.~{Coughlin}, C.W.
  {Stubbs}, L.~{Denneau}, A.~{Heinze}, B.~{Stalder}, H.~{Weiland}, {ATLAS},
  R.P. {Eatough}, M.~{Kramer}, A.~{Kraus}, H.~{Time Resolution Universe
  Survey}, E.~{Troja}, L.~{Piro}, J.~{Becerra Gonz{\'a}lez}, N.R. {Butler},
  O.D. {Fox}, H.G. {Khandrika}, A.~{Kutyrev}, W.H. {Lee}, R.~{Ricci},
  J.~{Ryan}, R.~E., R.~{S{\'a}nchez-Ram{\'\i}rez}, S.~{Veilleux}, A.M.
  {Watson}, M.H. {Wieringa}, J.M. {Burgess}, H.~{van Eerten}, C.J. {Fontes},
  C.L. {Fryer}, O.~{Korobkin}, R.T. {Wollaeger}, {RIMAS}, {RATIR}, F.~{Camilo},
  A.R. {Foley}, S.~{Goedhart}, S.~{Makhathini}, N.~{Oozeer}, O.M. {Smirnov},
  R.P. {Fender}, P.A. {Woudt}, S.~{South Africa/MeerKAT}, The Astrophysical
  Journal Letters \textbf{848}(2), L12 (2017).
\newblock \doi{10.3847/2041-8213/aa91c9}

\bibitem{yuan22018}
W.~{Yuan}, C.~{Zhang}, Z.~{Ling}, D.~{Zhao}, W.~{Wang}, Y.~{Chen}, F.~{Lu},
  S.N. {Zhang}, W.~{Cui}, in \emph{Space Telescopes and Instrumentation 2018:
  Ultraviolet to Gamma Ray}, \emph{Society of Photo-Optical Instrumentation
  Engineers (SPIE) Conference Series}, vol. 10699, ed. by J.W.A. {den Herder},
  S.~{Nikzad}, K.~{Nakazawa} (2018), \emph{Society of Photo-Optical
  Instrumentation Engineers (SPIE) Conference Series}, vol. 10699, p. 1069925.
\newblock \doi{10.1117/12.2313358}

\bibitem{Angel1979}
J.R.P. {Angel}, The Astrophysical Journal \textbf{233}, 364 (1979).
\newblock \doi{10.1086/157397}

\bibitem{hudec2015}
R.~{Hudec}, L.~{Pina}, A.~{Inneman}, V.~{Tichy}, in \emph{EUV and X-ray Optics:
  Synergy between Laboratory and Space IV}, \emph{Society of Photo-Optical
  Instrumentation Engineers (SPIE) Conference Series}, vol. 9510, ed. by
  R.~{Hudec}, L.~{Pina} (2015), \emph{Society of Photo-Optical Instrumentation
  Engineers (SPIE) Conference Series}, vol. 9510, p. 95100A.
\newblock \doi{10.1117/12.2177792}

\bibitem{yuan2016}
W.~{Yuan}, L.~{Amati}, J.K. {Cannizzo}, B.~{Cordier}, N.~{Gehrels},
  G.~{Ghirlanda}, D.~{G{\"o}tz}, N.~{Produit}, Y.~{Qiu}, J.~{Sun}, N.R.
  {Tanvir}, J.~{Wei}, C.~{Zhang}, in \emph{Gamma-Ray Bursts. Series: Space
  Sciences Series of ISSI}, vol.~61 (2016), pp. 237--279.
\newblock \doi{10.1007/978-94-024-1279-6_10}

\bibitem{zhao2014}
D.~{Zhao}, C.~{Zhang}, W.~{Yuan}, R.~{Willingale}, Z.~{Ling}, H.~{Feng},
  H.~{Li}, J.~{Ji}, W.~{Wang}, S.~{Zhang}, in \emph{Space Telescopes and
  Instrumentation 2014: Ultraviolet to Gamma Ray}, \emph{Society of
  Photo-Optical Instrumentation Engineers (SPIE) Conference Series}, vol. 9144,
  ed. by T.~{Takahashi}, J.W.A. {den Herder}, M.~{Bautz} (2014), \emph{Society
  of Photo-Optical Instrumentation Engineers (SPIE) Conference Series}, vol.
  9144, p. 91444E.
\newblock \doi{10.1117/12.2055434}

\bibitem{zhao2017}
D.~{Zhao}, C.~{Zhang}, W.~{Yuan}, S.~{Zhang}, R.~{Willingale}, Z.~{Ling},
  Experimental Astronomy \textbf{43}(3), 267 (2017).
\newblock \doi{10.1007/s10686-017-9534-5}

\bibitem{zhang2021}
T.~Zhang, F.~Yi, B.~Wang, J.~Liu, X.~Liang, M.~Yan, Y.~Wang, Z.~Zhao, S.N.
  Zhang, Microsystem Technologies \textbf{27}(5), 1895 (2021)

\bibitem{Li2021}
Z.~{Li}, H.~{Feng}, X.~{Huang}, J.~{Zhang}, J.~{Su}, X.~{Cai}, Z.~{Feng},
  Q.~{Liu}, H.~{Liu}, C.~{Gao}, L.~{Xiao}, X.~{Sun}, Nuclear Instruments and
  Methods in Physics Research A \textbf{1008}, 165430 (2021).
\newblock \doi{10.1016/j.nima.2021.165430}

\bibitem{Huang2021}
X.F. Huang, H.B. Liu, J.~Zhang, B.~Huang, W.J. Xie, H.B. Feng, X.C. Cai, X.W.
  Liu, Z.L. Li, J.Y. Gu, et~al., Nuclear Science and Techniques \textbf{32}(7),
  1 (2021)

\bibitem{Costa2001}
E.~{Costa}, P.~{Soffitta}, R.~{Bellazzini}, A.~{Brez}, N.~{Lumb}, G.~{Spandre},
  Nature \textbf{411}(6838), 662 (2001)

\bibitem{Feng2019}
H.~{Feng}, W.~{Jiang}, M.~{Minuti}, Q.~{Wu}, A.~{Jung}, D.~{Yang},
  S.~{Citraro}, H.~{Nasimi}, J.~{Yu}, G.~{Jin}, J.~{Huang}, M.~{Zeng}, P.~{An},
  L.~{Baldini}, R.~{Bellazzini}, A.~{Brez}, L.~{Latronico}, C.~{Sgr{\`o}},
  G.~{Spandre}, M.~{Pinchera}, F.~{Muleri}, P.~{Soffitta}, E.~{Costa},
  Experimental Astronomy \textbf{47}(1-2), 225 (2019).
\newblock \doi{10.1007/s10686-019-09625-z}

\bibitem{Soffitta2021}
P.~{Soffitta}, L.~{Baldini}, R.~{Bellazzini}, E.~{Costa}, L.~{Latronico},
  F.~{Muleri}, E.~{Del Monte}, S.~{Fabiani}, M.~{Minuti}, M.~{Pinchera},
  C.~{Sgro'}, G.~{Spandre}, A.~{Trois}, F.~{Amici}, H.~{Andersson},
  P.~{Attina'}, M.~{Bachetti}, M.~{Barbanera}, F.~{Borotto}, A.~{Brez},
  D.~{Brienza}, C.~{Caporale}, C.~{Cardelli}, R.~{Carpentiero},
  S.~{Castellano}, M.~{Castronuovo}, L.~{Cavalli}, E.~{Cavazzuti},
  M.~{Ceccanti}, M.~{Centrone}, S.~{Ciprini}, S.~{Citraro}, F.~{D'Amico},
  E.~{D'Alba}, S.~{Di Cosimo}, N.~{Di Lalla}, A.~{Di Marco}, G.~{Di Persio},
  I.~{Donnarumma}, Y.~{Evangelista}, R.~{Ferrazzoli}, A.~{Hayato},
  T.~{Kitaguchi}, F.~{La Monaca}, C.~{Lefevre}, P.~{Loffredo}, P.~{Lorenzi},
  L.~{Lucchesi}, C.~{Magazzu}, S.~{Maldera}, A.~{Manfreda}, E.~{Mangraviti},
  M.~{Marengo}, G.~{Matt}, P.~{Mereu}, A.~{Morbidini}, F.~{Mosti}, T.~{Nakano},
  H.~{Nasimi}, B.~{Negri}, S.~{Nenonen}, A.~{Nuti}, L.~{Orsini}, M.~{Perri},
  M.~{Pesce-Rollins}, R.~{Piazzolla}, M.~{Pilia}, A.~{Profeti}, S.~{Puccetti},
  J.~{Rankin}, A.~{Ratheesh}, A.~{Rubini}, F.~{Santoli}, P.~{Sarra},
  E.~{Scalise}, A.~{Sciortino}, T.~{Tamagawa}, M.~{Tardiola}, A.~{Tobia},
  M.~{Vimercati}, F.~{Xie}, The Astronomical Journal \textbf{162}(5), 208
  (2021).
\newblock \doi{10.3847/1538-3881/ac19b0}

\bibitem{Mar2011}
J.R. Mart{\'\i}nez, \emph{The adventures of Luis Alvarez: Identity politics in
  the making of an American science} (The University of Texas at Austin, 2011)

\bibitem{Sajid2015}
M.~{Sajid}, N.G. {Chechenin}, F.S. {Torres}, E.U. {Khan}, S.~{Agha}, Advances
  in Space Research \textbf{56}(2), 314 (2015).
\newblock \doi{10.1016/j.asr.2015.04.011}

\bibitem{2020SPIE}
J.~{{\v{R}}{\'\i}pa}, G.~{Dilillo}, R.~{Campana}, G.~{Galg{\'o}czi}, in
  \emph{Society of Photo-Optical Instrumentation Engineers (SPIE) Conference
  Series}, \emph{Society of Photo-Optical Instrumentation Engineers (SPIE)
  Conference Series}, vol. 11444 (2020), \emph{Society of Photo-Optical
  Instrumentation Engineers (SPIE) Conference Series}, vol. 11444, p. 114443P.
\newblock \doi{10.1117/12.2561011}

\bibitem{Zhao2019}
H.~Zhao, Y.~Huang, Astronomical Research \& Technology \textbf{16}(2), 244
  (2019)

\bibitem{Graham2019}
M.J. {Graham}, S.R. {Kulkarni}, E.C. {Bellm}, S.M. {Adams}, C.~{Barbarino},
  N.~{Blagorodnova}, D.~{Bodewits}, B.~{Bolin}, P.R. {Brady}, S.B. {Cenko},
  C.K. {Chang}, M.W. {Coughlin}, K.~{De}, G.~{Eadie}, T.L. {Farnham},
  U.~{Feindt}, A.~{Franckowiak}, C.~{Fremling}, S.~{Gezari}, S.~{Ghosh}, D.A.
  {Goldstein}, V.Z. {Golkhou}, A.~{Goobar}, A.Y.Q. {Ho}, D.~{Huppenkothen},
  {\v{Z}}.~{Ivezi{\'c}}, R.L. {Jones}, M.~{Juric}, D.L. {Kaplan}, M.M.
  {Kasliwal}, M.S.P. {Kelley}, T.~{Kupfer}, C.D. {Lee}, H.W. {Lin},
  R.~{Lunnan}, A.A. {Mahabal}, A.A. {Miller}, C.C. {Ngeow}, P.~{Nugent}, E.O.
  {Ofek}, T.A. {Prince}, L.~{Rauch}, J.~{van Roestel}, S.~{Schulze}, L.P.
  {Singer}, J.~{Sollerman}, F.~{Taddia}, L.~{Yan}, Q.Z. {Ye}, P.C. {Yu},
  T.~{Barlow}, J.~{Bauer}, R.~{Beck}, J.~{Belicki}, R.~{Biswas}, V.~{Brinnel},
  T.~{Brooke}, B.~{Bue}, M.~{Bulla}, R.~{Burruss}, A.~{Connolly}, J.~{Cromer},
  V.~{Cunningham}, R.~{Dekany}, A.~{Delacroix}, V.~{Desai}, D.A. {Duev},
  M.~{Feeney}, D.~{Flynn}, S.~{Frederick}, A.~{Gal-Yam}, M.~{Giomi},
  S.~{Groom}, E.~{Hacopians}, D.~{Hale}, G.~{Helou}, J.~{Henning}, D.~{Hover},
  L.A. {Hillenbrand}, J.~{Howell}, T.~{Hung}, D.~{Imel}, W.H. {Ip},
  E.~{Jackson}, S.~{Kaspi}, S.~{Kaye}, M.~{Kowalski}, E.~{Kramer}, M.~{Kuhn},
  W.~{Landry}, R.R. {Laher}, P.~{Mao}, F.J. {Masci}, S.~{Monkewitz},
  P.~{Murphy}, J.~{Nordin}, M.T. {Patterson}, B.~{Penprase}, M.~{Porter},
  U.~{Rebbapragada}, D.~{Reiley}, R.~{Riddle}, M.~{Rigault}, H.~{Rodriguez},
  B.~{Rusholme}, J.~{van Santen}, D.L. {Shupe}, R.M. {Smith}, M.T. {Soumagnac},
  R.~{Stein}, J.~{Surace}, P.~{Szkody}, S.~{Terek}, A.~{Van Sistine}, S.~{van
  Velzen}, W.T. {Vestrand}, R.~{Walters}, C.~{Ward}, C.~{Zhang}, J.~{Zolkower},
  Publications of the Astronomical Society of the Pacific \textbf{131}(1001),
  078001 (2019).
\newblock \doi{10.1088/1538-3873/ab006c}

\bibitem{Boller2016}
T.~{Boller}, M.J. {Freyberg}, J.~{Tr{\"u}mper}, F.~{Haberl}, W.~{Voges},
  K.~{Nandra}, Astronomy and Astrophysics \textbf{588}, A103 (2016).
\newblock \doi{10.1051/0004-6361/201525648}

\bibitem{Gezari2021}
S.~{Gezari}, Annual Review of Astronomy and Astrophysics \textbf{59}, 21
  (2021).
\newblock \doi{10.1146/annurev-astro-111720-030029}

\bibitem{Sazonov2021}
S.~{Sazonov}, M.~{Gilfanov}, P.~{Medvedev}, Y.~{Yao}, G.~{Khorunzhev},
  A.~{Semena}, R.~{Sunyaev}, R.~{Burenin}, A.~{Lyapin}, A.~{Meshcheryakov},
  G.~{Uskov}, I.~{Zaznobin}, K.A. {Postnov}, A.V. {Dodin}, A.A. {Belinski},
  A.M. {Cherepashchuk}, M.~{Eselevich}, S.N. {Dodonov}, A.A. {Grokhovskaya},
  S.S. {Kotov}, I.F. {Bikmaev}, R.Y. {Zhuchkov}, R.I. {Gumerov}, S.~{van
  Velzen}, S.~{Kulkarni}, Monthly Notices of the Royal Astronomical Society
  \textbf{508}(3), 3820 (2021).
\newblock \doi{10.1093/mnras/stab2843}

\end{thebibliography}

\end{document}